%% file: Puzia.tex
\documentclass{aa}
\usepackage{graphicx}
\usepackage[authoryear]{natbib}
\usepackage{psfig}

\bibpunct[, ]{(}{)}{;}{a}{}{,}

\begin{document}

\title{VLT Spectroscopy of Globular Cluster Systems\thanks{Based on
observations collected at the European Southern Observatory, Cerro
Paranal, Chile under programme ID P65.N-0281, P66.B-0068, and
P67.B-0034.}}

\subtitle{I. The Photometric and Spectroscopic Data Set.}

\author{Thomas H. Puzia \inst{1}, Markus Kissler-Patig \inst{2},
Daniel Thomas\inst{3}, Claudia Maraston \inst{3}, Roberto P. Saglia
\inst{1}, Ralf Bender \inst{1,3}, Tom Richtler \inst{4}, 
Paul Goudfrooij \inst{5}, \& Maren Hempel \inst{2}}
      
   \offprints{Thomas H. Puzia, \email{puzia@usm.uni-muenchen.de}}

   \institute{Sternwarte der Ludwig-Maximilians-Universit\"at,
        Scheinerstr. 1, 81679 M\"unchen, Germany, \\
        \email{puzia, saglia@usm.uni-muenchen.de}
      \and
      European Southern Observatory, 85749 Garching bei M\"unchen,
        Germany, \\\email{mkissler@eso.org}
      \and
      Max-Planck-Institut f\"ur extraterrestrische Physik,
        Giessenbachstrasse, 85748 Garching bei M\"unchen, Germany,
        \\\email{bender, maraston, dthomas@mpe.mpg.de}
      \and
      Grupo de Astronom\'{\i}a, Departamento de F\'{\i}sica, Casilla 160-C,
        Universidad de Concepci\'{o}n, Concepci\'{o}n, Chile,
        \\\email{tom@coma.cfm.udec.cl}
      \and 
      Space Telescope Science Institute, 3700 San Martin Drive,
        Baltimore, MD 21218, USA, \\\email{goudfroo@stsci.edu}
        }

   \authorrunning{Puzia et al.}  
   \titlerunning{VLT Spectroscopy of Globular Cluster Systems} 
      
   \date{Received ...; accepted ... }
   
   \abstract{We present Lick line-index measurements of extragalactic
   globular clusters in seven early-type galaxies (NGC~1380, 2434,
   3115, 3379, 3585, 5846, and 7192) with different morphological
   types (E--S0) located in field and group/cluster
   environments. High-quality spectra were taken with the FORS2
   instrument at ESO's Very Large Telescope. $\sim50$\% of our data
   allows an age resolution $\Delta t/t\approx0.3$ and a metallicity
   resolution $\sim0.25-0.4$ dex, depending on the absolute
   metallicity. Globular cluster candidates are selected from deep $B,
   V, R, I, K$ FORS2/ISAAC photometry with $80-100$\% success rate
   inside one effective radius. Using combined optical/near-infrared
   colour-colour diagrams we present a method to efficiently reduce
   fore-/background contamination down to $\la10$\%. We find clear
   signs for bi-modality in the globular cluster colour distributions
   of NGC~1380, 3115, and 3585. The colour distributions of globular
   clusters in NGC~2434, 3379, 5846, and 7192 are consistent with a
   broad single-peak distribution. For the analysed globular cluster
   systems the slopes of projected radial surface density profiles, of
   the form $\Sigma(R)\sim R\,^{-\Gamma}$, vary between $\sim 0.8$ and
   2.6. Blue and red globular cluster sub-populations show similar
   slopes in the clearly bi-modal systems. For galaxies with
   single-peak globular cluster colour distributions, there is a hint
   that the blue cluster system seems to have a more extended radial
   distribution than the red one. Using globular clusters as a tracer
   population we determine total dynamical masses of host galaxies out
   to large radii ($\sim 1.6 - 4.8\, R_{\rm eff}$). For the sample we
   find masses in the range $\sim8.8\cdot10^{10}M_\odot$ up to
   $\sim1.2\cdot10^{12} M_\odot$. The line index data presented here
   will be used in accompanying papers of this series to derive ages,
   metallicities and abundance ratios. A compilation of currently
   available high-quality Lick index measurements for globular
   clusters in elliptical, lenticular, and late-type galaxies is
   provided and will serve to augment the current data set.}
   
   \maketitle


\section{Introduction}
\label{ln:intro}

Compared to their host galaxies globular clusters are remarkably
simple stellar structures. They form throughout the lifetime of the
universe and are witnesses of major star-formation episodes
\citep[e.g.][]{ashman98, kissler-patig00, harris01}. As such, their
ages, metallicities, and chemical compositions can provide detailed
insights in the formation epochs, processes, and timescales which lead
to the assembly of the galaxies we observe in the local universe.

Photometry is one way to assess ages and metallicities of
extragalactic globular clusters \citep[among many
others][]{schweizer96, whitmore97, kissler-patig97, puzia99,
maraston01, jordan02, kissler-patig02, puzia02a, hempel02}. However,
the age-metallicity degeneracy of photometric colours hampers the
detailed reconstruction of star-formation histories
\citep[e.g.][]{faber72, oconnell76}.

Spectroscopy is an independent alternative to determine star-formation
histories and the basic chemistry of globular clusters. The Lick
system of absorption line indices \citep{burstein84, faber85,
gonzalez93, worthey94, worthey97, trager98}, although not free from
age-metallicity degeneracy, is a way to measure absorption features
which are sensitive to age and metallicity. In combination with
state-of-the-art simple stellar population (SSP) models that take the
effect of element abundance-ratio variations into account
\citep[e.g.][]{maraston03, thomas03} indices can shed light on
star-formation timescales and the chemical composition. With today's
8--10m class telescopes the mean Balmer-line index uncertainty for
individual extragalactic globular clusters can be reduced to values of
the order of the mean isochrone separation in SSP models ($\sim 0.1$
\AA\ between 12 and 13 Gyr). Data of this high-accuracy become
available, for the first time, and is in principle capable of
resolving star-formation histories even for very old stellar
populations.

Previous spectroscopy of globular cluster systems in early-type
galaxies aiming at the derivation of ages and metallicities of single
clusters was performed for M~87 \citep{cohen98}, NGC~1023
\citep{larsen02a}, NGC~1316 \citep{goudfrooij01}, NGC~1399
\citep{kissler-patig98, forbes01}, NGC~3115 \citep{kuntschner02},
NGC~3610 \citep{strader02}, NGC~4365 \citep{larsen03}, NGC~4472
\citep{beasley00}, and NGC~4594 \citep{larsen02b}. However, the data
quality allowed only in a few cases to determine the ages and
metallicities of {\it individual} globular clusters. The data of most
studies required summing the spectra of all or at least a given
sub-population of clusters to obtain meaningful results. Moreover, the
different choices of diagnostic plots (such as H$\beta$
vs. $\langle$Fe$\rangle$ or H$\gamma$ vs. Mg$b$) made the comparison
between galaxies difficult. The existence of SSP models with {\it
well-defined} abundance ratios only recently allows to account for
varying abundance ratios, such as [$\alpha$/Fe]. Inconsistent use of
index passband definitions between data and models introduced
additional uncertainties.

In this paper, we present photometry and Lick line-index measurements
from our on-going spectroscopic survey of globular cluster systems in
early-type galaxies. These high-quality spectroscopic data will be
used in subsequent papers of this series (Puzia et al. 2003, in
preparation) to derive accurate ages, metallicities, and [$\alpha$/Fe]
ratios in a self-consistent fashion. The sample presented here
includes photometric and spectroscopic data for 143 extragalactic
globular clusters.

The present paper is structured as follows: \S\ref{ln:photometry}
describes the photometric data set and the candidate
selection. \S\ref{ln:data} presents the spectroscopic sample, data
reduction, and the selection of candidates. Radial velocity
measurements, success rates of the candidate selection, and host
galaxy masses are discussed in \S\ref{ln:kin}. Sampled luminosities
are discussed in \S\ref{ln:lineindices}, followed by the description
of Lick-index measurements. A compilation of previously published Lick
indices for globular clusters in elliptical, lenticular, and late-type
galaxies other than studied here is given in \S\ref{ln:litdata}. The
work is summarised in \S\ref{ln:summary}.

\begin{table*}[ht!]
\centering
\caption[width=\textwidth]{Basic information on host galaxies. The
  references are: (1) \cite{RC3}; (2) NED$^{\rm a}$; (3) \cite{schlegel98}; (4)
  \cite{buta95}; (5) \cite{pahre99}; (6) \cite{tonry01}; (7)
\cite{tully88}; (8) \cite{elroy95}; (9) \cite{kissler-patig97}; (10)
\cite{ashman98}. }
\label{tab:galdat}
\begin{tabular}{l r r r r r r r l}
\hline\hline
\noalign{\smallskip}
Parameter          &NGC~1380      &NGC~2434      &NGC~3115      &NGC~3379      &   
                    NGC~3585      &NGC~5846      &NGC~7192      & Ref. \\
\noalign{\smallskip}
\hline
\noalign{\smallskip}
type               &  $-2$/LA     &  $-5$/E0+    &  $-3$/L$-$   &  $-5$/E1     &
                      $-5$/E6     &  $-5$/E0     &  $-4.3$/E+   &(1) \\
RA  (J2000)        &   03 36 27   &   07 34 51   &   10 05 14   &   10 47 50   &
                       11 13 17   &   15 06 29   &   22 06 50   &(2) \\
DEC (J2000)        &$-$34 58 34   &$-$69 17 01   &$-$07 43 07   &$+$12 34 55   &
                    $-$26 45 18   &$+$01 36 21   &$-$64 18 57   &(2)\\
v$_{\rm rad}$      &$1841\pm15$   &$1390\pm27$   & $670\pm12$   & $889\pm12$   &
                    $1399\pm27$   &$1710\pm12$   &$2897\pm32$   &(1)\\
E$_{B-V}$          & 0.017        & 0.248        & 0.047        & 0.024        &
                     0.064        & 0.055        & 0.034        &(3)\\
$(B-V)_{\rm eff,o}$& 0.92         & 1.09         & 0.94         & 0.98         &
                     0.99         & 1.03         & 0.97         &(1)\\
$(V-I)_{\rm eff,o}$& 1.21         & 1.42         & 1.25         & 1.24         &
                     1.26         & 1.28         & 1.24         &(4)\\
$(V-K)_{\rm eff,o}$& 3.36         & 3.10         & \dots        & 3.08         &
                     \dots        & 3.12         & \dots        &(5)\\
$(m-M)_V$          &31.23$\pm0.18$&31.67$\pm0.29$&29.93$\pm0.09$&30.12$\pm0.11$&
                    31.51$\pm0.18$&31.98$\pm0.20$&32.89$\pm0.32$&(6)\\ 
$M_B$              &$-20.04$      &$-19.48$      &$-19.19$      &$-19.39$      &
                    $-20.93$      &$-21.16$      &$-20.55$      &(7)\\
$a/b^{\rm b}$      & 0.56         & 0.94         & 0.49         & 0.93	       &
                     0.58         & 0.89         & 1.00         &(7)\\
$\sigma^{\rm c}$   & 225          & 204          & 264          & 209          &
                     218          & 252          & 184          &(8)\\
$\rho_{\rm xyz}^{\rm d}$
                   & 1.54         & 0.19         & 0.08         & 0.52         &
                     0.12         & 0.84         & 0.28         &(7)\\
$N_{\rm GC}^{\rm e}$
                   & $560\pm30$   & \dots        &$520\pm120$   &$300\pm160$   &
                     \dots        &$2200\pm1300$ & \dots        &(9),(10)\\
$S_{\rm N}^{\rm f}$& $1.5\pm0.5$  & \dots        &$1.6\pm0.4$   &$1.2\pm0.6$   &
                     \dots        &$3.5\pm2.1$   & \dots        &(9),(10)\\
\noalign{\smallskip}
\hline
\end{tabular}
\begin{list}{}{}
\item[$^{\mathrm{a}}$] http://nedwww.ipac.caltech.edu
\item[$^{\mathrm{b}}$] Ratio of semi-minor/semi-major axis
\item[$^{\mathrm{c}}$] Central velocity dispersion in km s$^{-1}$
\item[$^{\mathrm{d}}$] Environmental density of galaxies brighter than
$M_{\rm B}=-16$ in galaxies/megaparsec$^3$ \citep{tully88}
\item[$^{\mathrm{e}}$] Total number of globular clusters
\item[$^{\mathrm{f}}$] Specific frequency, $S_{\rm N}=N_{\rm GC}\cdot
10^{\, 0.4\cdot(M_V + 15)}$ \citep{harris81}
\end{list}
\end{table*}

\section{Pre-Imaging Data}
\label{ln:photometry}
The host galaxies (NGC~1380, NGC~2434, NGC~3115, NGC~3379, NGC~3585,
NGC~5846, and NGC~7192) were selected to sample a significant range in
environmental density at intermediate galaxy luminosity and velocity
dispersion in the range $-19.2\ga M_B \ga -21.2$ and $184 \la\sigma\la
264$ km s$^{-1}$, respectively. All galaxies are of early type
($T<-2$) according to the RC3 galaxy catalog \citep{RC3}. Our sample
includes five elliptical and two lenticular galaxies (see
Tab.~\ref{tab:galdat}). We used Tully's $\rho_{\rm xyz}$ parameter
\citep{tully88} to parameterize the environmental density per Mpc$^3$
of galaxies which are brighter than $M_{\rm B}=-16$ to separate field
from group/cluster environment. In this work, we define galaxies with
$\rho_{\rm xyz} < 0.5$ as field objects and galaxies with $\rho_{\rm
xyz} > 0.5$ as group/cluster members. Three galaxies of our sample
(NGC~1380, NGC~3379, and NGC~5846) are assigned group/cluster
membership, while the remaining four galaxies are considered field
members. Among other relevant parameters, $\rho_{\rm xyz}$ and M$_B$
are summarized in Table \ref{tab:galdat} for all our sample galaxies.

The imaging mode of FORS2 at ESO's Very Large Telescope was used to
obtain pre-imaging data for each galaxy in multiple filters to select
candidate globular clusters for spectroscopic follow-up. Exposure
times in each filter are summarized in Table
\ref{tab:jourphot}. Standard calibration routines in IRAF were applied
to bias and flatfield the images. Galaxy light was subtracted by,
first, removing stellar objects from the image by using SExtractor
\citep[v2.1.6][]{bertin96} and, second, smoothing the residual image
with a large median filter. The median filtered image was subsequently
subtracted from the original image. This procedure was iterated with a
smaller median filter to discard weak haloes around objects on steep
galaxy-light slopes near the central regions. SExtractor was used to
perform photometry in a 6-pixel-diameter aperture which was found to
yield the highest signal to noise of measured magnitudes. The residual
flux which falls outside the 6-pixel aperture was measured in a
growth-curve analysis for a handful of objects in each filter for each
single galaxy. Uncertainties for the aperture correction were found to
be of the order $\sim0.01$ mag. All instrumental magnitudes were
subsequently corrected with these corrections found.

Each data set was calibrated using standard-star observations for each
night provided by the quality control group of ESO. All observations
were performed under photometric conditions and could be calibrated to
an average intrinsic accuracy of $\sim0.03$ mag.

We augment our optical photometric data with the recently published
near-infrared data for NGC~3115 \citep{puzia02a} and for NGC~5846 and
NGC~7192 \citep{hempel02}.

\begin{table}[h!]
\centering
\caption[width=\textwidth]{Journal of photometric
observations. Exposure times are given in seconds. }
\label{tab:jourphot}
\begin{tabular}{l c c c c c }
\hline\hline
\noalign{\smallskip}
Galaxy  &    B  &    V &    R &   I &   K  \\
\noalign{\smallskip}
\hline
\noalign{\smallskip}
NGC 1380 &\dots &  700 &\dots & 700 & \dots \\
NGC 2434 &\dots &  700 &\dots & 700 & \dots \\
NGC 3115 &  160 &  300 &  160 & 300 & 15500$^{\mathrm{a}}$ \\
NGC 3379 &\dots &  300 &\dots & 300 & \dots \\
NGC 3585 &  800 &\dots &\dots & 800 & \dots \\
NGC 5846 &  900 &  300 &  160 & 300 & 10000$^{\mathrm{b}}$ \\
NGC 7192 &  900 &  600 &  900 & 600 & 12000$^{\mathrm{b}}$ \\
\noalign{\smallskip}
\hline
\end{tabular}
\begin{list}{}{}
\item[$^{\mathrm{a}}$] data were taken from \cite{puzia02a}.
\item[$^{\mathrm{b}}$] data were taken from \cite{hempel02}.
\end{list}
\end{table}

\subsection{Consistency Check with WFPC2 Photometry}

We use WFPC2/HST archive data which were obtained from ST-ECF in
Garching to check for consistency of our photometry with that of
WFPC2. For the sake of homogeneity, we use the pipeline-processed,
co-added (averaged), and cosmic-cleaned image cubes provided as
association files by the archive. Photometry was performed in the
standard \cite{holtzman95a} 0.5\arcsec\ radius aperture using
SExtractor and corrected for the y-CTE ramp as described in
\cite{holtzman95b}. We transform the WFPC2 filters F450W, F555W,
F702W, and F814W to Johnson-Cousins filters B, V, R, and I,
respectively, using the prescriptions in \cite{holtzman95a}.

At the distance of the two nearest galaxies in our sample, NGC~3115
and NGC~3379 (see Tab.~\ref{tab:galdat}), globular clusters are
resolved by HST. Therefore their photometry needs an additional
zero-point correction since the standard aperture corrections for
stellar profiles \citep[see][]{holtzman95a} do not apply. At the
distance of NGC 3115 a globular cluster with a typical half-light
radius of 3 pc appears with $\sim$0.06\arcsec\ on the chip and will be
resolved by the planetary camera (0.0455 \arcsec/pix). However, such
objects on the wide-field chips (0.0996 \arcsec/pix) are on the edge
of being resolved. $\sim90$\% of Milky Way globular clusters have
half-light radii smaller than 3 pc \citep{harris96} and their
counterparts in NGC~3115 and NGC~3379 are expected to have similar
size distributions \citep[e.g.][]{kundu98, larsen01}. Even if most
comparison objects are globular clusters with half-light radii $\sim3$
pc, we do {\it not} expect the aperture corrections to be larger than
the total uncertainty of the FORS2 and WFPC2 photometric
calibration. For the remaining sample galaxies most globular clusters
are not resolved by WFPC2. Hence, we do not apply any aperture
corrections to the WFPC2 photometry.

We find good agreement between the two photometric data sets with
offsets $\la |0.08|$ mag (see Table \ref{tab:offsets}) which were
calculated in the sense $\Delta m= m_{\rm FORS2} - m_{\rm WFPC2}$. On
average, the offsets are small and of the order of their uncertainties
showing no systematics with galaxy distance, background level,
etc. Hence, we do {\it not} apply these corrections to our FORS2
data.

\begin{table}[ht!]
\centering
\caption[width=\textwidth]{Photometric offsets between FORS2 and WFPC2
data. Offsets are defined in the sense $\Delta m= m_{\rm FORS2} -
m_{\rm WFPC2}$. The given uncertainties are the errors of the
mean. The last column shows the number of objects from which the
photometric offsets were calculated.}
\label{tab:offsets}
\begin{tabular}{l c c c c r}
\hline\hline
\noalign{\smallskip}
Galaxy  &$\Delta$B &$\Delta$V &$\Delta$R &$\Delta$I & $N_{\rm obj}$\\
\noalign{\smallskip}
\hline
\noalign{\smallskip}
NGC 1380 &\dots &$-0.035$  &\dots &\dots  & 142\\
         &      &$\pm0.018$&      &       & \\
NGC 2434 &\dots &$-0.012$  &\dots &$-0.044$ & 113\\
         &      &$\pm0.021$&      &$\pm0.022$ & \\
NGC 3115 &\dots &$+0.078$  &\dots &$+0.043$ & 79\\
         &      &$\pm0.017$&      &$\pm0.017$ &\\
NGC 3379 &\dots &$-0.005$  &\dots &$+0.079$ & 46\\
         &      &$\pm0.020$&      &$\pm0.030$ &\\
NGC 3585 &\dots &\dots     &\dots &$-0.080$ & 89\\
         &      &          &      &$\pm0.025$ &\\
NGC 5846 &\dots &$+0.011$  &$+0.066$  &$-0.055$ & 89\\
         &      &$\pm0.023$&$\pm0.039$&$\pm0.019$ & \\ 
NGC 7192 &$-0.049$  &$+0.004$  &\dots&$-0.035$ & 74\\
         &$\pm0.033$&$\pm0.025$&     &$\pm0.022$ &\\
\noalign{\smallskip}
\hline
\end{tabular}
\end{table}

\subsection{Colour-Magnitude Diagrams}
\label{ln:cmd}
In the following two Sections we present the photometric selection of
globular cluster candidates for follow-up spectroscopy. For this
purpose we use colour-magnitude and colour-colour diagrams.

\begin{figure*}[!ht]
\centering 
    \includegraphics[width=8.5cm]{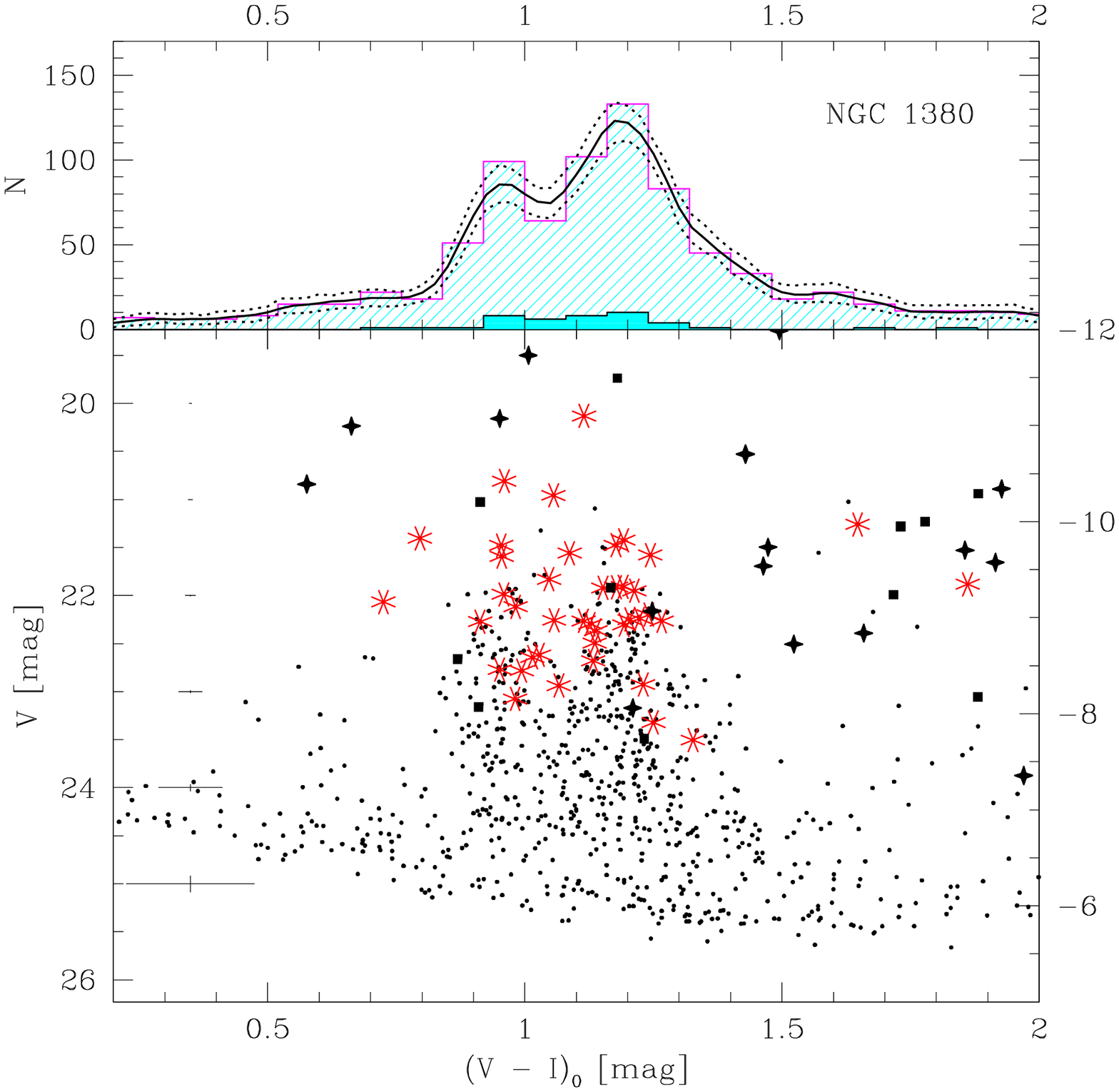}
    \includegraphics[width=8.5cm]{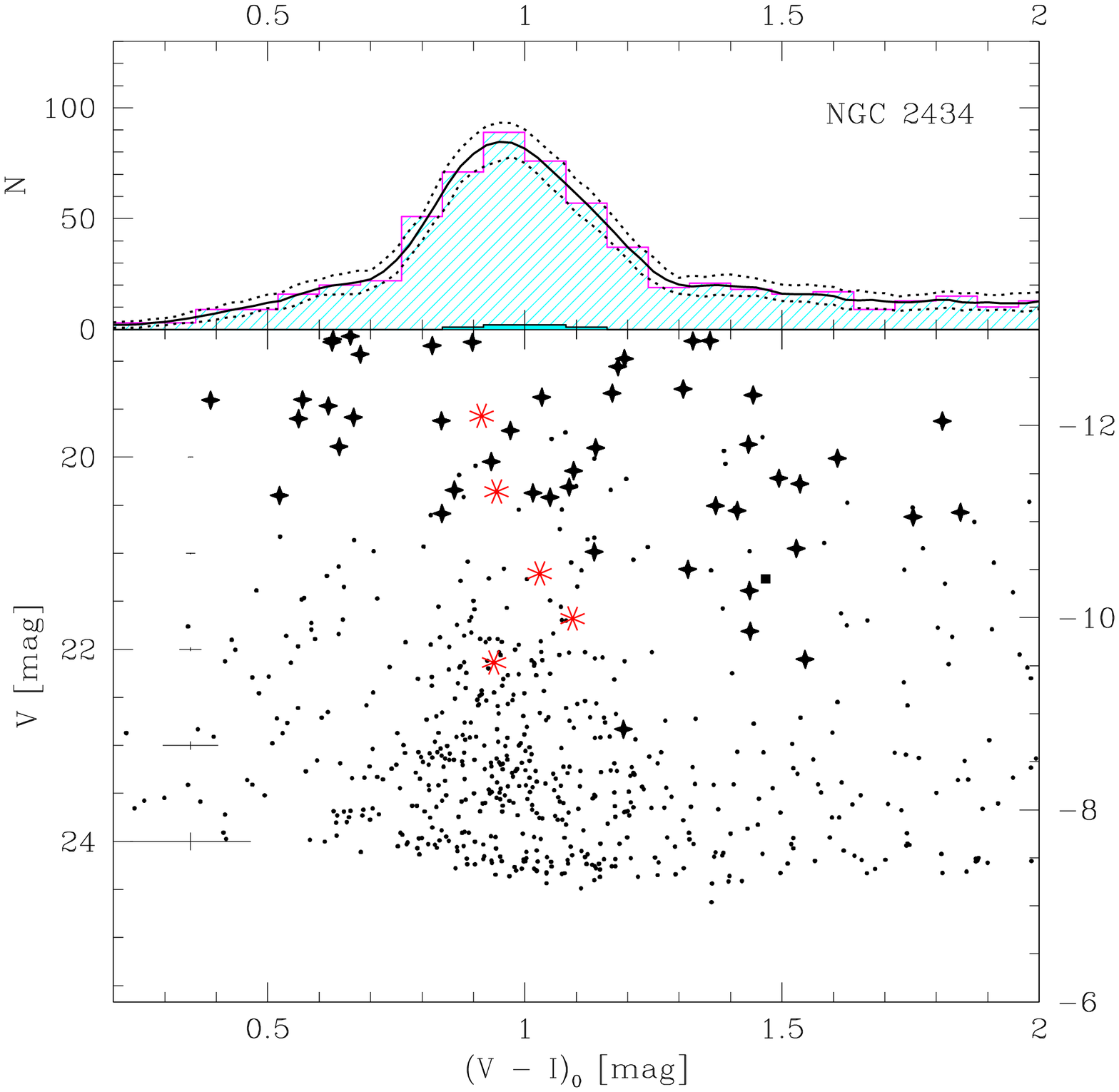}
    \includegraphics[width=8.5cm]{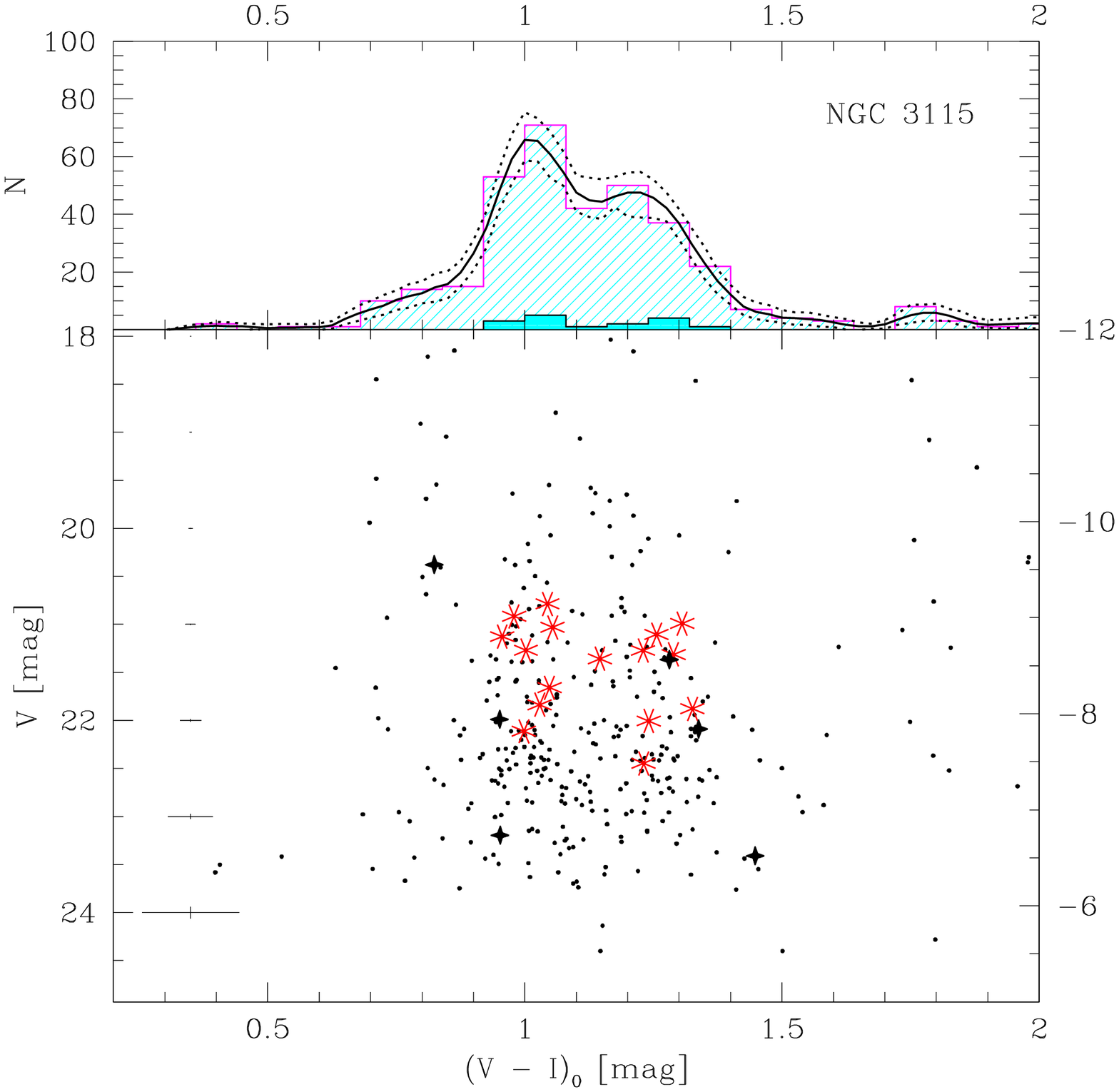}
    \includegraphics[width=8.5cm]{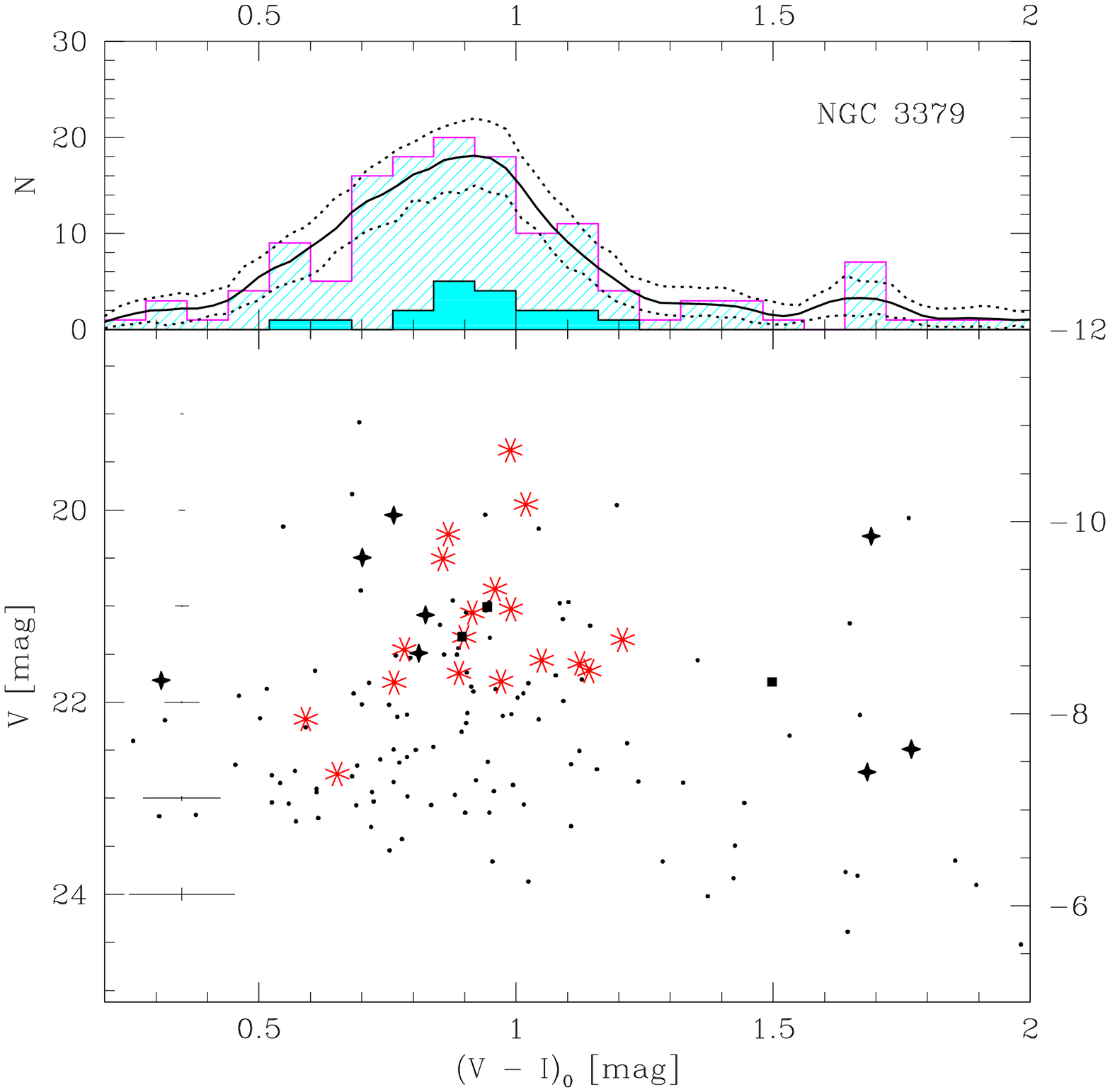}

    \caption{$V$ vs. $V-I$ CMDs for globular cluster systems of
    NGC~1380, NGC~2434, NGC~3115, and NGC~3379. Asterisks indicate
    spectroscopically confirmed globular clusters; 4-prong stars and
    filled squares show objects whose redshifts are consistent with
    foreground stars and background galaxies, respectively. Left
    ordinates show apparent magnitudes, right ordinates indicate
    absolute magnitudes calculated using distance moduli from
    Table~\ref{tab:galdat}. Each panel shows average photometric error
    bars near the left ordinate. The upper sub-panels show histograms
    of the colour distributions. Hatched histograms were created from
    the entire photometric data, solid histograms show colour
    distributions of spectroscopically confirmed globular
    clusters. The solid and dotted lines are probability density
    estimates with their bootstraped 90\% confidence limits \citep[for
    details see][]{silverman86}. The bin size of the histograms was
    adjusted to 0.08 mag which roughly corresponds to the mean
    photometric error.}
\label{ps:cmd}
\end{figure*}
\addtocounter{figure}{-1}
\begin{figure*}[!ht]
\centering 
    \includegraphics[width=8.5cm]{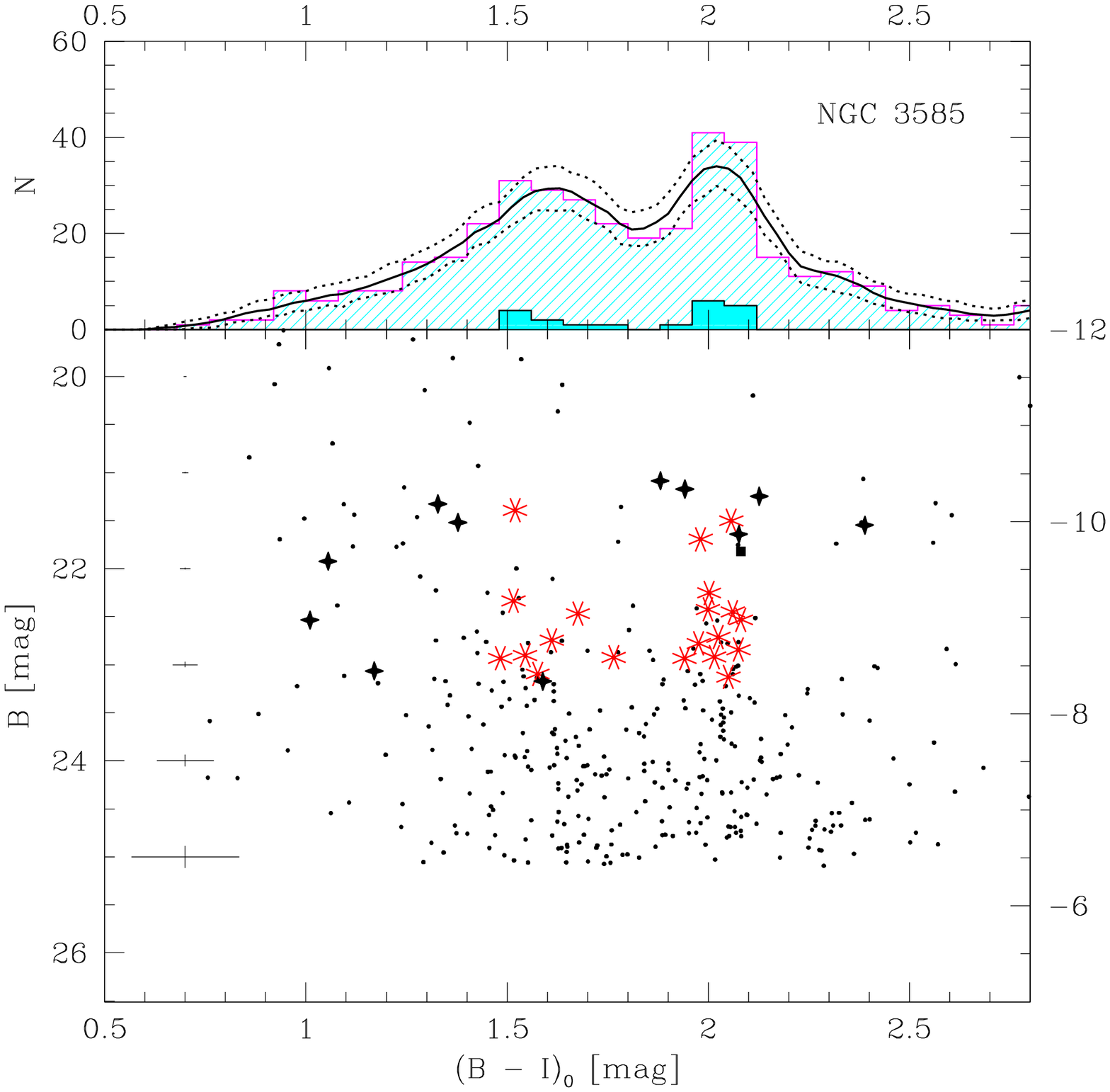}
    \includegraphics[width=8.5cm]{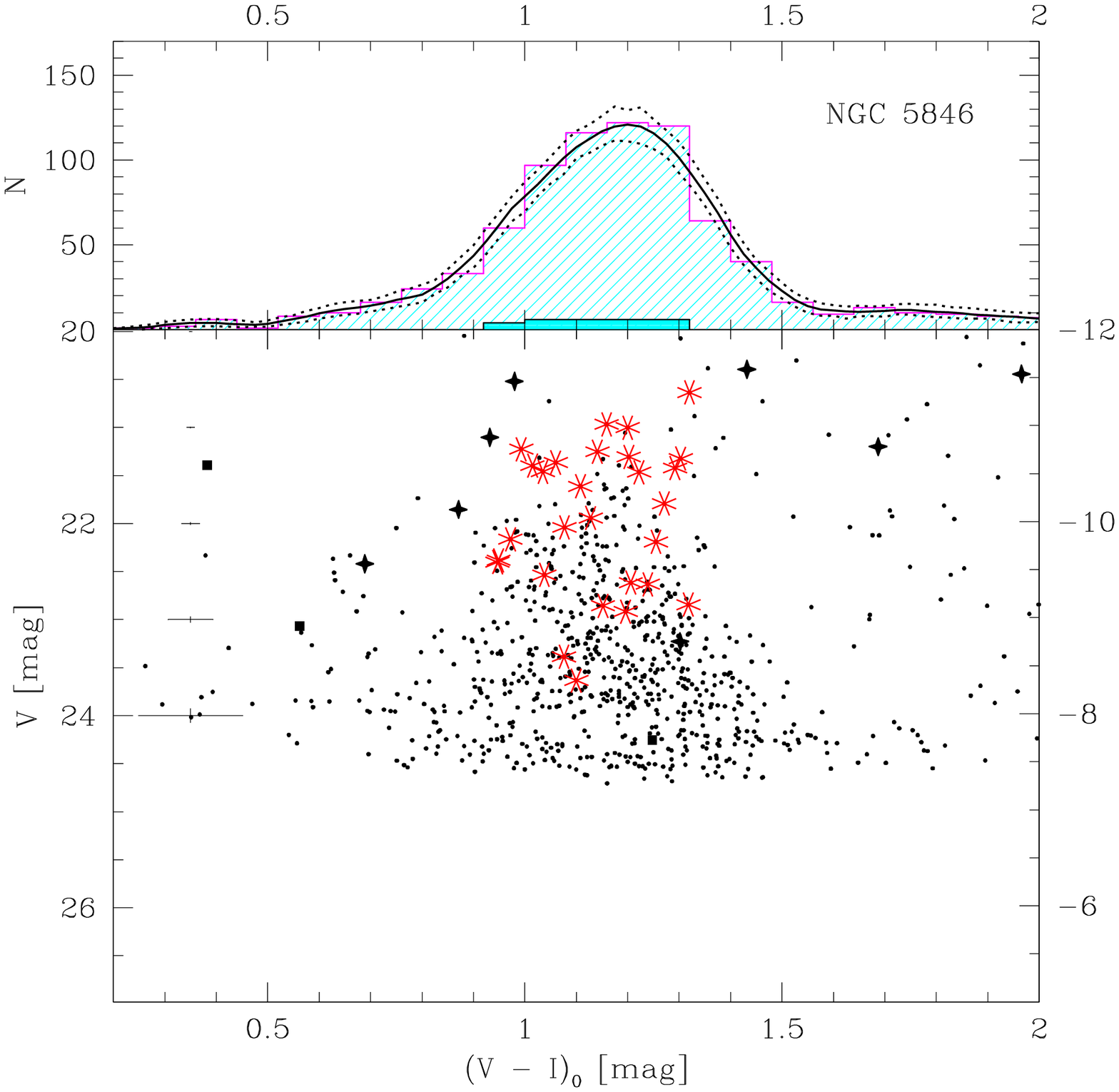}
    \includegraphics[width=8.5cm]{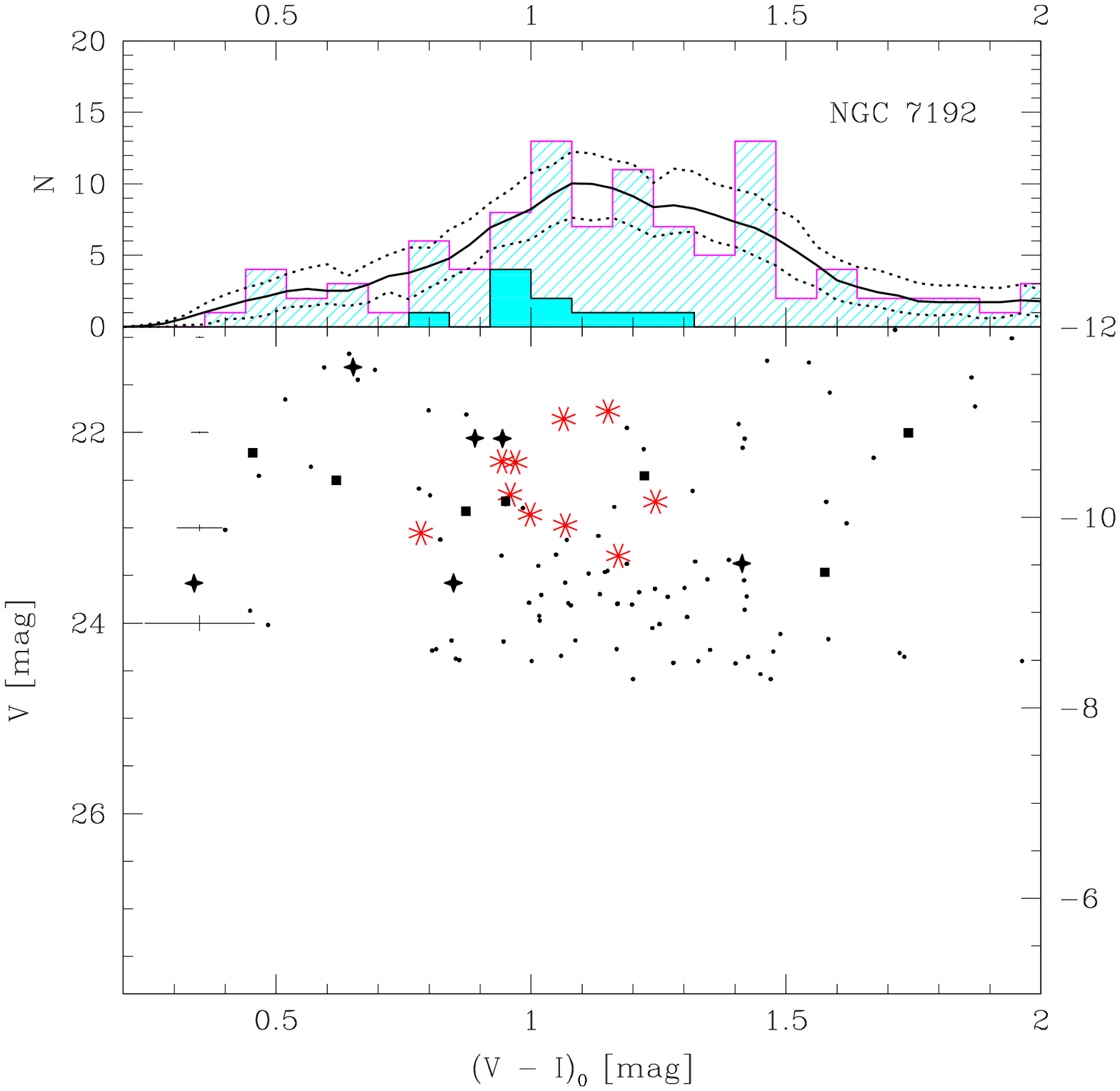}
    \includegraphics[width=8.5cm]{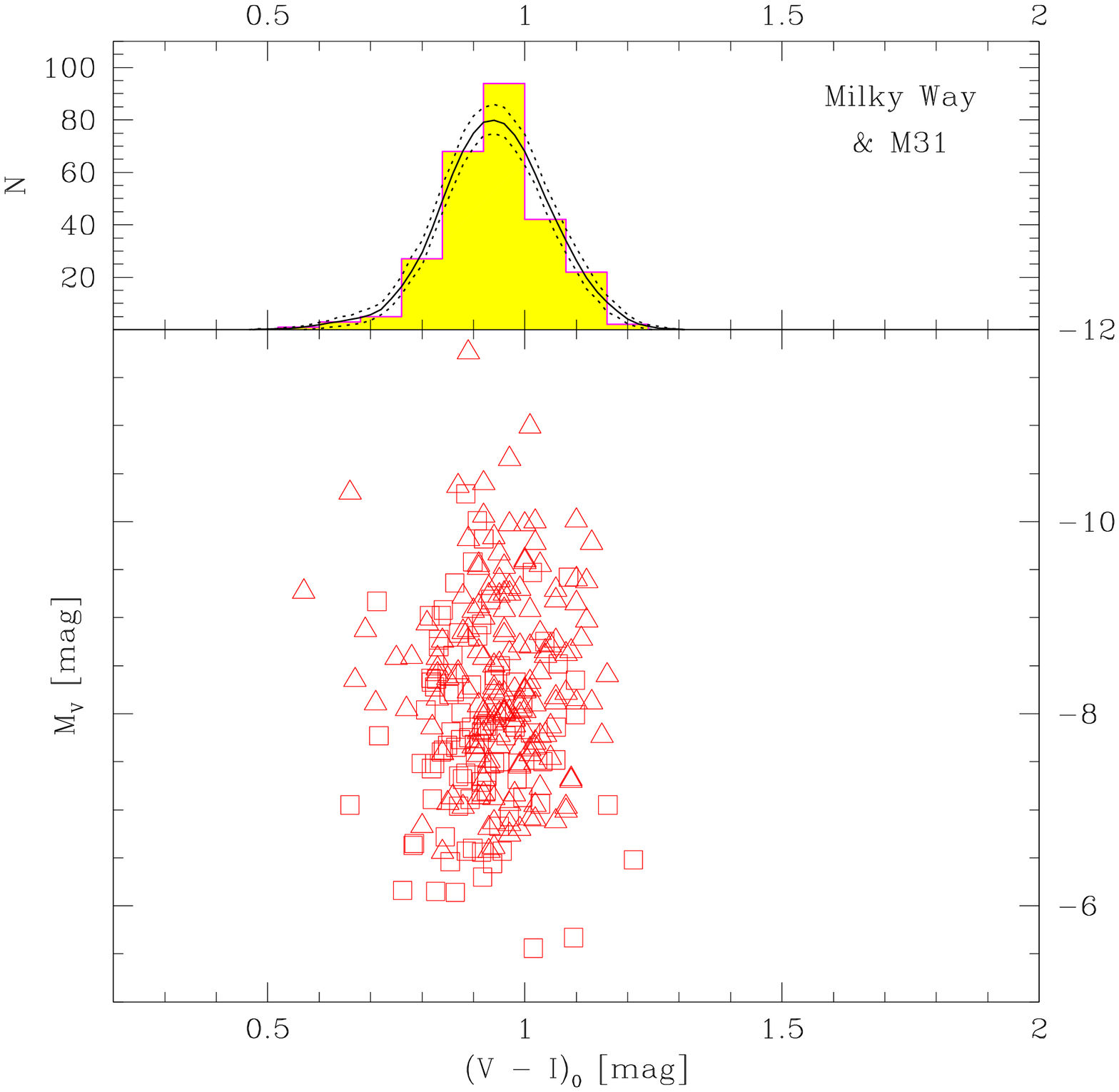}

    \caption{-- {\it continued}. CMDs for NGC~3585, NGC~5846 and
    NGC~7192. Please, note that due to the lack of $V$-band data, we
    plot $B$ vs. $B-I$ for NGC~3585. For comparison, the CMD for
    globular clusters in the Milky Way ({\it squares}) and M31 ({\it
    triangles}) is plotted in the lower right panel. The Milky Way
    data were taken from the 1999 update of the McMaster catalog
    \citep{harris96}, while M31 data are from \cite{barmby00}.}
\end{figure*}

Figure~\ref{ps:cmd} shows colour-magnitude diagrams (CMDs) for our
sample globular cluster systems using the FORS2 pre-imaging
data. Small dots indicate the entire data extracted from our
pre-imaging fields. Asterisks, squares, and 4-prong stars show
magnitudes and colours of spectroscopically confirmed globular
clusters (see Sect.~\ref{ln:rv}), background galaxies, and foreground
stars, respectively.

Globular cluster candidates were pre-selected by their photometric
error ($\Delta m\leq 0.2$ mag), FWHM ($\leq1.5\,
\langle$FWHM$\rangle$), PSF ellipticity ($\epsilon \leq 0.6$), and the
SExtractor star/galaxy classifier ($>0.0$, i.e. only clearly extended
sources were rejected). Due to the good spatial coverage of the FORS2
field of view, each sample is the most comprehensive compilation of
candidate globular cluster colours so far. We find clear bi-modalities
in NGC~1380, NGC~3115, and NGC~3585 (note that we lack $V$ band
photometry for NGC~3585 and use $B$ magnitudes instead). Weak
indications for possible multi-modality are found in each of the
former three colour distributions. To test these distributions for
{\it bi}-modality, we apply the KMM algorithm \citep{ashman94} to the
constrained samples with a colour range $0.8<V-I<1.4$ for NGC~1380,
$0.9<V-I<1.4$ for NGC~3115, and $1.2<B-I<2.4$ for NGC~3585. The code
yields peaks at $V-I=0.94\pm0.01$ and $1.20\pm0.01$ mag for NGC~1380
(with a number ratio blue/red$=0.46$), $V-I=1.01\pm0.01$ and
$1.21\pm0.01$ mag for NGC~3115 (blue/red$=0.87$), and
$B-I=1.56\pm0.01$ and $2.05\pm0.01$ for NGC~3585 (blue/red$=1.02$).

The colour histograms for NGC~2434, NGC~3379, NGC~5846, and NGC~7192
are consistent with single-peak distributions. However, it is
interesting to note that these peaks are systematically broader than
colour peaks of sub-groups in clearly multi-modal distributions. One
reason for a broad single-peak colour distribution (in the optical)
might be that the gap between two old metal-poor and metal-rich
globular cluster populations is filled by metal-rich intermediate-age
clusters. A large spread in age and metallicity in the underlying
globular cluster system would naturally produce a sequence in colours
rather than multiple distinct peaks. In the context of the
hierarchical galaxy formation scenario, {\it clearly} bi-modal
globular cluster colour distributions may have to be considered as
special cases of a wide range of colour distribution morphologies.

Indeed, using a combination of optical and near-infrared photometry it
was recently shown that NGC~4365 hosts an intermediate-age globular
cluster sub-population which produces a single-peak $V-I$ colour
distribution \citep{puzia02a}. \cite{hempel02} find intermediate-age
globular clusters in NGC~5846 and NGC~7192, although the case of
NGC~7192 is less conclusive. For details on near-infrared-optical
colours of globular clusters in NGC~3115, NGC~5846, NGC~7192 we refer
the reader to \cite{puzia02a} and \cite{hempel02}.

Most of the colour distributions are consistent with previous
colour-distribution studies \citep{gebhardt99, larsen01, kundu01a,
kundu01b}, but only where the latter had large enough sample size. Our
photometry goes deep (reaching the GCLF turn-over in most cases, see
Fig.~\ref{ps:vhisto}) and our field sampling
(6.7\arcmin$\times6.7$\arcmin ) is large enough to cover a
representative fraction ($\geq51$\%\footnote{The fraction was
determined with the surface density profiles found in
Sect.~\ref{ln:radprof}.})  of the observed globular cluster
system. For instance, the $V-I$ colour distribution of NGC~3379 gained
a substantial amount of blue globular clusters ($\sim15$\% of the
entire population down to $V=23.5$) which have not been included in
previous HST/WFPC2 studies \citep[e.g.][]{larsen01}. This is likely to
be due to a significant difference in spatial distribution of red and
blue globular clusters in this galaxy. There is evidence that red
clusters are more concentrated towards the center than the blue
globular cluster sub-population which rather resides in the halo (see
Sect.~\ref{ln:radprof}). This illustrates that colour distributions
which were created from photometric data of limited field size
(e.g. HST/WFPC2) might be misleading if significant differences in
spatial distributions of globular cluster sub-populations are present.

\subsection{Optical/Near-Infrared Colour-Colour Diagrams}
\label{ln:ccm}

\begin{figure*}[!ht]
\centering 
          \includegraphics[width=8.5cm]{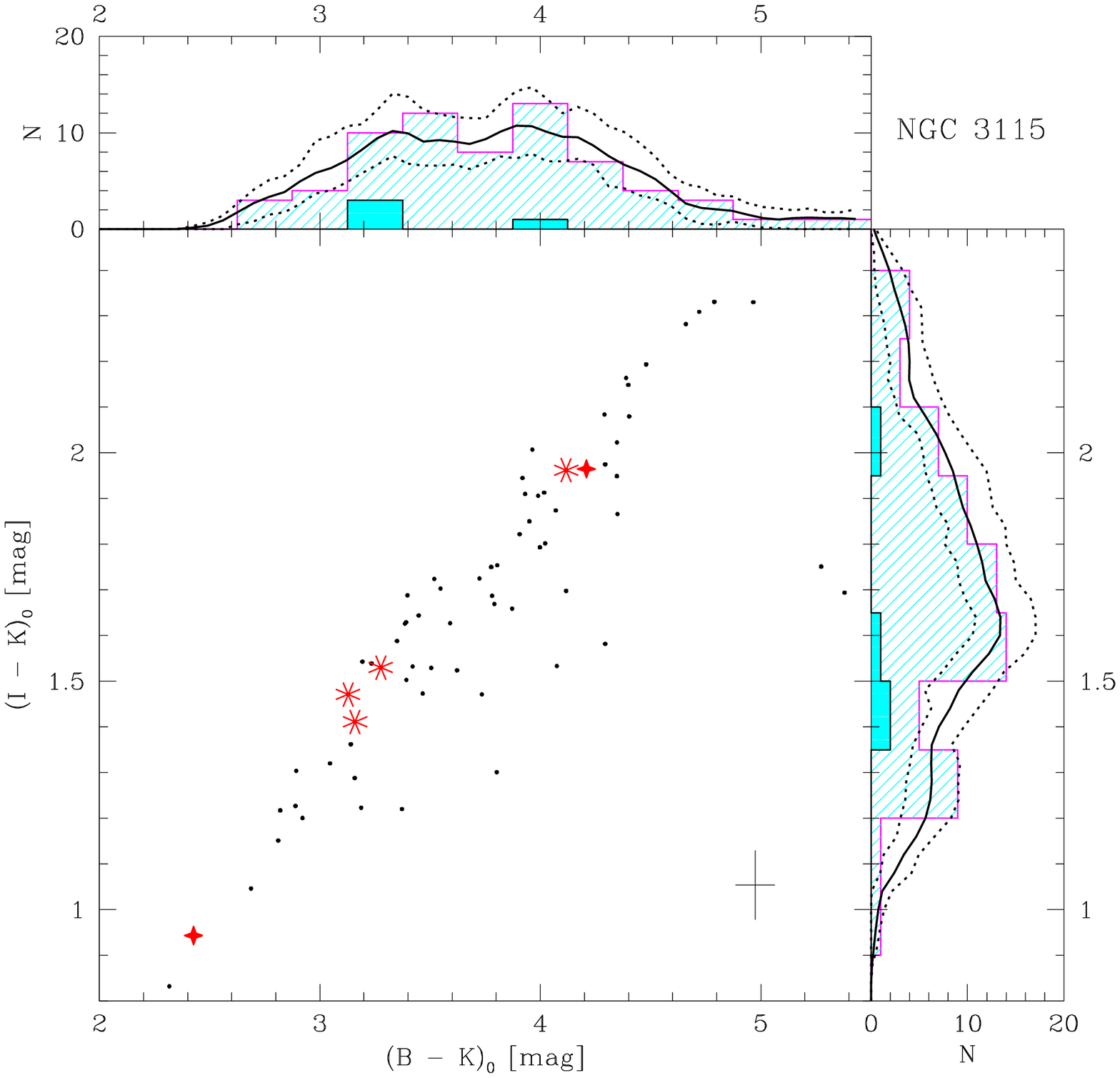}
	\includegraphics[width=8.5cm]{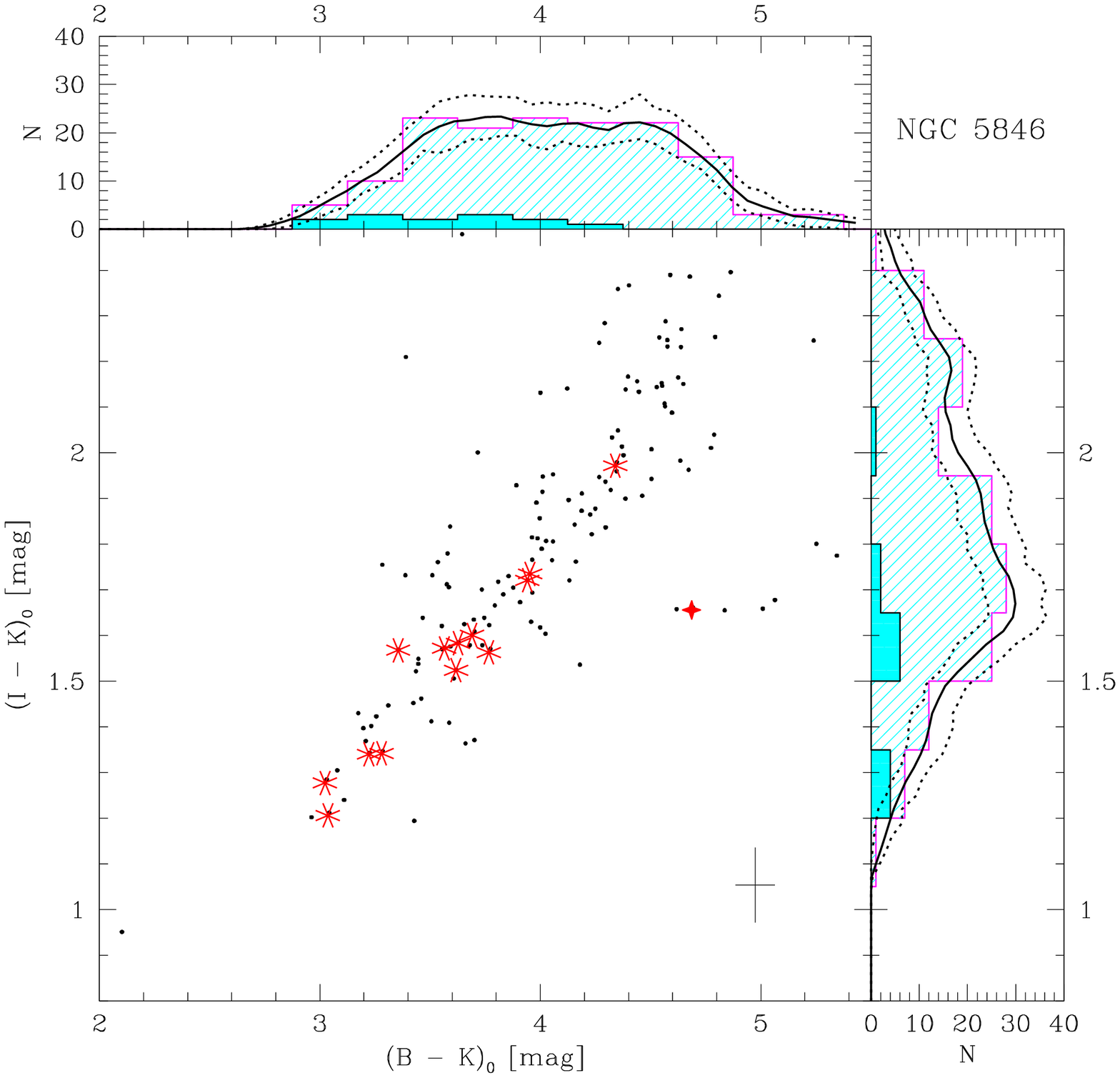}
	\includegraphics[width=8.5cm]{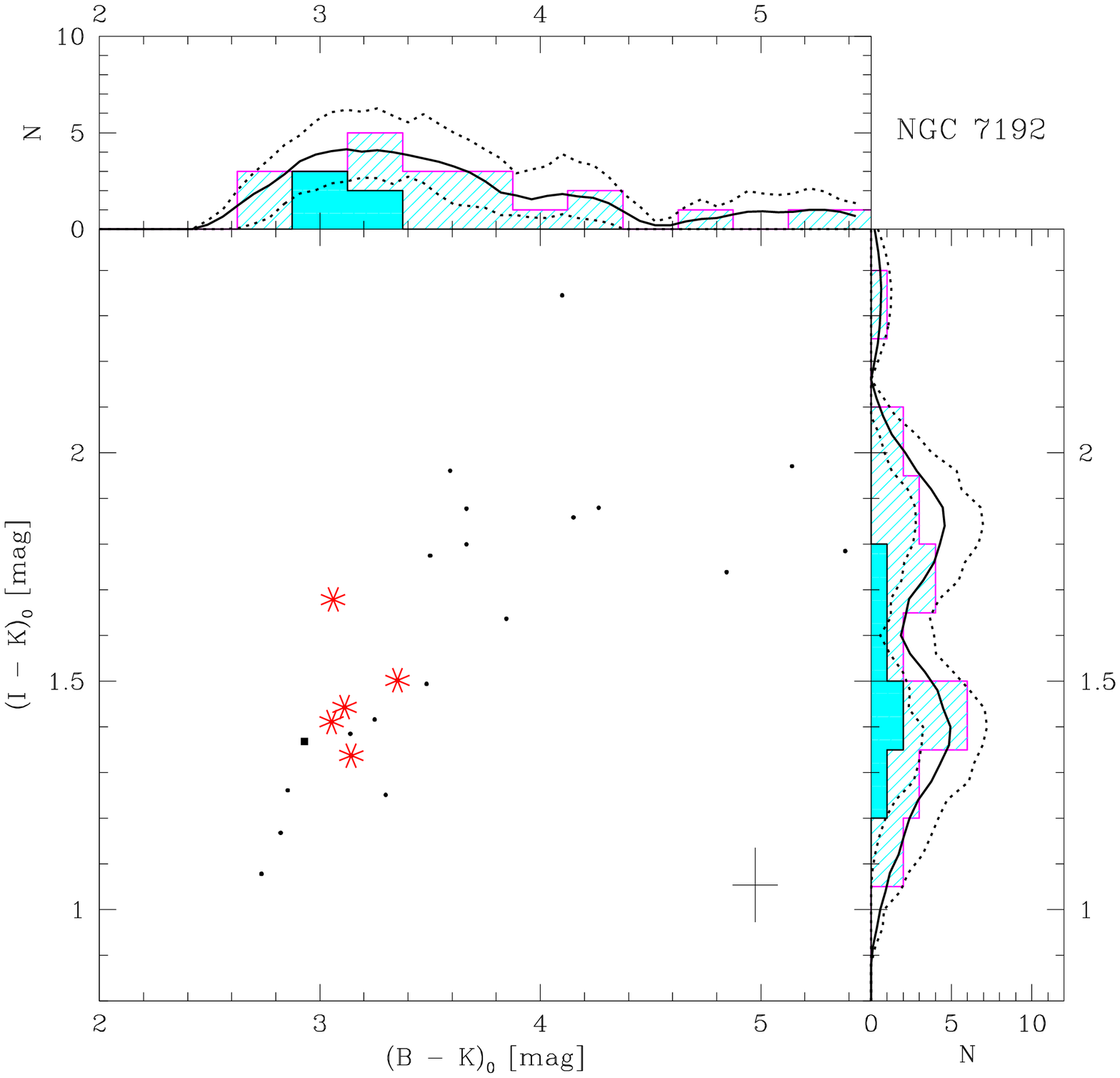}
	\includegraphics[width=8.5cm]{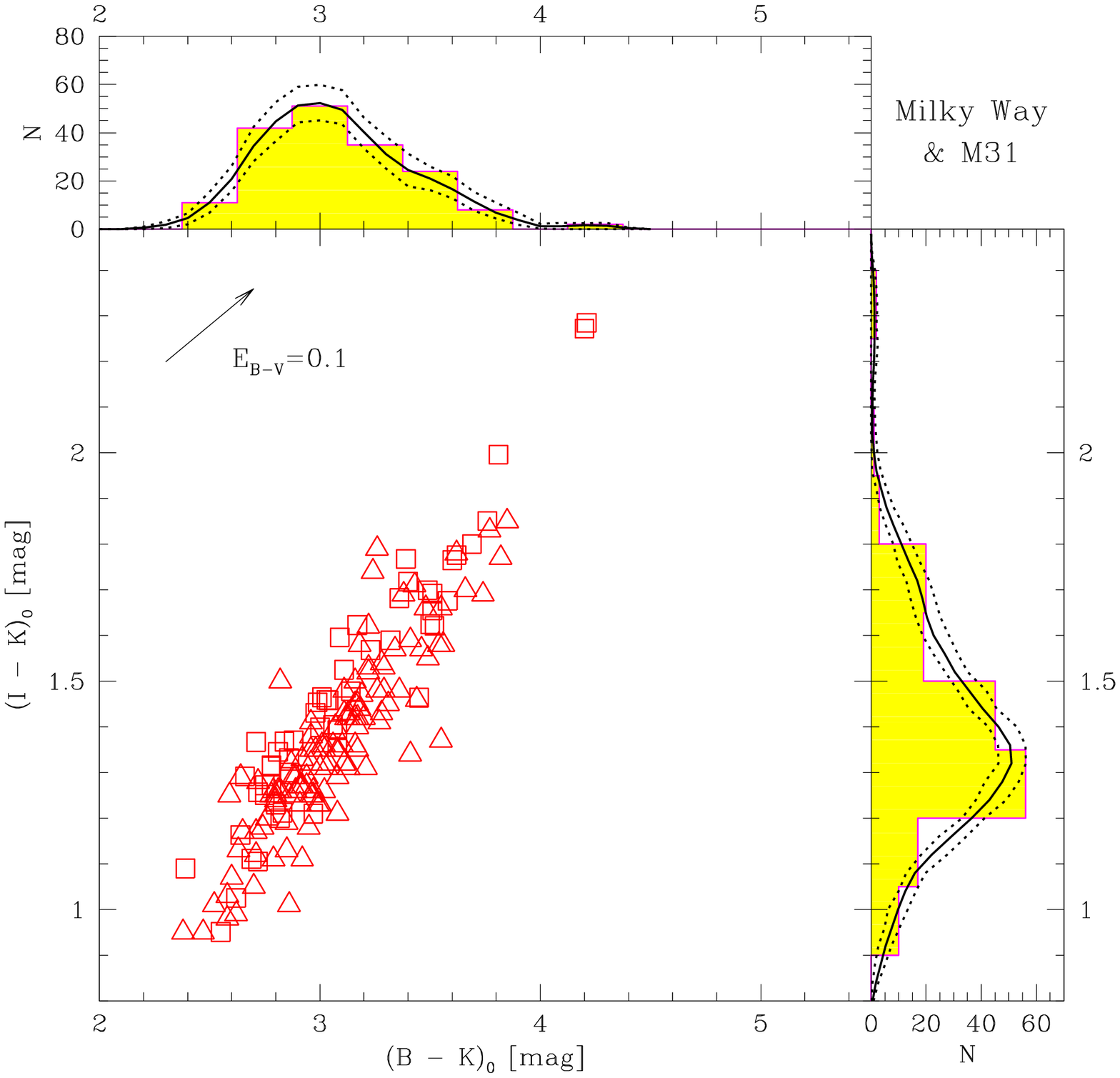}

	\caption{Optical/near-infrared $I-K$ vs. $B-K$ two-colour
	diagrams for objects in NGC~3115, NGC~5846, and
	NGC~7192. Near-infrared data for NGC~3115 were taken from
	\cite{puzia02a} while the near-infrared data for NGC~5846 and
	NGC~7192 were taken from \cite{hempel02}. Average photometric
	errors are indicated in the lower right corner of each
	diagram. Asterisks indicate spectroscopically confirmed
	globular clusters. 4-prong stars and solid squares show
	colours of confirmed foreground stars and background galaxies,
	respectively. Hatched and solid histograms in the sub-panels
	show the colour distributions of all objects and of
	spectroscopically confirmed globular clusters. Solid lines
	within the sub-panels are probability density estimates with
	their 90\% confidence limits (dotted lines). In the lower
	right panel we show colours of globular clusters in M31 ({\it
	triangles}) and the Milky Way ({\it squares}). Optical colours
	for Milky Way globular clusters were taken from
	\cite{harris96}, near-infrared colours were adopted from
	Aaronson \& Malkan (in preparation). The data for M31 globular
	clusters are from \cite{barmby00}.}
\label{ps:ccm}
\end{figure*}

We combine now our optical FORS2 photometric data with recently
published near-infrared data for NGC~3115 \citep{puzia02a}, NGC~5846,
and NGC~7192 \citep{hempel02} and construct optical/near-infrared
two-colour diagrams of candidate and confirmed globular clusters. All
near-infrared data were obtained with the ISAAC instrument attached to
ESO's VLT with a $2.5$\arcmin$\times2.5$\arcmin\ field of
view. Figure~\ref{ps:ccm} shows $I-K$ vs. $B-K$ diagrams with
spectroscopically confirmed globular clusters marked as asterisks.
Due to the smaller field of view of the near-IR data, these two-colour
diagrams are restricted to the central regions of each galaxy and do
not cover a representative of the entire globular cluster system (see
discussion in \S\ref{ln:cmd}). Although optical/near-infrared colours
are powerful metallicity discriminators \citep[e.g.][]{puzia02b}, they
are of limited use for the slit mask design due to the very
constrained field of view.

However, {\it a posteriori} it is worthwhile to compare
optical/near-infrared colours of globular clusters, background
galaxies and foreground stars in order to minimise contamination of
the candidate selection. Based on our set of photometric passbands
($B,V,R,I,K$), we find that the combination of $I-K$ and $B-K$
separates globular clusters from stars and galaxies most reliably. In
the following we describe how to reduce the contamination of globular
cluster candidate samples by foreground stars and background galaxies.

\begin{figure*}[!ht]
\centering 
   \includegraphics[width=8.9cm]{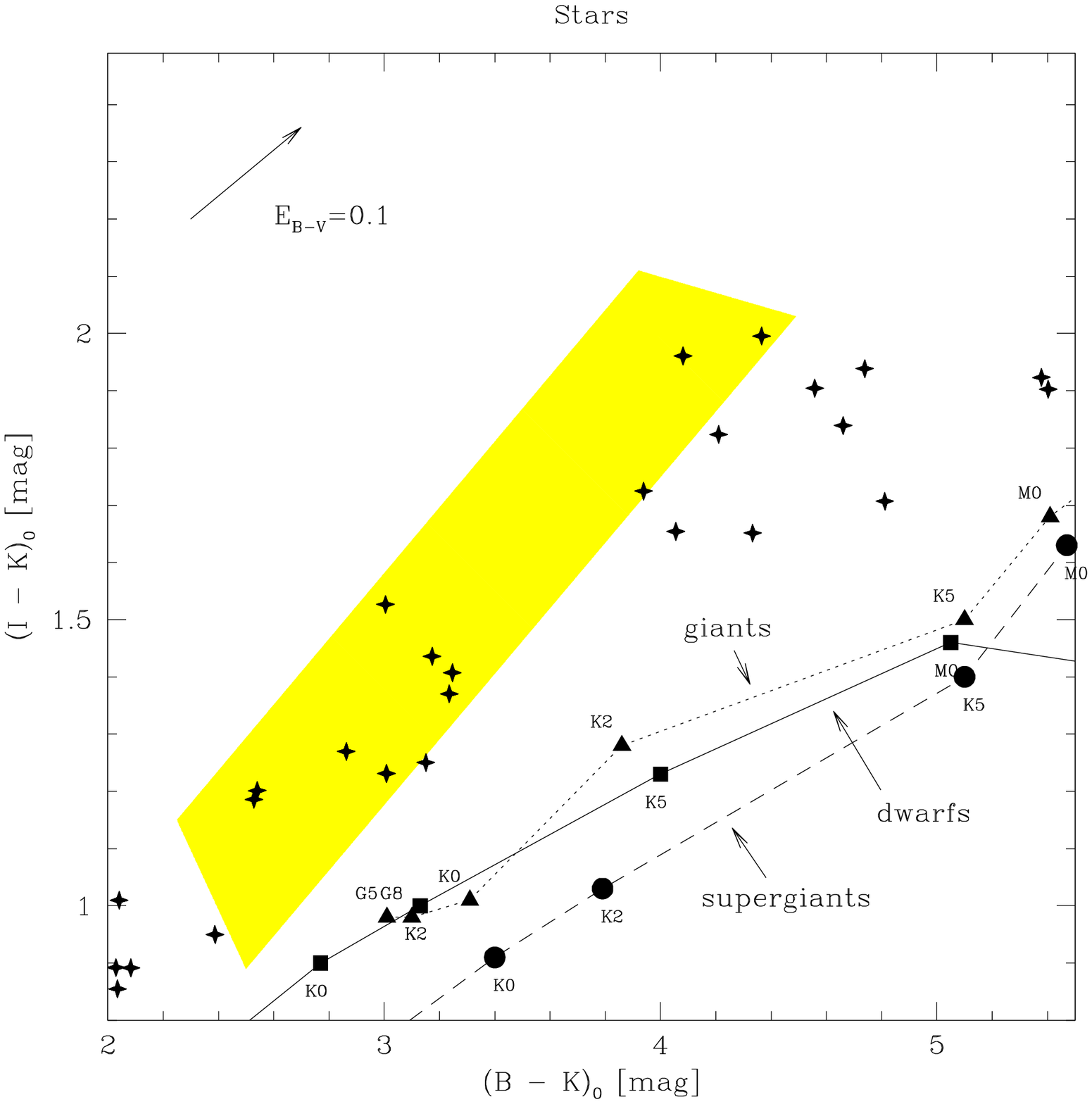}
   \includegraphics[width=8.9cm]{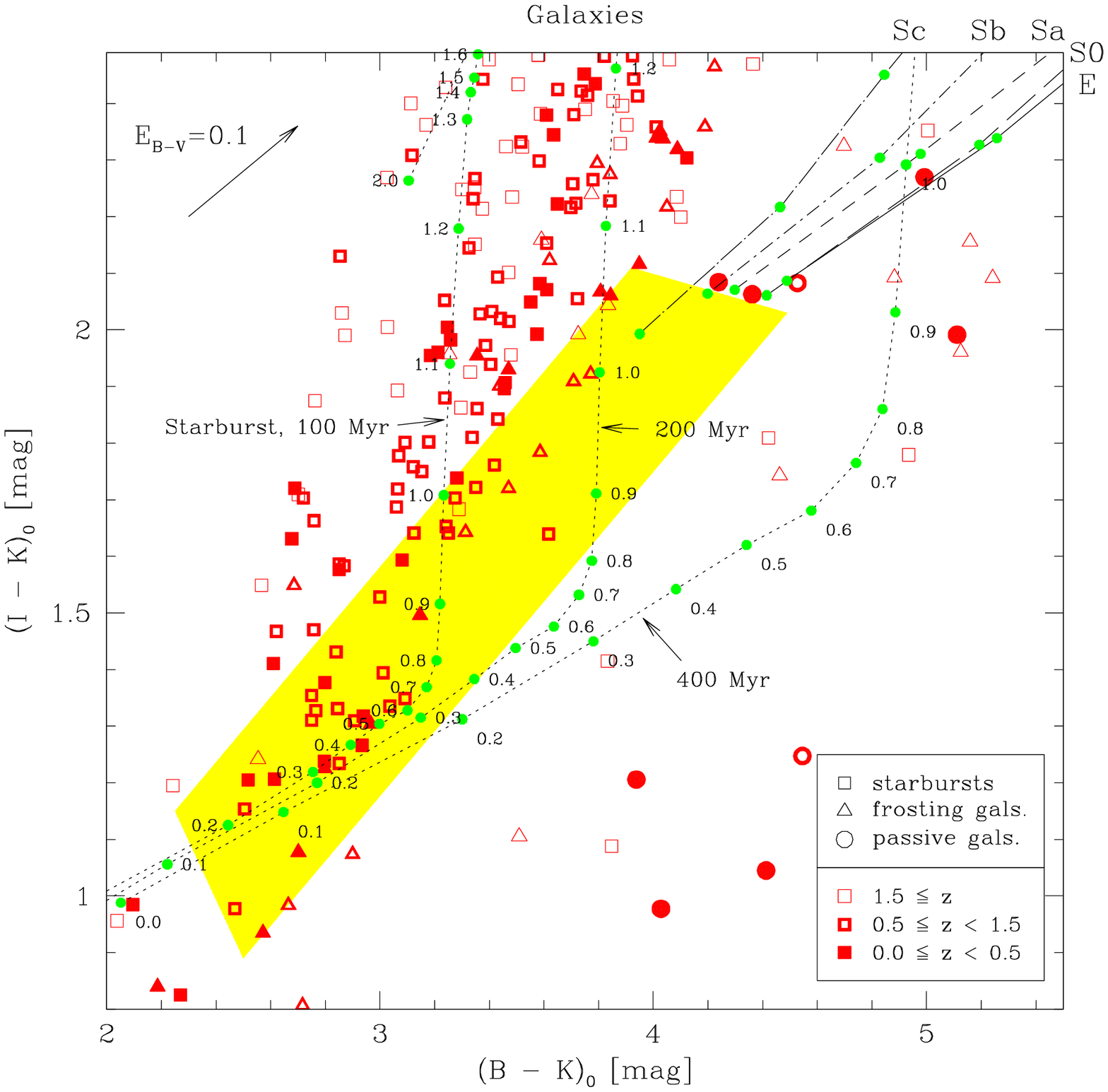}
	\caption{This figure compares in the two-colour diagram
          the position of globular clusters (indicated by a shaded
          region, see also Fig.~\ref{ps:ccm}) versus the mean locus of
          possible contaminating objects, such as foreground stars
          (left panel) and background galaxies (right panel). {\bf
          Left panel:} Measured colours of stars were taken from the
          FORS Deep Field \citep{heidt03, gabasch03} and are indicated
          by 4-prong stars. Due to their small surface density, we use
          the full set of stars found in the FDF which has a field-of
          view of $\sim39$ square arcminutes. Number counts have to be
          corrected when contamination estimates are performed for our
          sample (field of view 6.25 square arcminutes for the
          $K$-selected data, $\sim45$ square arcminutes for the
          optical data). Spectral sequences of dwarf, giant, and
          supergiant stars in the disk are shown as solid, dotted, and
          dashed lines, respectively, and were taken from
          \cite{cox00}. Spectral types are indicated along each
          sequence. The offset between FDF stars and the spectral
          sequences is due to a higher metallicity of the latter which
          were constructed from stars in the solar neighbourhood. {\bf
          Right panel:} Open squares mark starbursts, open triangles
          are galaxies in which star-formation ceased a few Gyrs ago
          (so-called ``frosting'' galaxies), while open circles show
          colours of old, passively evolving elliptical galaxies. The
          full data set from the FDF was used ($I_{\rm 50\%\,
          compl.}\approx26.37$, \citealt{heidt03}) and thinned so that
          expected number counts in the field-of-view of the
          $K$-selected data can be directly read off. The
          photometrically determined redshift of galaxies is indicated
          by the filling factor of symbols (see legend in the lower
          right corner). Additionally, we k-correct the colours for
          E, S0, Sa, Sb, and Sc galaxies using template spectra of
          \cite{mannucci01} and \cite{kinney96}. The sequences are
          labeled in the upper right corner of the right panel. Shaded
          dots along the sequences indicate redshifts which increase
          from $z=0.0$ in steps of $\Delta z=0.1$ towards redder
          colours. Dotted lines show the evolution of k-corrected
          colours for starburst galaxies with an age of 100, 200, and
          400 Myr using templates from Maraston (2003, in
          preparation). Redshifts are marked along each sequence.}
\label{ps:fdf}
\end{figure*}

\subsubsection{Eliminating Foreground Stars}
Globular clusters fall on a rather narrow sequence in the $I-K$
vs. $B-K$ diagram (hatched region in Figure~\ref{ps:fdf}). In general,
at similar $I-K$ colours, cool giant stars lack $B$ band flux compared
to the integrated light of globular clusters and have therefore redder
$B-K$ colours. As the effective temperature of red giants is a
sensitive function of metallicity, more metal-poor (warmer) stars will
be more likely to resemble mean globular cluster colours. This is
shown in the left panel of Figure~\ref{ps:fdf}. It is important to
keep in mind that age and metallicity of contaminating stars depend on
the sampled galactic coordinates. In order to illustrate the
difference in colour between disk and halo stars in the $I-K$
vs. $B-K$ diagram we plot colours of stars found in the FORS Deep
Field (FDF, galactic coordinates $l=191.40^o$, $b=-86.46^o$;
\citealt{heidt03}, \citealt{gabasch03}) and representative colours of
disk dwarf, giant, and supergiant stars in the solar neighborhood
\citep{cox00}. While the former sample is likely to be dominated by
old metal-poor halo stars\footnote{The FDF line of sight is almost
perpendicular to the Galactic disk.}, the latter data resemble colours
of metal-rich disk stars. To show the metallicity offset between
metal-poor and metal-rich stars more clearly we plot all stars found
in the FDF field with a field-of view of $\sim39$ square arcminutes
(indicated by stars in Fig.~\ref{ps:fdf}). Thus, number counts have to
be rescaled as the field of view of our combined optical/near-infrared
photometry is only 6.25 square arcminutes. Metal-rich disk stars can
be reliably separated from globular cluster candidates as the former
are significantly redder in $B-K$. The colours of metal-poor halo
stars, on the other hand, are more similar to globular cluster
colours. However, the surface density of these stars with
metallicities [Fe/H]~$\la-1.5$ is relatively low, of the order of one
star per field-of-view in our optical data ($\sim45$ square
arcminutes) in the range $18\leq V\leq22.5$, with little dependence on
galactic coordinates \citep{robin96}.

In general, the combination of optical and near-infrared photometry
provides a good discriminator to distinguish between globular clusters
and metal-rich foreground stars. However, the colours of metal-poor
globular clusters can be mimicked by metal-poor halo stars. These
stars need to be sorted out by other selection criteria such as
magnitude. Based on the FDF data and the Galactic stellar population
model \citep{robin96}, we expect a total stellar contamination of $\la
1-10$\% (depending on the richness of the globular cluster system) at
galactic latitudes $|b|\ga40^o$. At lower latitudes the foreground
contamination is rising.

\subsubsection{Eliminating Background Galaxies}
Another source of contamination are background galaxies. To estimate
their $I-K$ and $B-K$ colours, we use a sub-sample of the FDF data
which corresponds to our combined optical/near-infrared data in
field-of-view size and photometric completeness (mainly limited by the
near-infrared photometry at $K\sim21.5$). In the right panel of
Figure~\ref{ps:fdf} we plot $I-K$ vs. $B-K$ colours of galaxies with
high star-formation rate (open squares), galaxies in which star
formation ceased a few Gyrs ago (so-called ``frosting'' galaxies, open
triangles), and passively evolving ellipticals (open circles)
\citep[see][for a quantitative classification]{heidt03,
gabasch03}. Judging from the FDF data in Figure~\ref{ps:fdf}, we find
that the blue part of the mean globular clusters locus, indicated by
the shaded region, is mainly contaminated by starburst galaxies while
the red part is prone to contain ``frosting'' galaxies.  Depending on
the boundary definitions, we find $\sim20-30$ starbursts and
$\sim10-15$ ``frosting'' galaxies inside the region where globular
clusters are preferentially found. However, most of these galaxies
would be resolved in our photometry (typical seeing $\leq 1$\arcsec )
and rejected by the FWHM (or size) selection.

Potentially problematic objects are distant starburst galaxies which
are barely resolved and still bright enough to be classified as
globular cluster candidates. At redshift unity, one arcsecond
corresponds to $\sim8$ kpc in a flat $\Lambda$-universe with
$\Omega_m=0.3$ and $H_0=70$ km s$^{-1}$ Mpc$^{-1}$ (at $z=2$, one
arcsecond covers $\sim8.4$ kpc). Typical sizes of distant starburst
galaxies range between a few hundred pc to a few kpc \citep[e.g.][]{
guzman98, soifer01} and can be reliably resolved with ground-based
photometry up to $z\approx0.1$. At $z\approx1.0$ even the brightest
starbursts with typical absolute magnitudes $M_V\approx-21$ up to
$-22$ mag are already too faint to enter our magnitude selection (cut
at $V=23$ mag). We conclude that provided good photometric quality
(seeing $\leq1$\arcsec ), starbursts are reliably rejected by the
combination of FWHM and magnitude selection below $z\approx0.1$ and
above $z\approx1.0$. Optical/near-infrared colours can be a good
additional discriminator for the remaining intermediate-redshift space
as shown in the following.

We simulate the redshift evolution (both k-correction and luminosity
corrections) of optical/near-infrared colours of a 100, 200, 400, and
800 Myr old starburst using template spectra taken from Maraston (2003,
in preparation). The latter templates include the stellar evolutionary
phase of thermally pulsing AGB stars, that dominates the infrared and
bolometric flux at these ages \citep[e.g.][]{maraston01}. We use
spectra of instantaneous star-formation with a metallicity
[Z/H]~$=-0.33$. We find that young starbursts between 100 and 200 Myr
produce colours which are consistent with globular cluster colours in
the redshift range between $z=0.1$ and $z=1.0$. As most starbursts are
intrinsically located in high-reddening regions, we note that bright
unresolved starbursts with ages $\leq100$ Myr and reddening values
$E_{B-V}>0.1$ can show typical globular cluster colours, as well. Those
objects could contaminate the blue sub-sample of globular cluster
candidates (see reddening vector in the right panel of
Fig.~\ref{ps:fdf}). Starburst older than $\sim300$ Myr have
colours that are inconsistent with globular cluster colours beyond a
redshift $z\approx0.2$. Indeed, at these ages the AGB-phase transition
boosts both colours to $I-K\ga2.5$ and $B-K\ga4.0$.

The k-corrected colours for galaxies are simulated with empirical
template spectra of \cite{mannucci01} and \cite{kinney96} for
elliptical, lenticular, and spiral galaxies. To account for the
luminosity evolution requires to adopt a model for these galaxies and
goes far beyond the aim of this exercise. The k-correction paths
are shown in the right panel of Figure~\ref{ps:fdf} with redshifts
indicated by filled dots starting at $z=0.0$ and increasing in steps
of $\Delta z=0.1$ to redder colours. It is obvious that
non-starforming early-type galaxies entirely avoid the mean colour
locus of globular clusters at all redshifts. Colours of low-$z$
later-type galaxies are only marginally consistent with the reddest
globular clusters. Sc galaxies below $z=0.1$ intersect the shaded
region where red globular clusters are found. However, these galaxies
are efficiently rejected by the FWHM selection.

We conclude that the $I-K$ vs. $B-K$ diagram allows one to reliably
disentangle globular cluster candidates from foreground disk stars as
well as early-type and spiral galaxies with no or little on-going
star-formation. Remaining potential contaminants are unresolved
starbursts with ages $\la300$ Myr at intermediate redshifts between
$z\approx0.1$ and $\sim1.0$. Based on FDF data we expect $\sim35-45$
background galaxies down to $I=22.5$ within the FORS field-of-view
($\sim45$ square arcminutes) with colours resembling those of globular
clusters. The majority of these galaxies is resolved and rejected by
the PSF selection. Indeed, the only background galaxy found in our
$K$-selected sample with optical/near-infrared colours ($4$\% of the
sample; a fill-in object with $V=22.83$, $B-K=3.06$, and $I-K=1.42$)
is consistent with a young unresolved starburst galaxy at
$z\approx0.14$. The spectrum has too low S/N to allow a more accurate
classification based on spectral features. The corresponding absolute
magnitude of the object is $M_V\approx-16.3$ and is consistent with a
SMC-type galaxy \citep[$M_V=-16.2$ mag][]{binney98}.

The overall surface density of contaminating objects with
globular-cluster colours and magnitudes which are bright enough for
integrated-light spectroscopy ($V\ga22-23$ at 8--10m-class telescopes
with colours as indicated in Fig.~\ref{ps:fdf}) is negligible
($\la10$\%) where the surface density of globular clusters is high,
that is within $\sim1\, R_{\rm eff}$. Multi-object spectroscopic
studies are therefore expected to have a highest success rates in
central regions of a globular cluster system. We refer to
Section~\ref{ln:success} for an analysis of the success rate of the
candidate selection.

\subsection{Radial Surface Density Profiles of Globular Cluster Systems}
\label{ln:radprof}
\begin{figure}[!t]
\centering 
   \includegraphics[width=4.2cm]{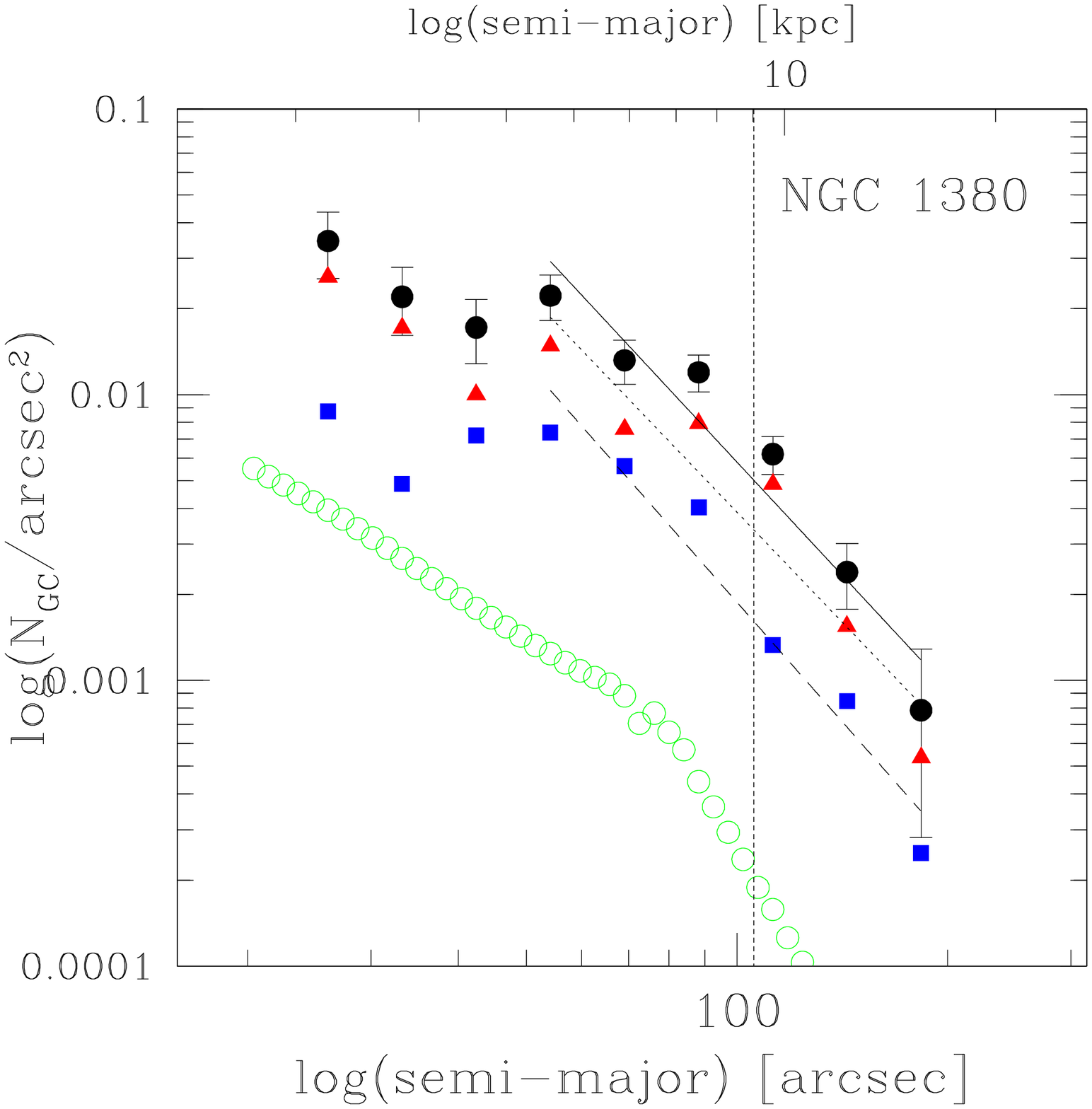}
   \includegraphics[width=4.2cm]{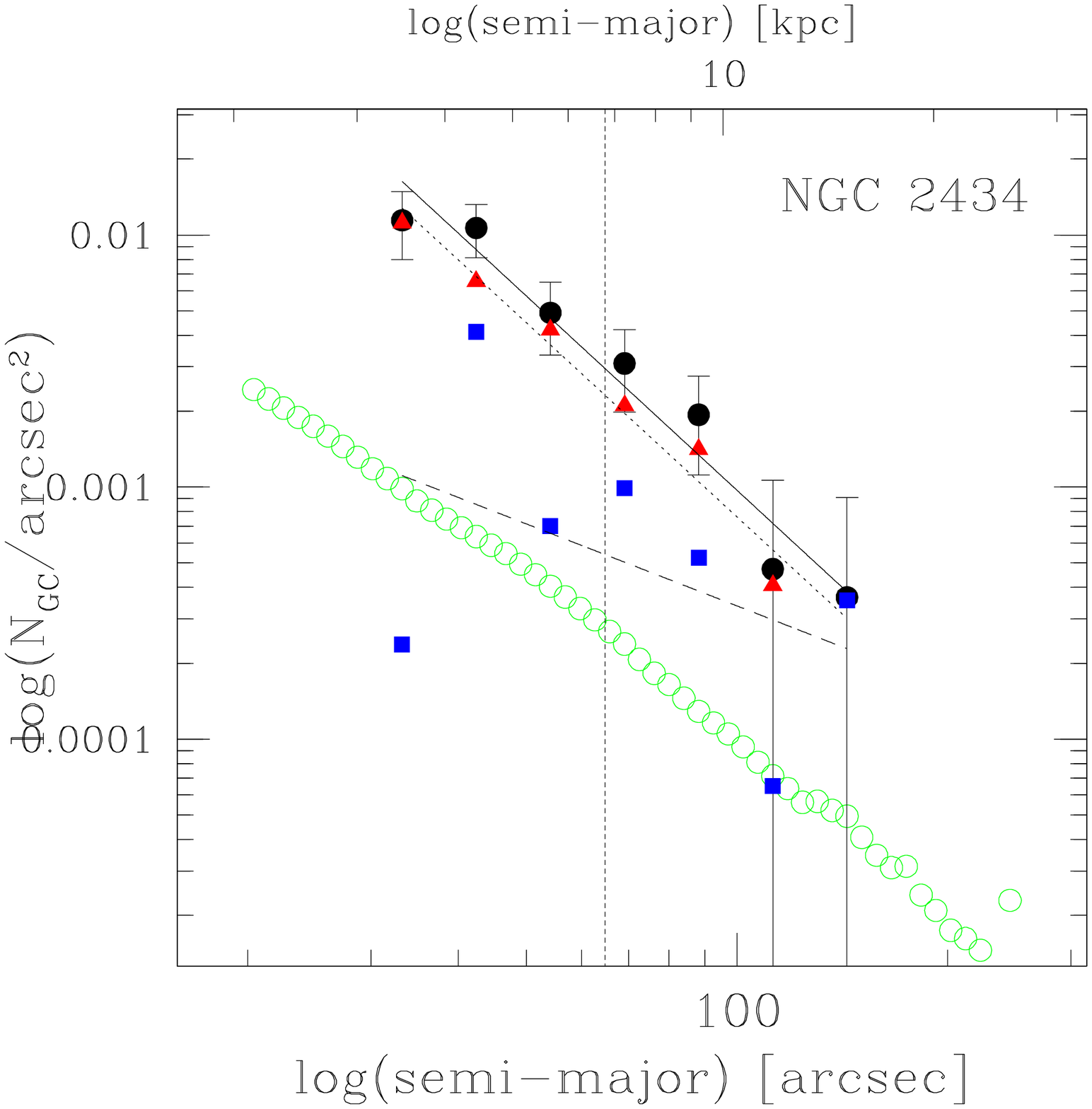}
   \includegraphics[width=4.2cm]{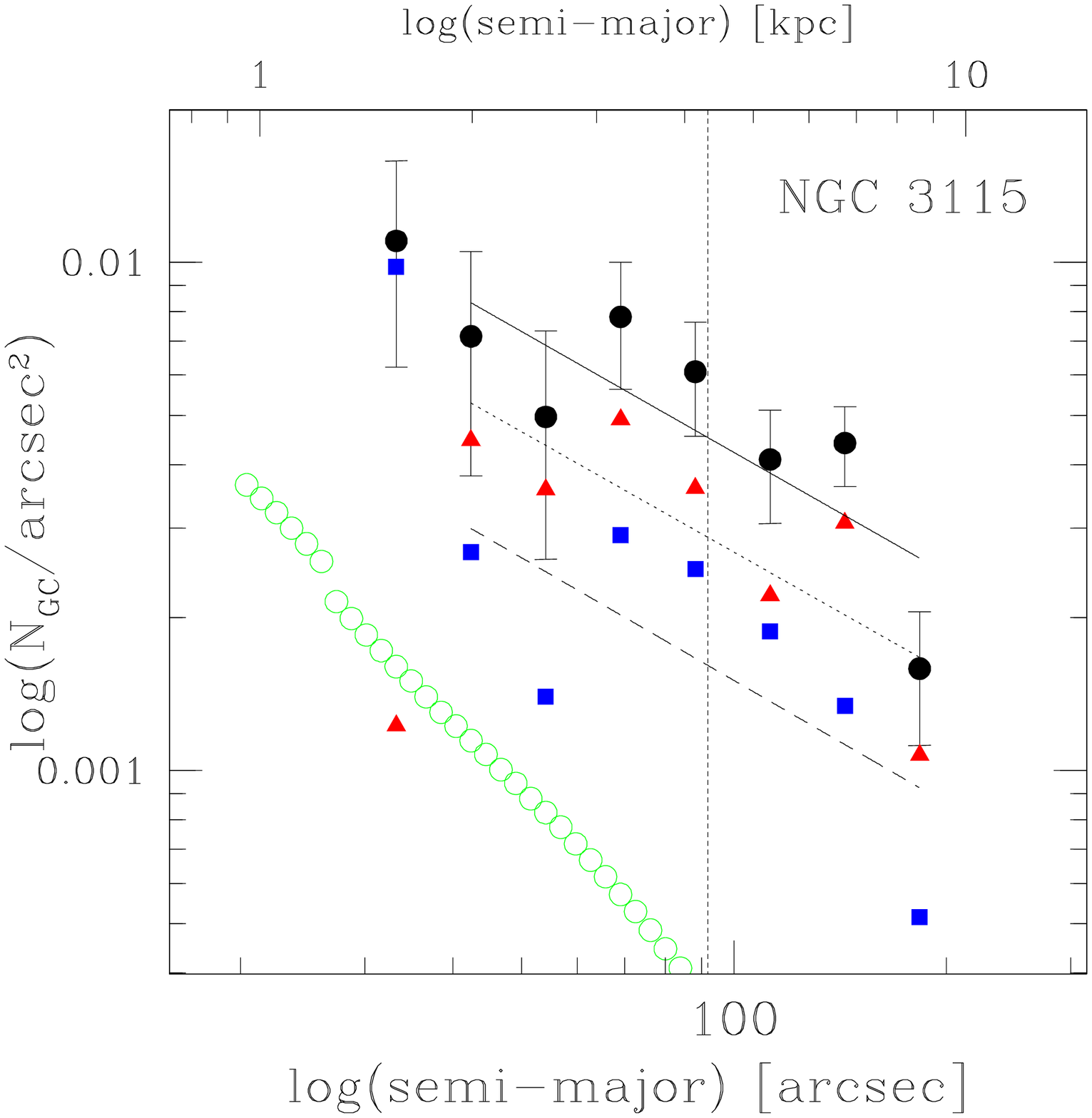}
   \includegraphics[width=4.2cm]{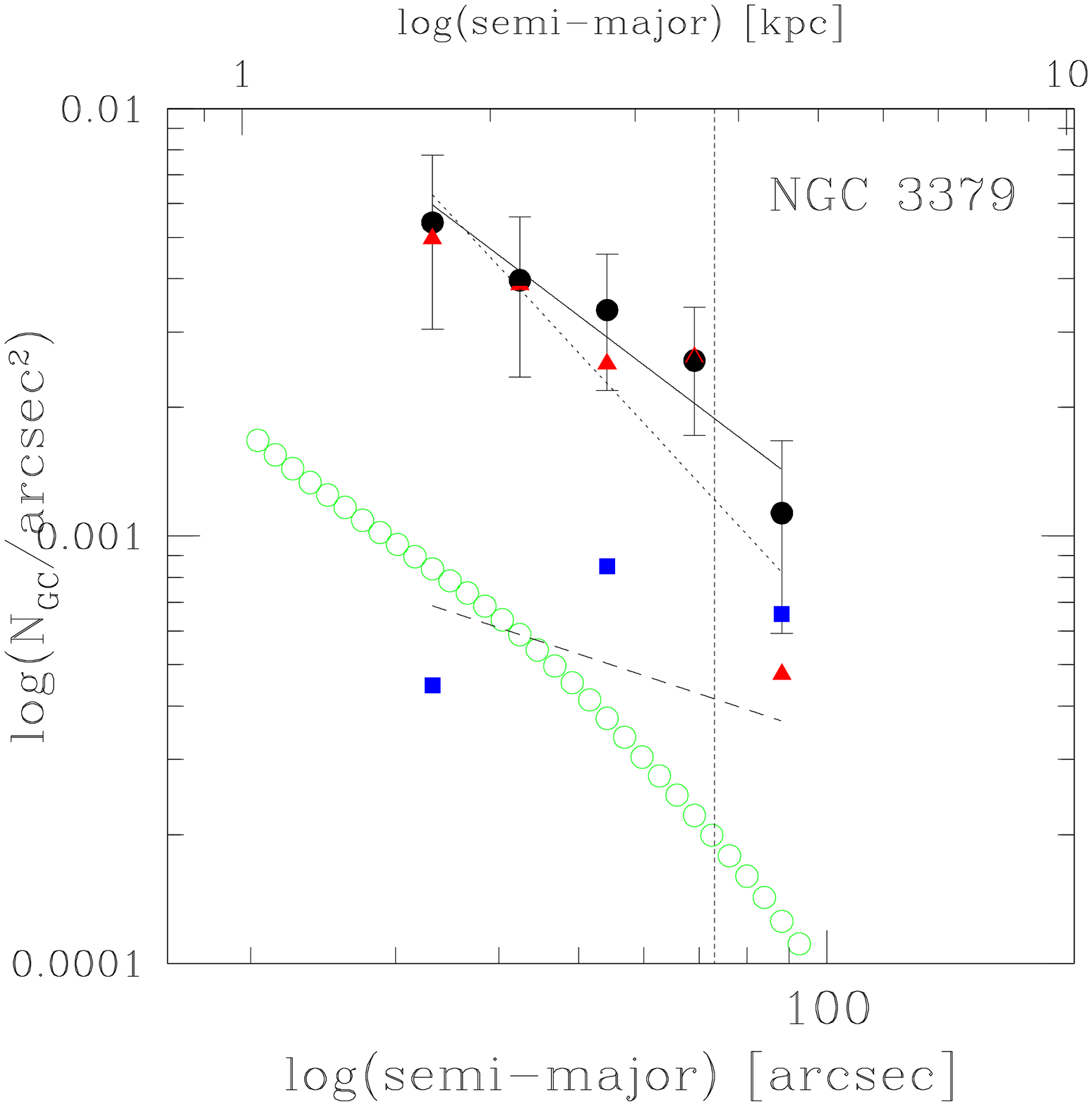}
   \includegraphics[width=4.2cm]{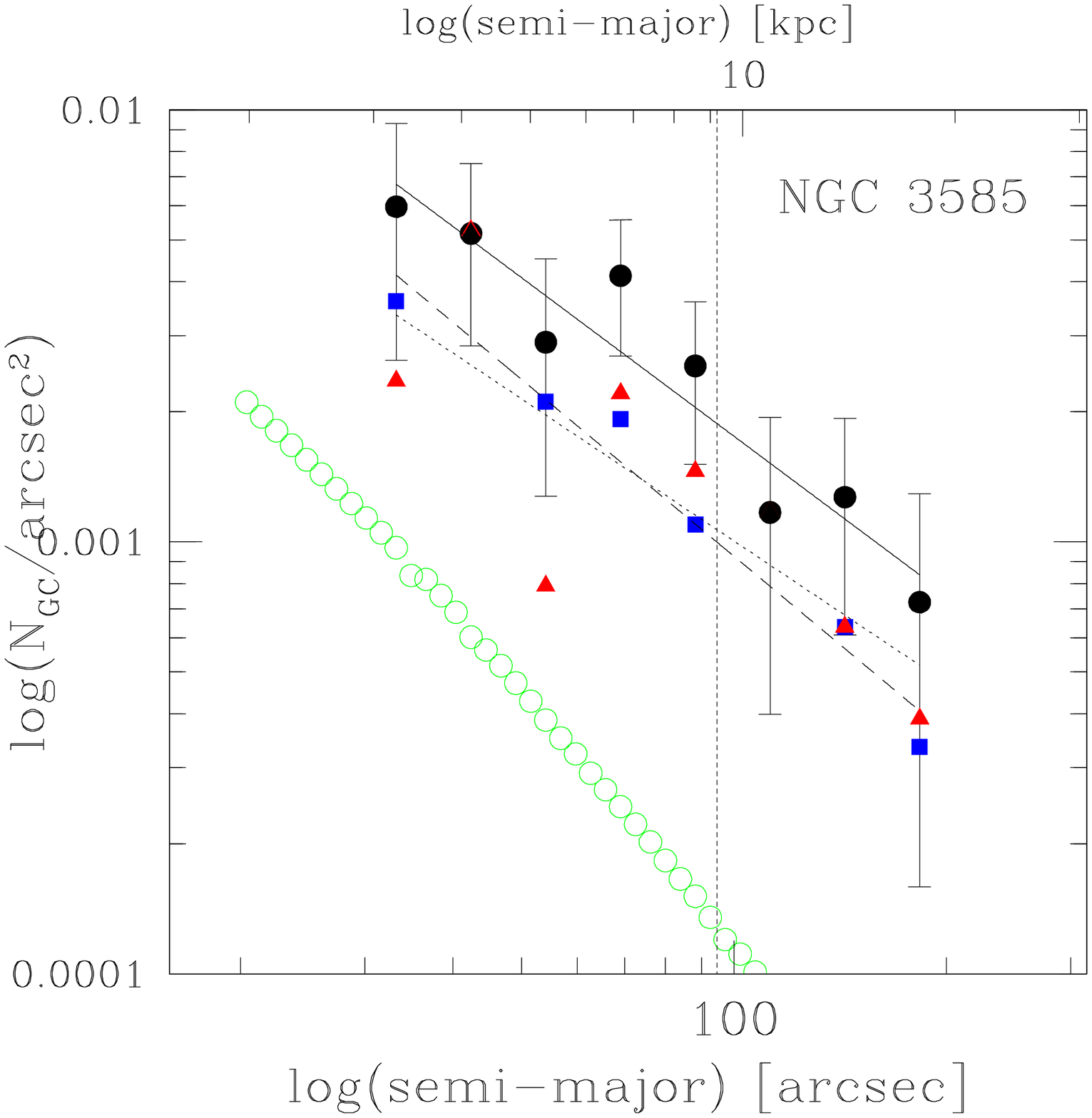}
   \includegraphics[width=4.2cm]{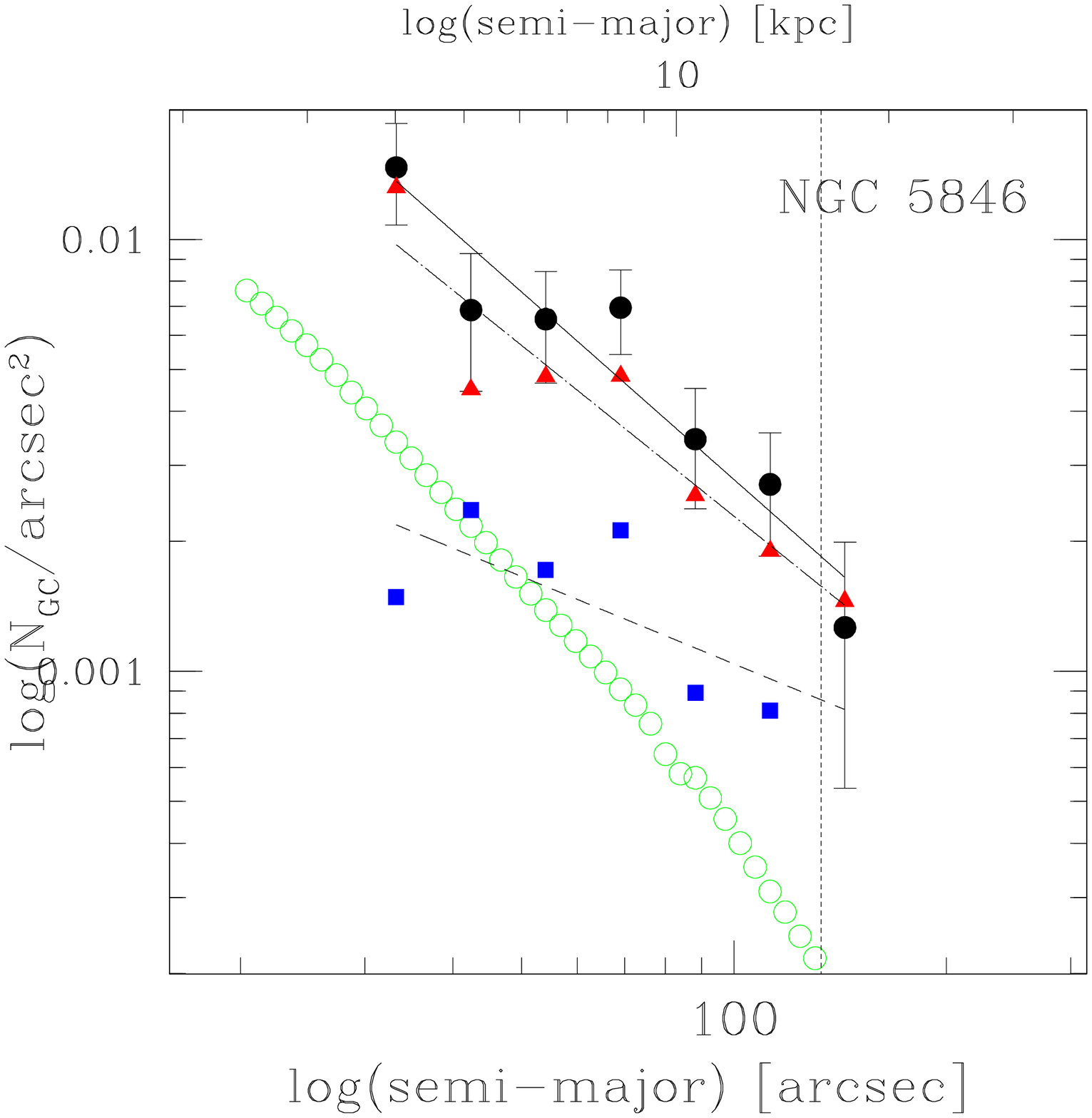}
   \includegraphics[width=4.2cm]{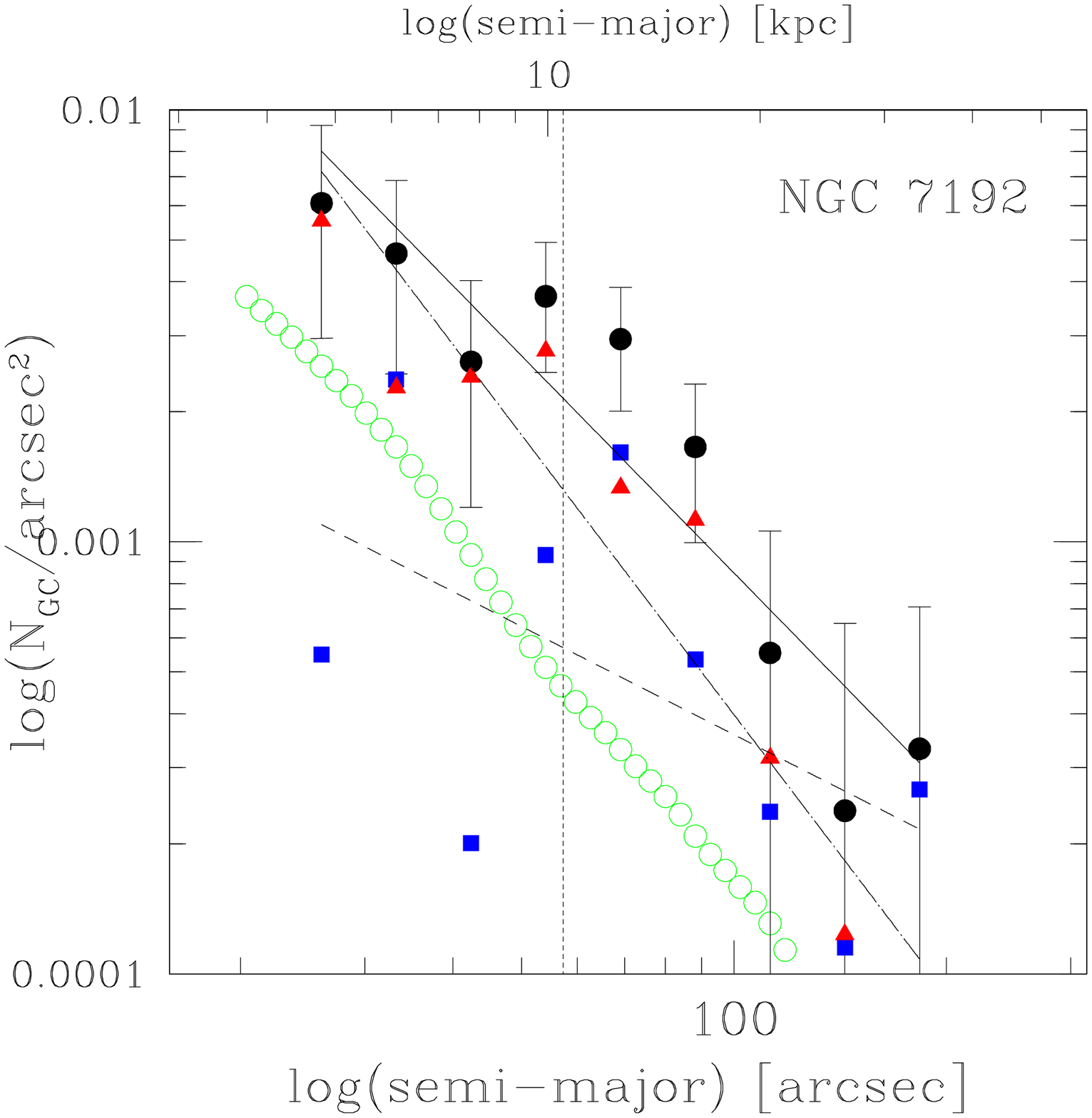} 

   \caption{Surface density and surface brightness profiles of
	globular cluster systems and their host galaxies. The surface
	density of all globular cluster candidates is indicated by
	black dots with a power-law fit indicated as a solid
	line. Multi-modal colour distributions were divided at
	$V-I=1.1$ or $B-I=1.8$ into blue and red globular
	clusters. Power-law fits to the blue and red sub-populations
	are shown as dashed and dotted lines, respectively. Open
	circles show the scaled surface brightness profile of the host
	galaxy derived from $I$ band photometry. Both profiles were
	calculated using the same isophotes. A vertical line indicates
	the effective radius of each system taken from
	\cite{RC3}. Poisson errors are indicated for the total surface
	density profiles.}
\label{ps:radprof}
\end{figure}

\begin{table}[h!]
\centering
\caption[width=\textwidth]{Power-law exponents, $\Gamma$, of
globular-cluster surface density profiles, $\Sigma(R)\sim
R\,^{-\Gamma}$, of globular cluster systems and sub-populations. The
second and third column shows the slopes for the profiles of the blue
and red sub-populations, respectively. In the same order, these slopes
are indicated as dashed and dot-dashed lines in
Figure~\ref{ps:radprof}. The last column are the slopes of the entire
cluster system. All given errors are uncertainties of the fit. Numbers
in parentheses indicate a brute-force division of globular cluster
systems which show a single-peak colour distribution in a blue and red
globular cluster sub-population (see text for details).}
\label{tab:radslopes}
\begin{tabular}{l c c c }
\hline\hline
\noalign{\smallskip}
Galaxy   & blue    & red     & all \\
\noalign{\smallskip}
\hline
\noalign{\smallskip}
NGC 1380 & $2.78\pm0.33$ & $2.56\pm0.42$ & $2.63\pm0.37$ \\
NGC 2434 &($1.08\pm0.98$)&($2.56\pm0.24$)& $2.56\pm0.25$ \\
NGC 3115 & $0.80\pm0.36$ & $0.79\pm0.26$ & $0.79\pm0.28$ \\
NGC 3379 &($0.64\pm0.84$)&($2.09\pm0.67$)& $1.46\pm0.28$ \\
NGC 3585 & $1.36\pm0.12$ & $1.09\pm0.35$ & $1.22\pm0.20$ \\
NGC 5846 &($0.67\pm0.35$)&($1.31\pm0.21$)& $1.44\pm0.33$ \\
NGC 7192 &($0.82\pm0.48$)&($2.15\pm0.28$)& $1.67\pm0.24$ \\
\noalign{\smallskip}
\hline
\end{tabular}
\end{table}

In future papers we will study the properties of globular cluster
systems as a function of their host's properties. To derive host
galaxy masses we use globular clusters as tracer particles of the
galaxy potential in Section~\ref{ln:galaxymasses}. The tracer mass
estimator requires the knowledge of the surface-density profiles of
the tracer-particle population. Our photometric data probe a large
enough range in galactocentric radius to reliably sample the globular
cluster surface density profile. Moreover, in some globular cluster
system formation scenarios the surface density profile of blue and red
clusters is expected to change differently as a function of radius
\citep[e.g.][]{ashman92, cote98}. It is therefore important to study
the profiles of each globular cluster sub-sample to control radial
biases of further analyses.

In order to derive surface-density profiles we use the IRAF task {\sc
ellipse} and the pre-selected FORS2 photometric sample (see
Sect.~\ref{ln:cmd}). First, we model the surface-brightness profiles
of the galaxy light on images which feature the best seeing and which
were cleaned of point sources. The same elliptical isophotes are used
for surface-brightness profiles in other passbands and to construct
the surface-density profile of globular-cluster candidates.
Colour-magnitude diagrams in Figure~\ref{ps:cmd} show that photometric
completeness as a function of colour does not greatly affect the
radial surface-density profiles as the change in completeness level is
negligible within the colour region from which globular clusters are
selected.

In Figure~\ref{ps:radprof} we plot surface density profiles of
globular cluster candidates. Prior to creating the profiles, we
excluded extremely blue objects (which are likely contaminating
fore-/background sources, see Sect.~\ref{ln:ccm}) by applying cuts at
$V-I=0.3$ (NGC~3379), $V-I=0.7$ (NGC~2434, NGC~7192), $V-I=0.8$
(NGC~1380, NGC~3115, NGC~5846), and $B-I=1.3$ (NGC~3585). Objects with
colours redder than $V-I=1.3$ (NGC~2434, NGC~3379), $V-I=1.4$
(NGC~1380, NGC~3115), $V-I=1.5$ (NGC~5846), $V-I=1.6$ (NGC~7192), and
$B-I=2.3$ (NGC~3585) were also rejected. Outermost isophotes were used
to subtract the background light and the surface density of background
objects. Depending on the distance of the galaxy, the outermost
accessible radii vary between $\sim10$ and $\sim50$ kpc, which
corresponds to $R\geq2\, R_{\rm eff}$ for all sample galaxies.
Henceforth we compare only the slopes of the surface brightness
profiles of the galaxy and the surface density profiles of globular
cluster systems. Within the range of our globular cluster data
colour-index changes of the diffuse light are negligible and we
representatively use the $I$ band surface brightness profile, since
$I$-band data is available for all our sample galaxies. In order to
have a fair comparison of the slopes we plot $\log(N_{\rm
GC}/$arcsec$^2)$ and $\mu_I/2.5$ versus the logarithmic semi-major
axis galactocentric distance (Fig.~\ref{ps:radprof}). However, this
procedure compares globular cluster number counts with a luminosity
density of the host galaxy and is only valid if the mean M/L ratio of
globular clusters remains constant as a function of radius. The mean
M/L ratio is subject to change when young globular clusters are
concentrated at a given radius. However, we expect that the majority
of the globular cluster system is old and that the comparison of the
slopes is acceptable to the first order.

The surface-density profile of each globular-cluster system appears
generally comparable to or less steep than the galaxy light
profile. In the case of the two S0 galaxies, NGC~1380 and NGC~3115,
both globular-cluster profiles might be influenced towards the center
by the presence of a stellar disk \citep{bothun90}. Disk shocking and
disk instabilities might be responsible for enhanced cluster
destruction and could decrease the globular cluster surface density
close to the center \citep{gnedin97}. Another consequence of the disk
is a reduced photometric completeness in the central parts of the
galaxy. We exclude central regions inside $\sim0.5\, R_{\rm eff}$ from
the profile fitting for these two galaxies. In fact, the
surface-density profile of globular clusters in NGC~1380 tends to be
less steep inside $R_{\rm eff}$ compared to radii larger than one
effective radius. Globular cluster systems in elliptical galaxies are
less affected by dynamical erosion or by a varying photometric
completeness.

We fit power-law profiles of the form $\Sigma(R)\sim R\,^{-\Gamma}$ to
the selected globular cluster data (solid circles in
Fig.~\ref{ps:radprof}). Where colour multi-modality is apparent (see
Sect.~\ref{ln:cmd}), we divide each candidate sample into a blue
(solid squares) and red (solid triangles) globular clusters with a cut
at $V-I=1.1$ (NGC~1380, NGC~3115) and $B-I=1.8$ (NGC~3585) and fit
power-law profiles to each sub-sample and to the whole globular
cluster system. Profiles for blue sub-samples are plotted as dashed
lines. Dotted lines indicate profiles for red globular cluster
candidates. Table~\ref{tab:radslopes} summarises the slopes for each
galaxy.

The power-law exponents of the globular cluster surface-density
profiles cover a wide range from $\Gamma\approx 0.8$ to steep profiles
with $\Gamma\approx 2.6$. Blue and red globular cluster
sub-populations appear to have similar profiles in all multi-modal
galaxies. Although less significant, different power-law exponents are
found for blue and red globular clusters (cuts at $V-I=0.95$ for
NGC~2434, $V-I=0.8$ for NGC~3379, and $V-I=1.1$ for NGC~5846 and
NGC~7192) in galaxies with a single-peak colour distribution. In all
cases the metal-poor globular-cluster system is more spatially
extended than its metal-rich counterpart. The absolute cluster number
densities of blue and red clusters reach comparable values at large
radii $\ga1\, R_{\rm eff}$. This inevitably leads to the fact that in
unimodal galaxies red globular clusters dominate our spectroscopic
samples close to the center, while blue clusters are preferentially
selected in the halo at large radii. This very interesting result
requires a more detailed analysis and must be considered when radial
analyses of globular cluster systems are performed.

\subsection{Selection of Globular Cluster Candidates for Spectroscopy}
\label{ln:selection}

For the selection of globular cluster candidates for spectroscopic
follow-up we focus on objects with colours representative of
high-density regions of a given colour distribution (see
Fig.~\ref{ps:cmd}). Compliant with the restrictions of the slit-mask
design (non-uniform spatial coverage of galactocentric radii, minimum
slit length for good sky subtraction, limited deviation from the mask
meridian for sufficient wavelength coverage, etc.) we representatively
sample the underlying colour distributions of each globular cluster
system. Furthermore, we focused primarily on objects inside one
effective radius where the surface density of clusters is relatively
high compared with surface densities of foreground stars and
background galaxies.

Another constraint results from the faint magnitudes of the cluster
candidates. To increase the likelihood of selecting a globular cluster
we assigned a high priority to objects with magnitudes around the
expected turn-over of the globular cluster luminosity function
(GCLF). This however, has to be traded-off with the minimum S/N of
$\sim20$ \AA$^{-1}$ which is required for reliable index
measurements. Overall the limiting magnitude was adjusted to $V=23$
mag and was exceeded only in a few cases where the slit-mask design
forced it.

In particular, our first-choice targets were drawn from the
pre-selected sample (see Sect.~\ref{ln:cmd}) in the colour range
$0.8\la V-I\la 1.3$, and where colour information was available, from
$1.5\la B-I\la 2.5$ and $1.0\la B-R\la 1.7$. According to simple
stellar-population models of \cite{maraston03} these colour ranges are
expected for stellar populations with metallicities [Z/H]$\ga-1.0$
between $\sim1$ and $\sim15$ Gyr. These cuts exclude clusters with
very low metallicities ($\la-1.5$ dex) which have ages less than
$\sim5$ Gyr. Before the final selection, all colours of cluster
candidates were corrected for the respective foreground extinction
taken from \cite{schlegel98}\footnote{Even for the highest-z galaxy in
our sample, NGC~7192, the k-corrections of colours used for candidate
selection are of the order of a few hundredths mag. Hence, we do not
consider these negligible colour corrections.}.

In general, the upper selection criteria favour globular clusters
which are brighter than the GCLF turn-over. If young ($\la5$ Gyr)
globular clusters are present, they will be preferentially selected
compared to old globular clusters due to their brighter magnitudes. In
that case, our sample is likely to be biased towards young metal-rich
clusters at ages $\la5$ Gyr.

First-choice candidates are used to create the slit masks. Remaining
gaps in-between two slits are filled by objects which suffice slightly
relaxed selection criteria. To fill the slit masks most efficiently,
we relaxed the magnitude limit and the FWHM cut to include also faint
objects. More than 50\% of these fill-in objects was found to be
genuine globular clusters.

\section{Spectroscopic Data}
\label{ln:data}

\begin{table*}[!ht]
\begin{center}
\caption{Journal of spectroscopic observations. Exposure times are
  given in seconds. Two masks were used for NGC~1380 and NGC~2434. All
  slit-mask observations were performed with the FORS mask exchange
  unit (MXU), except for NGC~3115 where we used 19 movable slits of
  the FORS instrument to create a slit mask (MOS mode). Note that no
  longslit spectroscopy (LSS) was obtained for NGC~3115.}
\label{tab:obslog}
\begin{tabular}[angle=0]{lllll}
\hline\hline
\noalign{\smallskip}
 Galaxy & Program No. & Nights & MOS/MXU Exptime & LSS Exptime \\
\noalign{\smallskip}
\hline
\noalign{\smallskip} 
NGC~1380 & P66.B-0068 & 28th -- 31st Dec 2000& MXU mask1: 8$\times$1800 & 4$\times$1800\\
         &            &                      & MXU mask2: 6$\times$1800 & \\
NGC~2434 & P66.B-0068 & 28th -- 31st Dec 2000& MXU mask1: 8$\times$1800 & 4$\times$1800\\
         &            &                      & MXU mask2: 6$\times$1800 & \\
NGC~3115 & P65.N-0281 & 5th \& 6th May 2000  & MOS mask:  6$\times$1800 & \dots \\
NGC~3379 & P66.B-0068 & 28th -- 31st Dec 2000& MXU mask:  8$\times$1800$+$1200 & 3$\times$1800\\
NGC~3585 & P67.B-0034 & 26th -- 29th May 2001& MXU mask: 15$\times$1800 & 5$\times$1800\\
NGC~5846 & P67.B-0034 & 26th -- 29th May 2001& MXU mask: 18$\times$1800 & 5$\times$1800\\
NGC~7192 & P67.B-0034 & 27th -- 29th May 2001& MXU mask: 17$\times$1800$+$900 & 6$\times$1800$+$900\\
\noalign{\smallskip}
\hline
\end{tabular}
\end{center}
\end{table*}

We created two slit masks for NGC~1380 and NGC~2434 with 98 and 100
objects in total, respectively. For NGC~3115 we used the MOS unit of
FORS2 with the 19 slits aligned to cover 22 objects. For NGC~3379,
NGC~3585, NGC~5846, and NGC~7192 we designed one slit mask each to
take spectra of 34, 35, 39, and 34 globular cluster candidates,
respectively. In total we obtained spectra for 362 globular cluster
candidates in seven early-type galaxies.

All data were obtained with the FORS2 instrument at UT2 (unit
telescope 2, Kueyen) of ESO's VLT. The data of period 65 (NGC~3115)
were taken with the multi-object slit (MOS) unit with 19 movable
slits. In period 66 and 67 spectroscopic observations were carried out
with the mask-exchange unit (MXU). A MOS mask with its 19 movable
slits restricts the observations to a limited amount of objects per
exposure. The MXU unit, instead, increases the multiplexity by at
least a factor of two allowing for simultaneous spectroscopy of up to
$\sim40$ objects per frame. The total exposure time for each
individual mask was adjusted according to the observing conditions and
the magnitudes of selected objects. Typical exposure times vary
between $\sim$3 and $\sim9$ hours per mask. The observations were
split into sub-integrations of 1800 seconds (see
Table~\ref{tab:obslog} for details). All exposures used the 600B+22
grism with 600 grooves per mm resulting in a dispersion of 1.2 \AA\
per pixel (R $\sim$ 780) on a 2048$\times$ 2048 pix$^2$ thinned
Tektronix CCD chip with 24$\mu$m pixels. The readout was done in a
single-channel mode. In period 65 this resulted in 5.2 $e^{-}$ readout
noise with a gain 1.85 $e^{-}$/ADU. Observations in period 66 and 67
had 5.41 $e^{-}$ readout noise and a 1.91 $e^{-}$/ADU gain. All
observations use a slit width of 1\arcsec . The mean wavelength
coverage of the system is $\sim3450-5900$ \AA\ with a final resolution
of $\sim5$ \AA.

\subsection{Spectroscopic Data Reduction}
All spectra were processed with standard reduction techniques
implemented in {\sc IRAF}\footnote{IRAF is distributed by the National
Optical Astronomy Observatories, which are operated by the Association
of Universities for Research in Astronomy, Inc., under cooperative
agreement with the National Science Foundation.}. In summary, after
subtracting a masterbias all frames were divided by a normalized
flat-field image. The residual gradients on the normalized flat were
found to be smaller than 0.5\%. The resulting images were cleaned off
cosmics with the routine by \cite{goessl02} employing a 9-$\sigma$
threshold and a characteristic cosmic-ray FWHM of 1.1 pixel. Optical
field distortions in the FORS field-of-view bend spectra which lie
away from the optical axis. This effect complicates an accurate sky
subtraction as tilted sky lines would be incorrectly subtracted when a
central wavelength solution is applied to the object's slit
aperture. We have, therefore, calculated a wavelength solution for
each pixel row from arclamp spectra and rectified all slit spectra
according to this distortion mask. Subsequently, the IRAF task {\sc
apall} was applied to the rectified MOS/MXU images and used to define
object and background apertures. The upper and lower boundaries of an
object aperture were adjusted so that the object flux was still higher
than the adjacent sky noise level. Limits at $\sim15$\% of the peak
flux were found to be optimal for all aperture boundaries. The same
task was also used to trace the apertures along the dispersion axis
and optimally extract the object flux according to
\cite{horne86}. During the extraction procedure the sky is modeled in
one dimension perpendicular to the dispersion axis by a linear
relation with a $\kappa$-$\sigma$ clipping to remove residual bad
pixels. Finally, all spectra are transformed into wavelength space
with an accuracy better than 0.1 \AA\ using a low-order
spline. Subsequently, we used the flux standards EG21, Feige56,
Feige110, LTT377, LTT1020, LTT1788, and LTT3864 to transform raw
counts into flux units.

From the different exposures we average all single spectra of each
object. Due to varying observing conditions (seeing, atmospheric
transparency, alignment of slit masks, etc.), the spectra of each
sub-integration series have changing signal-to-noise ratios (S/N). To
obtain a final spectrum with the highest possible S/N, we average all
single spectra of each given object and weight them by their
individual S/N. To determine the weights we calculate the S/N for each
spectrum in the range around 5000 \AA. The change in S/N between the
final spectra with and without weighting is $\sim 10$\%. The following
analysis steps make use of the optimally combined spectra.

\section{Kinematics}
\label{ln:kin}
\subsection{Radial Velocities}
\label{ln:rv}
\begin{figure}[!t]
\centering 
   \includegraphics[width=4.2cm]{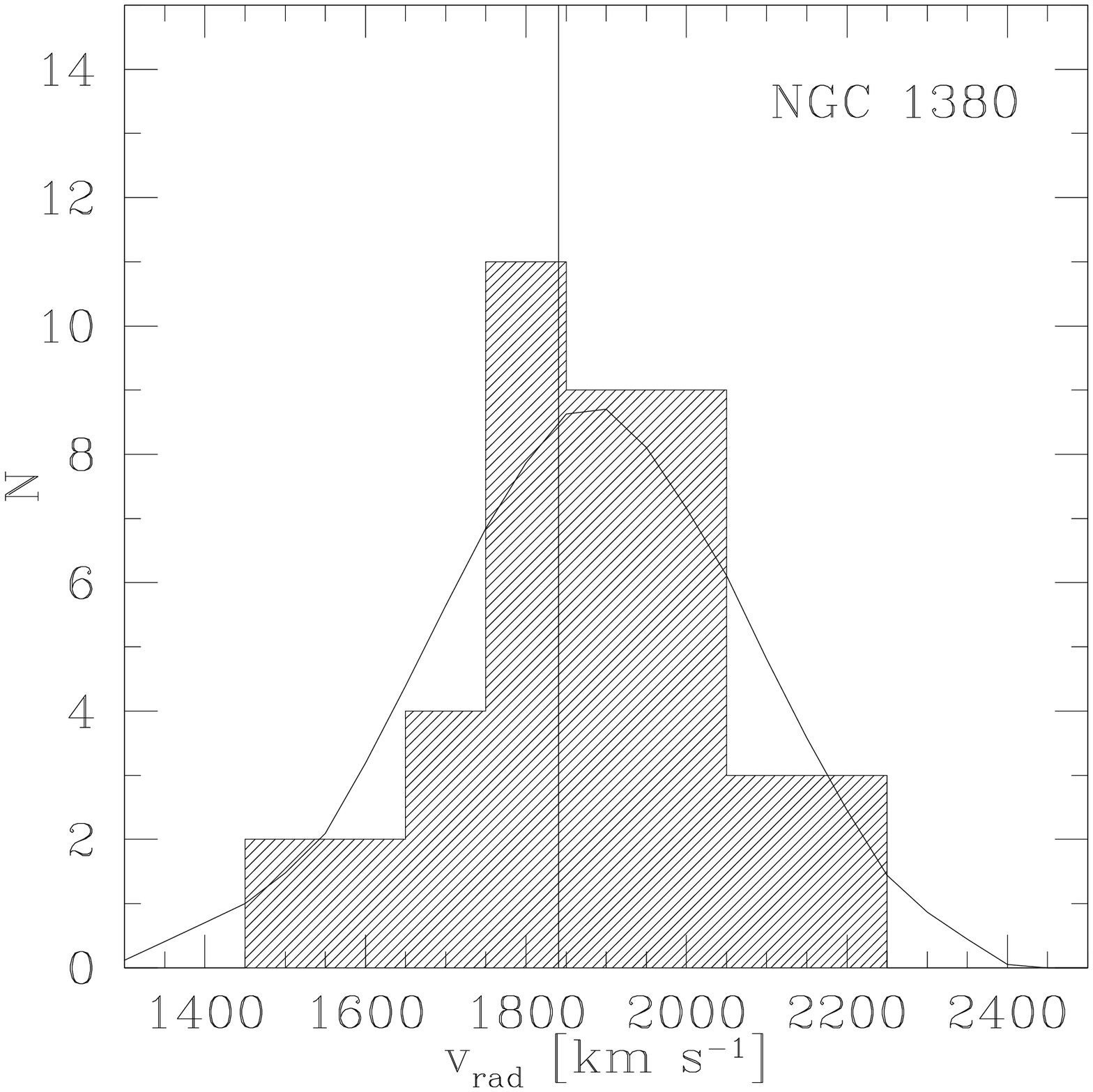}
   \includegraphics[width=4.2cm]{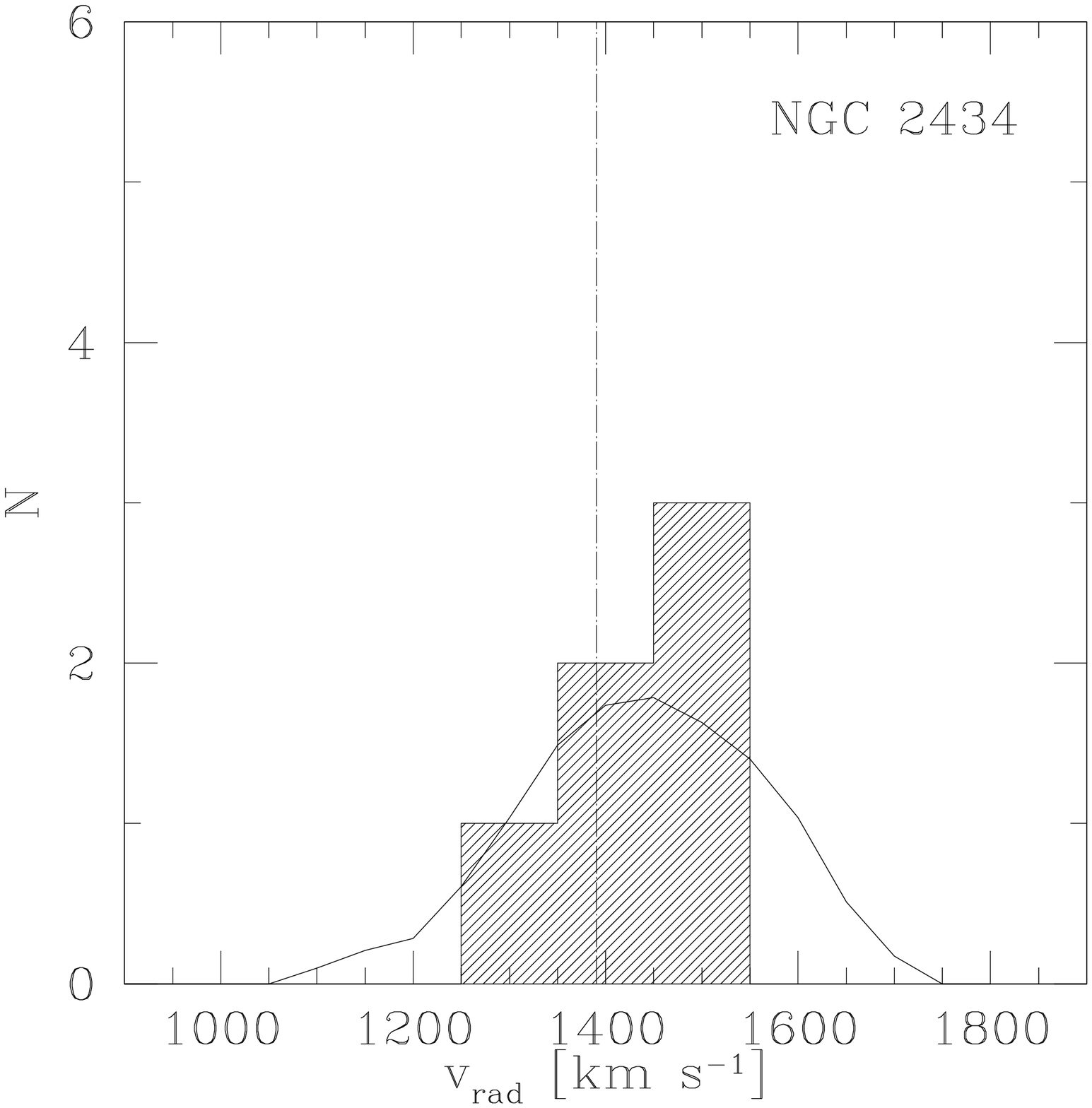}
   \includegraphics[width=4.2cm]{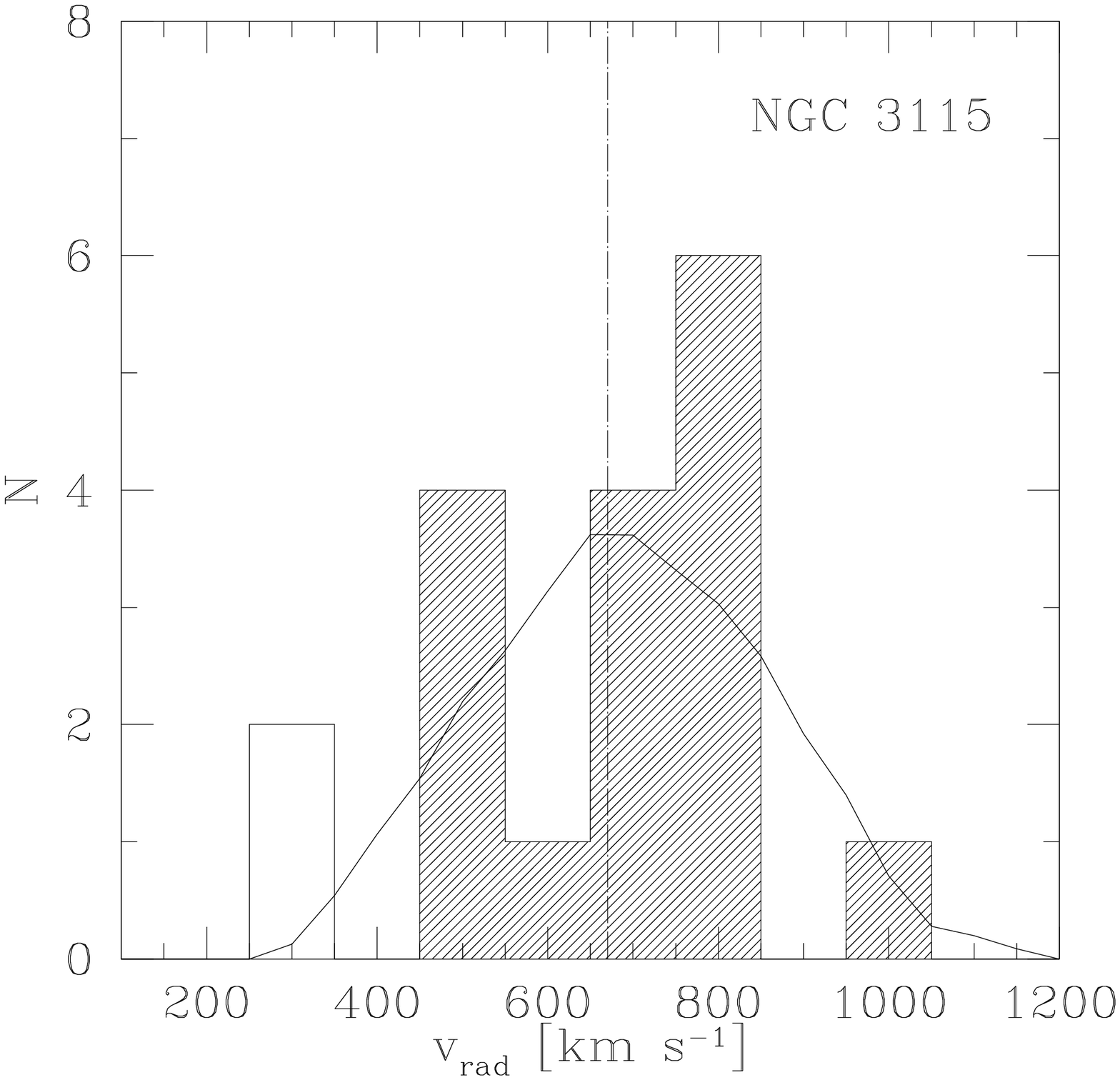}
   \includegraphics[width=4.2cm]{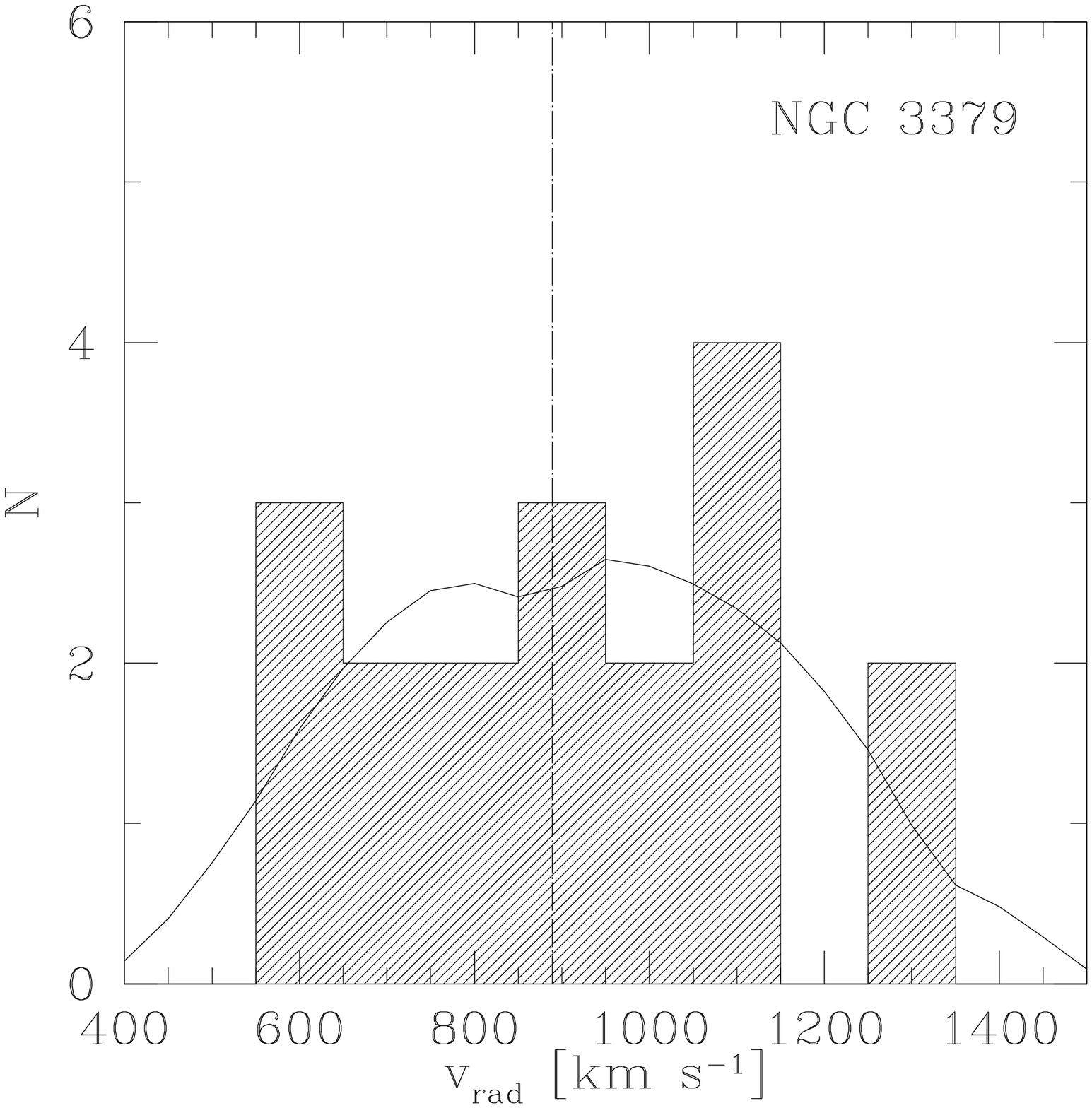}
   \includegraphics[width=4.2cm]{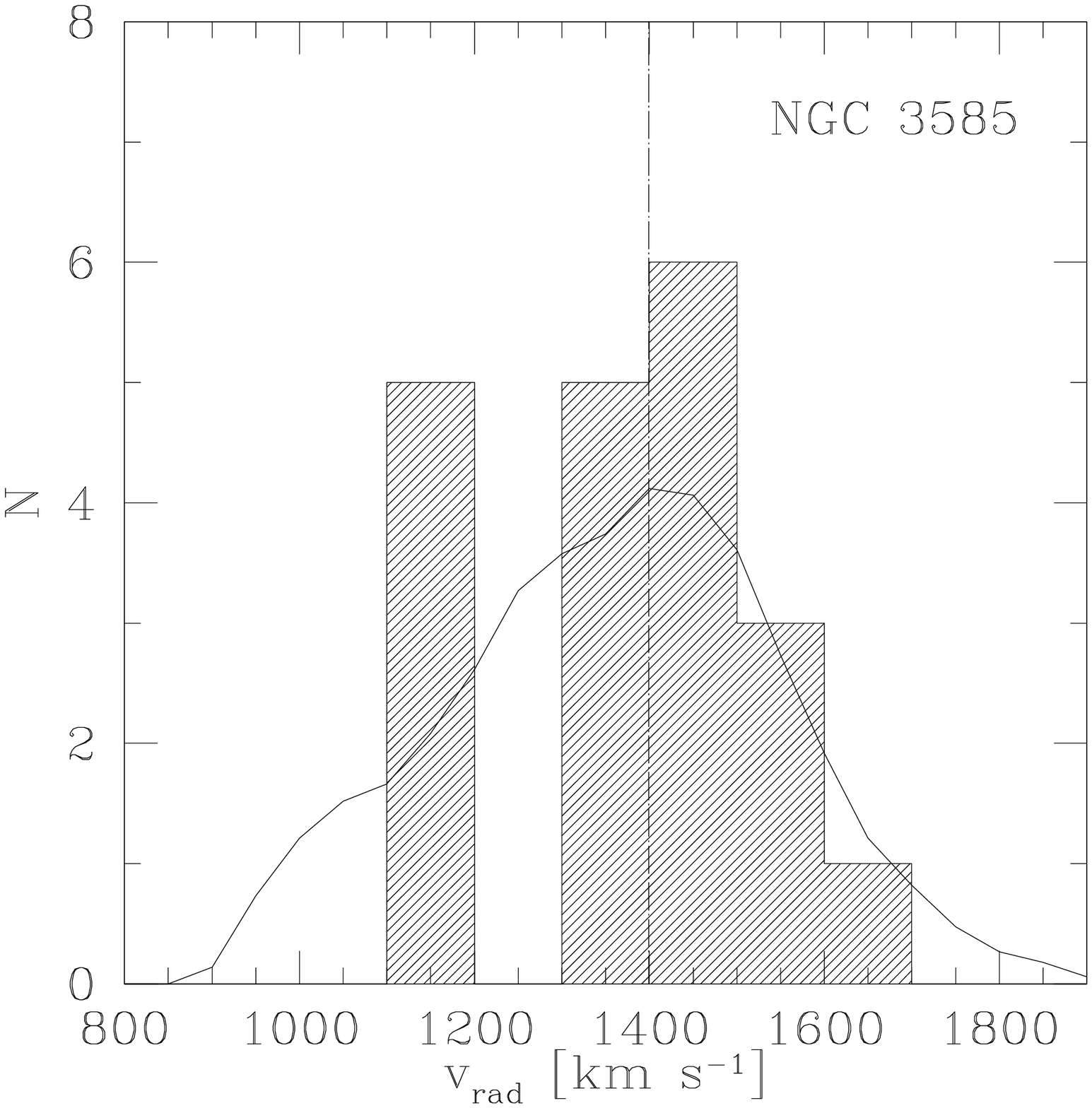}
   \includegraphics[width=4.2cm]{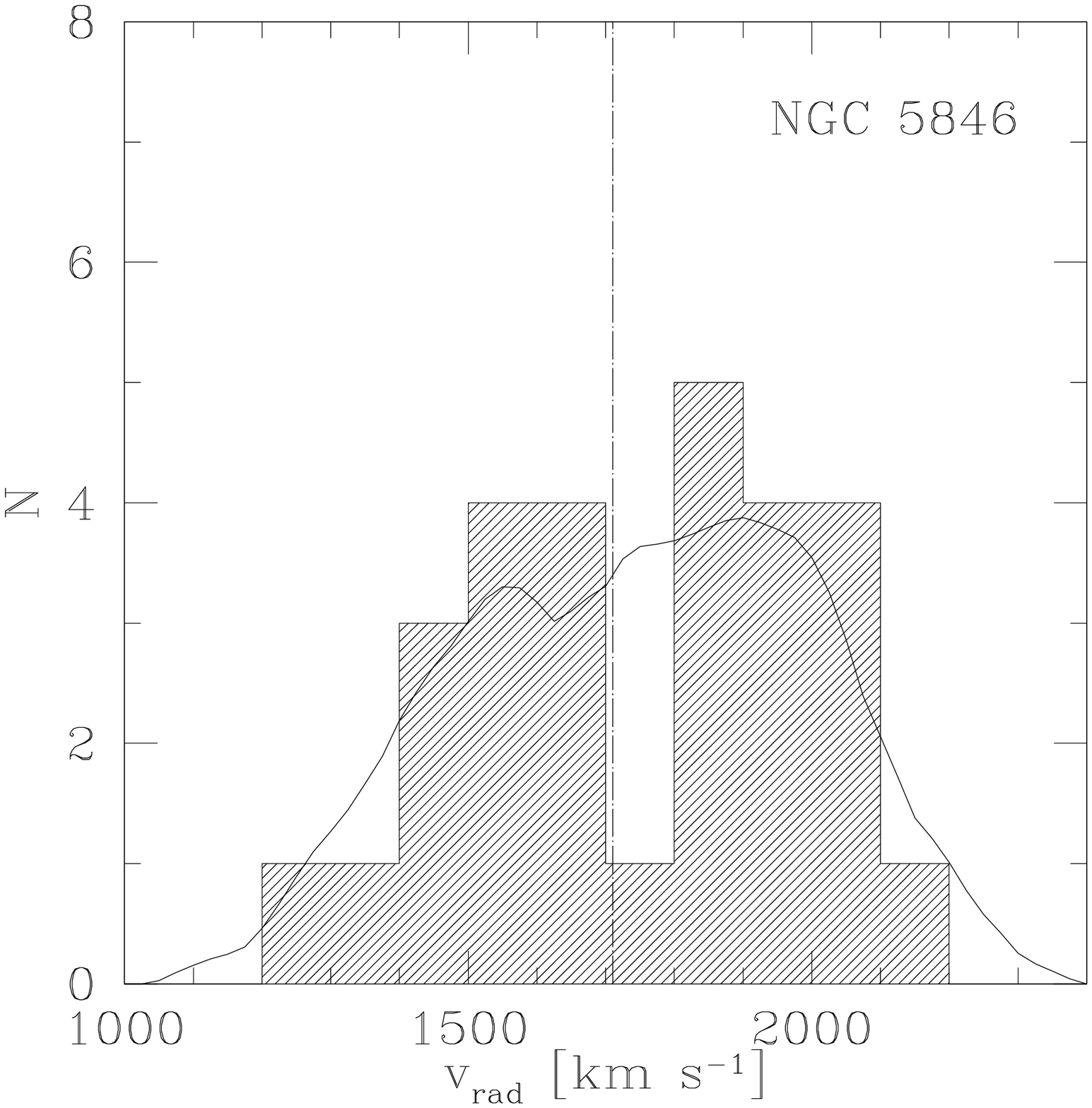}
   \includegraphics[width=4.2cm]{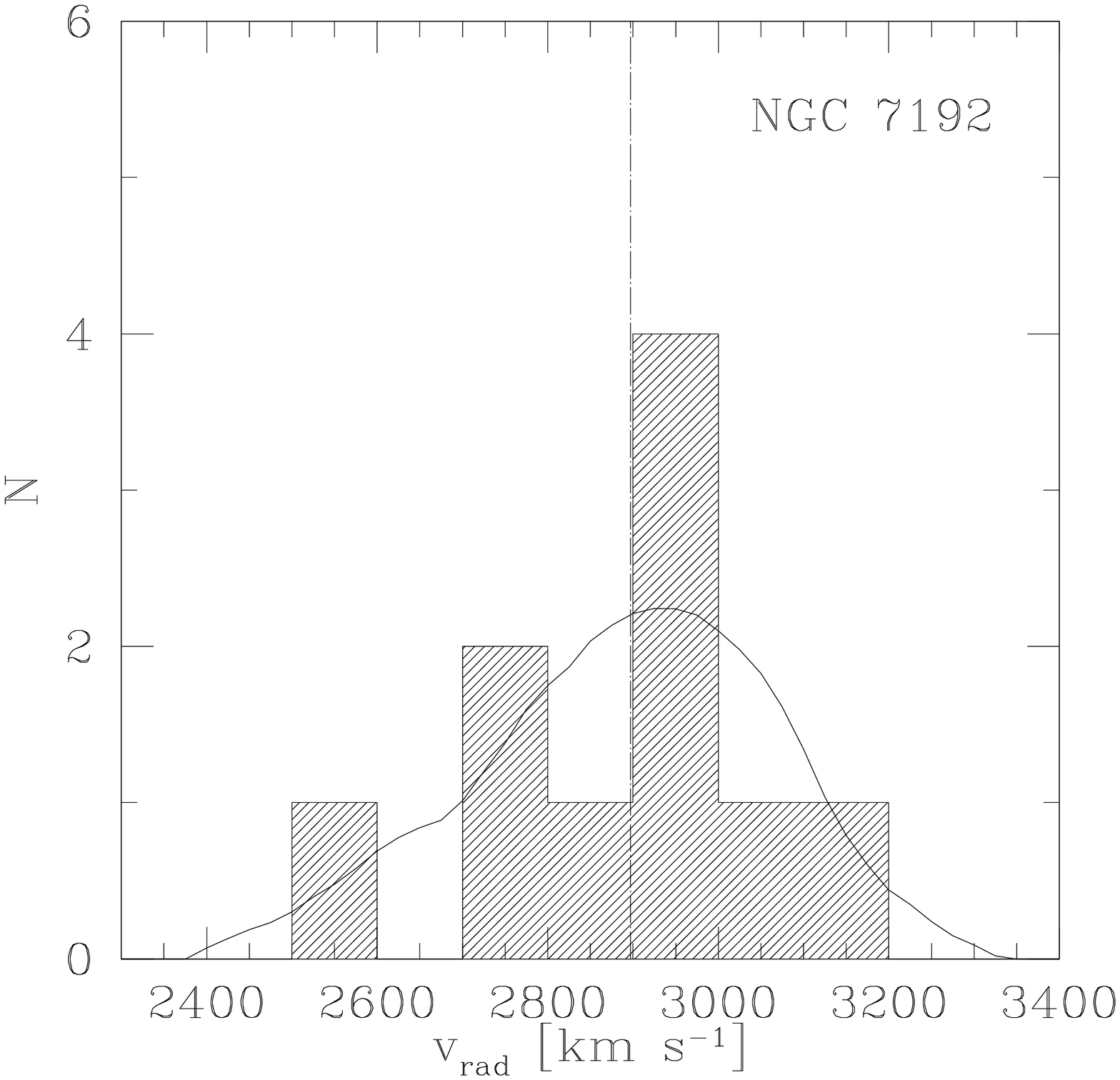} 

   \caption{Radial velocity histograms for globular clusters in all
	studied galaxies. The mean heliocentric radial velocity of the
	host was determined in the optical and is shown as a vertical
	dot-dashed line taken from the RC3 catalog \citep{RC3}. The
	solid line is a probability density estimate using an
	Epanechnikov kernel with a fixed width $\delta v_{\rm rad}=
	100$ km s$^{-1}$ \citep[for details see][]{silverman86}. Two
	ambiguous objects in NGC~3115 are indicated by an open
	histogram (see also Sect.~\ref{ln:rv}).}
\label{ps:vradhisto}
\end{figure}

Radial velocities ($v_r$) are measured by cross-correlating the
combined candidate spectra with high-S/N spectra of two M31 globular
clusters, 158-213 and 225-280 \citep{huchra82} using the IRAF task
{\sc fxcor}. The resulting heliocentric radial-velocity histograms are
shown in Figure~\ref{ps:vradhisto}. The plotted distributions are
clearly concentrated around the mean systemic radial velocity measured
from the diffuse light \citep{RC3}. We define bona-fide globular
clusters with radial velocities which are off by maximally $\pm400$ km
s$^{-1}$ of the mean systemic $v_r$. For most galaxies the distinction
between globular clusters and contaminants such as foreground stars
and background galaxies is unequivocal since the latter have either
much lower or higher $v_r$. Exceptions are two objects in
NGC~3115.

In the case of NGC~3115, objects \#10 and \#15 have relatively low
radial velocities $v_{\rm rad,10}=344\pm48$ and $v_{\rm
rad,15}=285\pm19$ km s$^{-1}$, respectively. Assuming a simple
rotation model of the Milky Way the mean streaming velocity of
foreground stars in the direction of NGC~3115
($l=247.78^o$,$b=36.78^o$) is $v_{\rm
rad}=220\cdot\sin(2l)\cos^2(b)\approx100$ km s$^{-1}$
\citep{vandekamp67}. It is therefore unlikely that the two objects are
foreground disk stars. Both are, however, not completely inconsistent
with high-velocity stars in the galactic halo. Within the colour
limits of our selection, 32 stars with magnitudes between
$V\approx21.5$ and $\sim22.5$ are expected\footnote{Model predictions
were calculated using the code available at
www.obs-besancon.fr/www/modele/modele\_ang.html} in the direction of
NGC~3115 within the FORS2 field of view \citep{robin96}. To decide
more conclusively whether the two objects are foreground stars or
genuine globular clusters their spectra deserved a detailed
investigation. However, the result was inconclusive (see
Appendix~\ref{ln:app1015}).

We conclude that the two objects cannot be assigned confidently to
either of the two groups; stars or globular clusters. Both spectra
are, therefore, kept in the globular cluster sample but we flag them
as problematic.

Using radial velocities we confirm 43 globular clusters in NGC~1380, 6
in NGC~2434, 18 in NGC~3115, 18 in NGC~3379, 20 in NGC~3585, 28 in
NGC~5846, and 10 globular clusters in NGC~7192.

\subsection{Success Rates of Globular-Cluster Candidate Selection}
\label{ln:success}

In the following, we calculate the success rate of our candidate
selection. For each galaxy, the success rate varies with changing
surface density of globular clusters and therefore with galactocentric
radius. It is instructive to calculate the success rate for our entire
sample and inside one effective radius $R_{\rm eff}$, {\it with} and
{\it without} our colour pre-selection. In other words, {\it with} and
{\it without} fill-in objects. All values are summarised in
Table~\ref{tab:succ}. In general the success rates drop for larger
galactocentric radii.

The major fraction of contaminants are foreground stars, thus it is
not surprising that the success rate correlates with galactic
latitude. High success rates ($\ga 80$\%) are guaranteed if colour
selection is applied and the field of view of the spectrograph covers
about one to two effective radii, provided that the host galaxy is
located at high galactic latitudes ($|b|\ga40^o$). Only NGC~2434
suffers from severe foreground contamination. In the case of NGC~1380,
a background galaxy cluster is placed right behind the galaxy and
contaminates the candidate selection. Unfortunately, the HST
photometry does not cover a large enough area to efficiently weed out
resolved candidates. All other galaxies have very good success rates
inside $1\, R_{\rm eff}$ typically between $\sim80$ and 100\% which
validates the efficient candidate selection.

\begin{table}[ht!]
\centering
\caption[width=\textwidth]{Success rates of photometric globular
cluster selection as a function of galactocentric radius for the
entire sample and inside one effective radius. The numbers give the
fraction of confirmed globular clusters with respect to the total
number of objects for which spectroscopy was obtained {\it without}
and {\it including} colour selection of candidates (marked by the
index ``sel''). The effective radii were taken from the RC3
catalog. Success rates for NGC~3115 were calculated including object
\#10 and \#15.}
\label{tab:succ}
\begin{tabular}{l c c c c c}
\hline\hline
\noalign{\smallskip}
Galaxy  & total & $<1\, R_{\rm eff}$ & total$_{\rm sel}$ & $<1\, R_{\rm eff,sel}$  \\
\noalign{\smallskip}
\hline
\noalign{\smallskip}
NGC 1380 & 0.44 & 0.54 & 0.75 & 0.80 \\
NGC 2434 & 0.07 & 0.06 & 0.19 & 0.11 \\
NGC 3115 & 0.82 & 1.00 & 0.88 & 1.00 \\
NGC 3379 & 0.53 & 0.77 & 0.78 & 0.90 \\
NGC 3585 & 0.57 & 0.90 & 0.73 & 0.90 \\
NGC 5846 & 0.72 & 0.78 & 0.86 & 0.90 \\
NGC 7192 & 0.29 & 0.83 & 0.50 & 0.80 \\
\noalign{\smallskip}
\hline
\end{tabular}
\end{table}

\subsection{Host Galaxy Masses}
\label{ln:galaxymasses}
The mass of a galaxy can be probed by its globular cluster system out
to large radii ($\ga2\, R_{\rm eff}$) including a significant fraction
of the halo mass. In the past, several simple mass estimators based on
the virial theorem, such as the {\it projected mass estimator}
\citep{bahcall81, heisler85}, have been developed to derive masses of
galaxy groups. Unfortunately, a key assumption of these mass
estimators is that the tracer population follows the mass density of
the probed potential. While to zeroth order this is true for galaxy
groups, the assumption fails when globular clusters are used as
tracers for galaxy potentials (see also Sect.~\ref{ln:radprof}).

Recently, a mass estimator was generalised to cases where the tracer
population does not follow the mass profile \citep{evans03}. We use
this {\it tracer mass estimator} to derive masses for our sample
galaxies using radial velocities and projected radii of our globular
cluster samples. For an isothermal potential\footnote{A basic
underlying assumption of all mass estimators is a steady state
equilibrium.}, the general form of the estimator is
\begin{equation}
M_{\rm press}=\frac{C}{GN}\sum_i (v_{i, {\rm los}}-\langle v\rangle)^2
R_i
\end{equation}
where 
\begin{equation}
C=\frac{16(\gamma-2\beta)}{\pi
(4-3\beta)}\cdot\frac{4-\gamma}{3-\gamma} \cdot\frac{1-(r_{\rm
in}/r_{\rm out})^{3-\gamma}}{1-(r_{\rm in}/r_{\rm out})^{4-\gamma}}
\end{equation}

Here, $\langle v\rangle$ is the system's mean radial velocity and
$\beta$ the anisotropy parameter $1-\sigma_t^2/\sigma_r^2$ which is
unity for purely radial orbits and $-\infty$ for a system with solely
tangential orbits \citep{binney81}. The exponent of the {\it
three-dimensional} density profile of the globular cluster population,
defined through $\rho(r)=\rho_0\cdot r^{-\gamma}$, is not known a
priori. However, to a good approximation the power-law rule $\gamma =
1+d\log\Sigma / d\log R$ \citep{gebhardt96} can be used to derive
$\gamma$ from the surface density profiles in Section~\ref{ln:radprof}
assuming spherical symmetry. The projected radii $R_{\rm in}$ and
$R_{\rm out}$ are taken as the 3-dimensional minimal and maximal
galactocentric distances $r_{\rm in}$ and $r_{\rm out}$.

The mass estimator applies only to a pressure-supported tracer
population. That is, any net rotation has to be subtracted from the
sample before the tracer mass estimator is applied. We eliminate the
mean rotational component by fitting a rotation curve to the entire
globular cluster sample following \cite{gebhardt00}. Total masses are
calculated by adding the rotational component (assuming a flat
rotation curve at large radii) to the pressure component from the
tracer mass estimator
\begin{equation}
M_{\rm total}=\frac{R_{\rm out}\langle v_{\rm max}\rangle^2}{G}+M_{\rm
press},
\end{equation}
Total masses are calculated for isotropic globular cluster orbits
($\beta=0$). For reasonably extreme anisotropies these mass estimates
were found to vary by at most $\sim30$\% \citep{evans03}. Taking into
account the uncertainties in the mean system velocity and the
rotational mass component as well as statistical fluctuations due to
the limited sample size, we expect a $30-50$\% uncertainty in the
total mass estimate. Table~\ref{tab:mass} summarizes the results for
all sample galaxies.

\begin{table*}[ht!]
\centering
   \caption[width=\textwidth]{Host galaxy masses in units of
     $10^{11}M_\odot$. The total mass was determined from the full set
     of globular clusters. Inner and outer projected radii which are
     defined by the projected radial spread of the sample are given in
     kpc. The rotational and pressure component of the total mass
     estimate are given separately. The expected uncertainty of the
     total mass estimate is $\sim30-50$\%. The last two columns show
     the total mass estimate inside $1\, R_{\rm eff}$ and the number
     of test particles. The galactocentric radius (in kpc) of the most
     distant globular cluster for the mass estimate inside one
     effective radius is given in column $R_{\rm eff,out}$.}
\label{tab:mass}
\begin{tabular}{l c c c c c c c c}
\hline\hline
\noalign{\smallskip}
Galaxy  & $R_{\rm in}$ & $R_{\rm out}$ & $R_{\rm out, <1\, R_{\rm eff}}$ & 
$M_{\rm rot}$ & $M_{\rm press}$ & $M_{\rm total}$ & $M_{\rm total, <1\,
  R_{eff}}$ & N$_{\rm GC, <1\, R_{eff}}$\\ 
\noalign{\smallskip}
\hline
\noalign{\smallskip}
NGC 1380 & 1.12 & 17.13 & 2.53 & 0.18 & 8.44 & 8.62 & 1.75 & 19\\
NGC 2434 & 1.40 & 13.80 & 2.10 & 0.09 & 0.79 & 0.88 &
0.31$^{\mathrm{a}}$ & 5\\ 
NGC 3115$^{\mathrm{b}}$ & 1.33 & 14.55 & 4.82 & 0.26 & 2.93 & 3.19 &
0.70 & 18\\ 
NGC 3379 & 0.84 & 10.41 & 2.88 & 0.09 & 2.76 & 2.85 & 0.96 & 10\\
NGC 3585 & 2.26 & 19.14 & 2.73 & 0.21 & 2.62 & 2.83 & 1.41 & 9\\
NGC 5846 & 2.31 & 24.65 & 1.63 & 0.27 & 11.6 & 11.9 & 6.38 & 18\\
NGC 7192 & 3.95 & 38.02 & 3.62 & 1.21 & 4.47 & 5.68 & 0.89 & 5\\
\noalign{\smallskip}
\hline
\end{tabular}
\begin{list}{}{}
\item[$^{\mathrm{a}}$] No data are available inside $1\, R_{\rm eff}$;
  the given mass was calculated inside $2\, R_{\rm eff}$.
\item[$^{\mathrm{b}}$] Masses were calculated using globular cluster
  data from our study and those of \cite{kavelaars98} and
  \cite{kuntschner02}.
\end{list}
\end{table*}

\section{Line Indices}
\label{ln:lineindices}
\subsection{Sampled Luminosities}
\label{ln:sampllum}
\begin{figure}[!t]
\centering 
   \includegraphics[width=4.2cm]{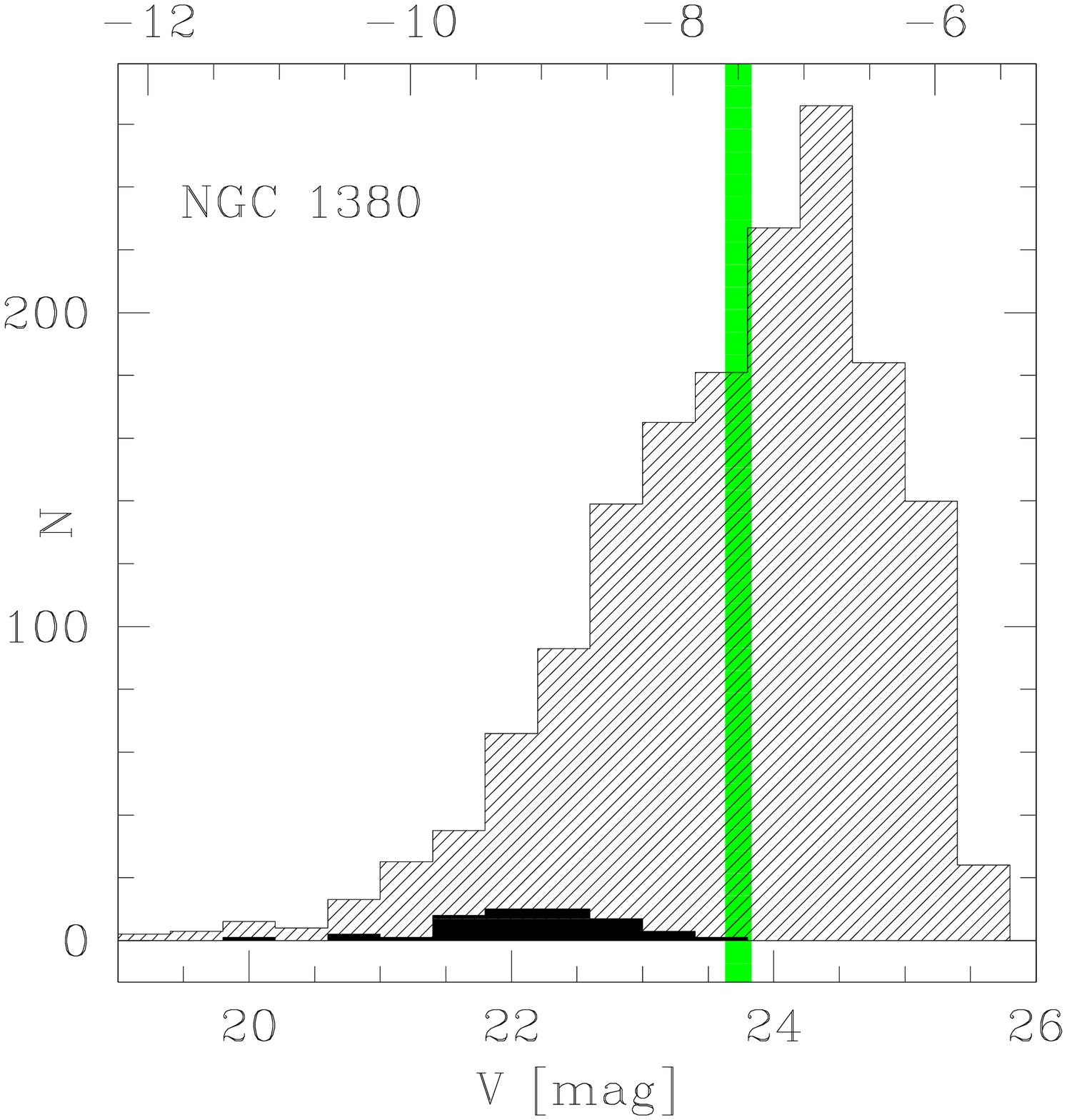}
   \includegraphics[width=4.2cm]{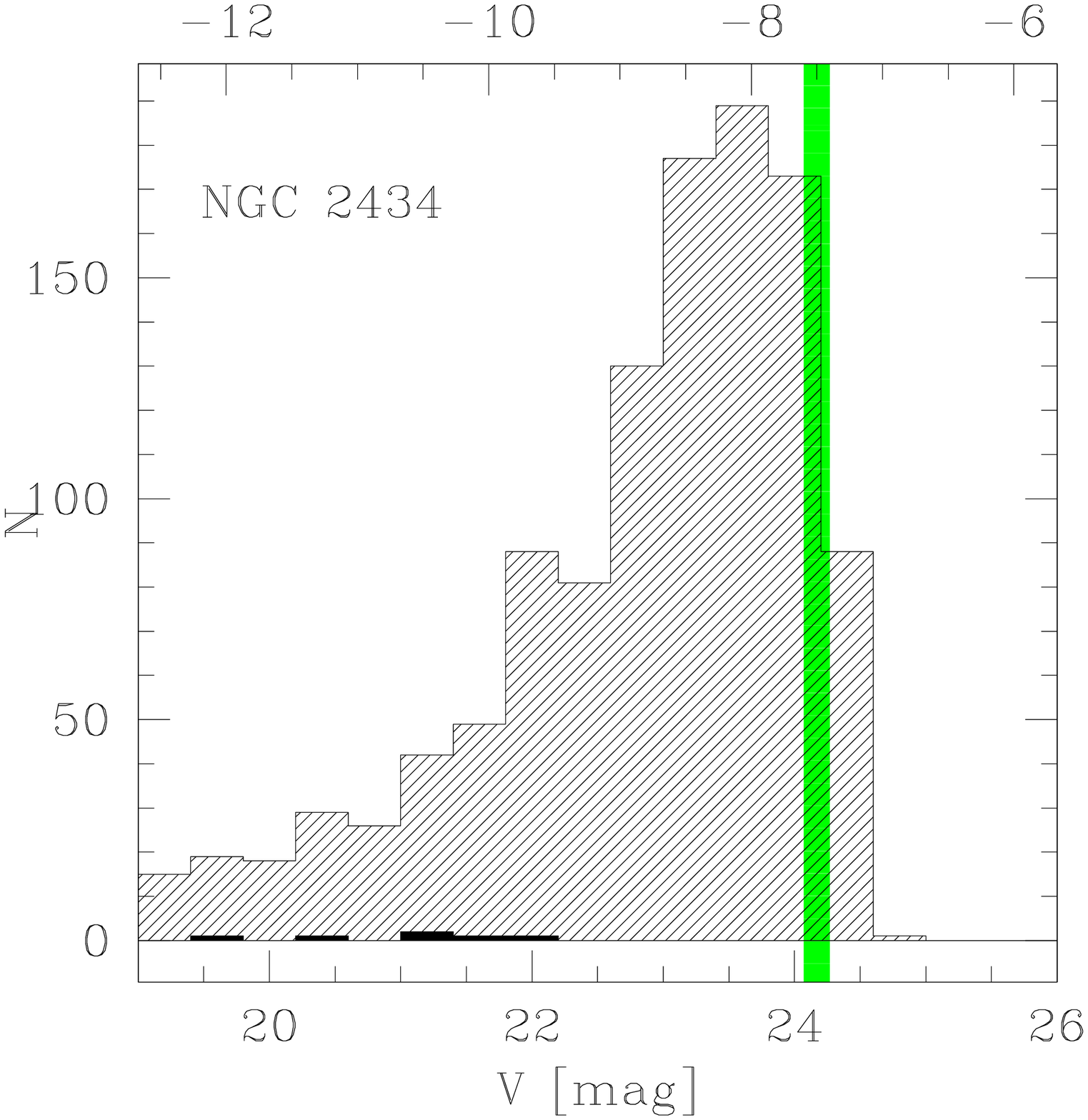}
   \includegraphics[width=4.2cm]{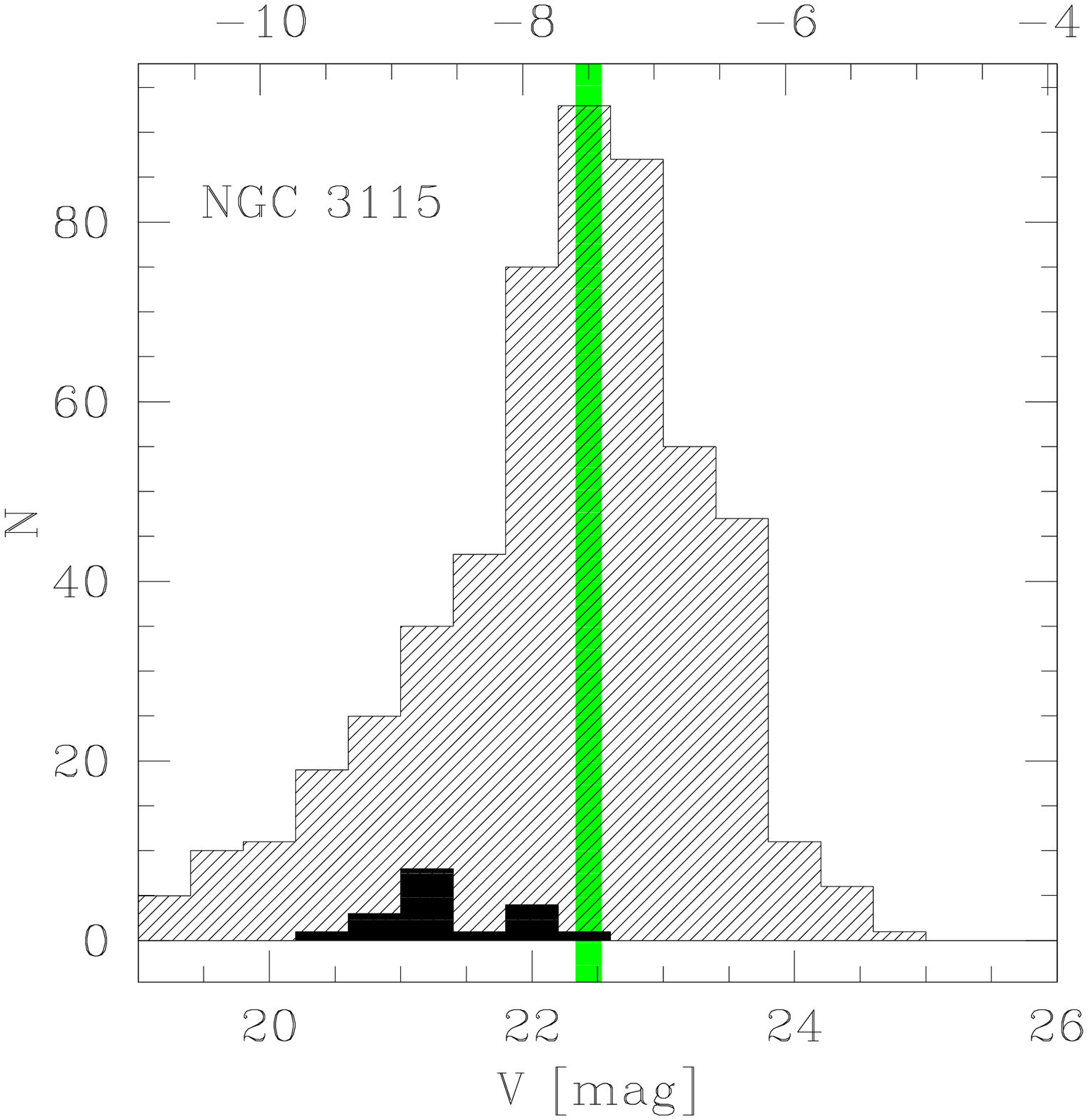}
   \includegraphics[width=4.2cm]{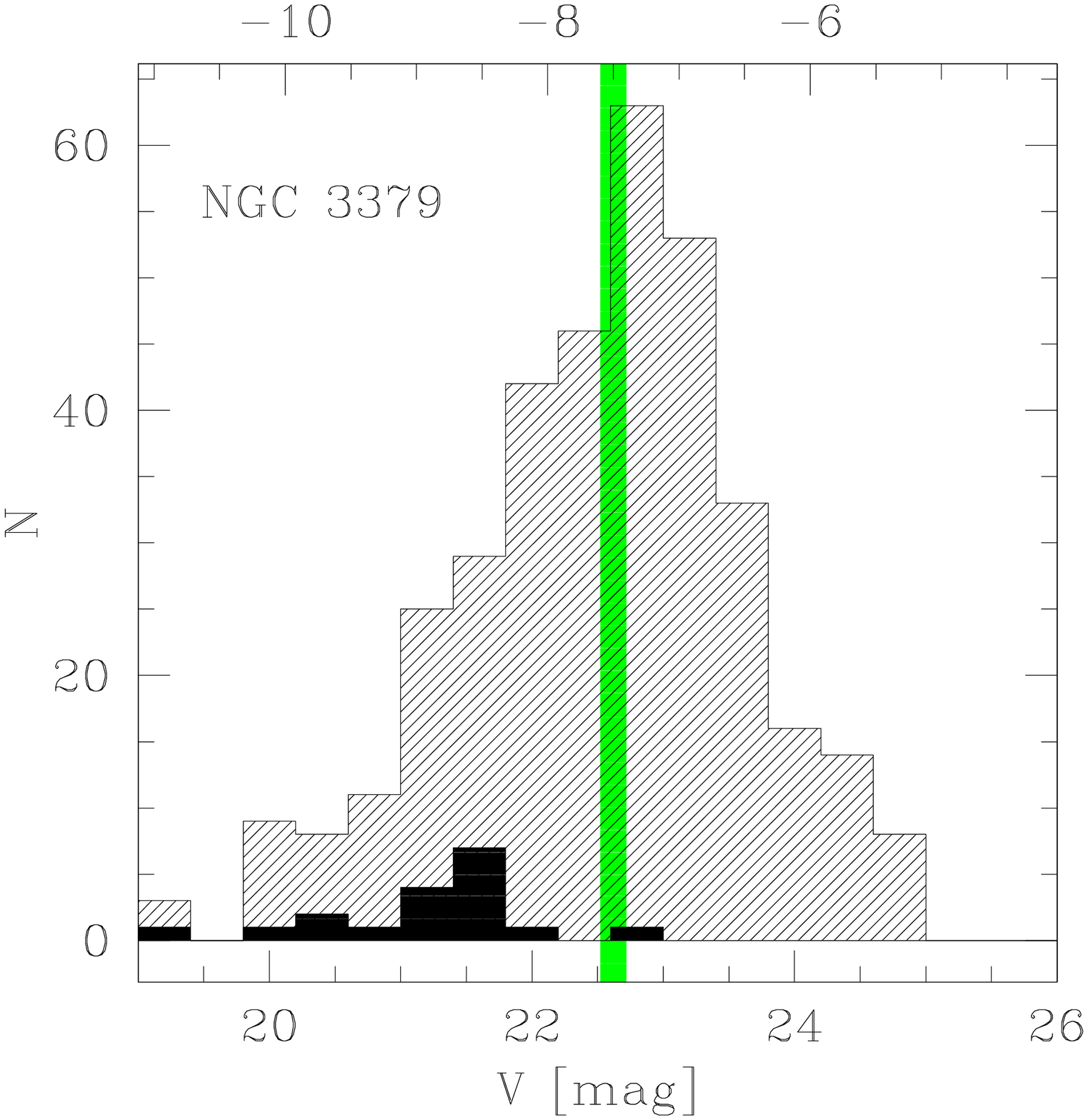}
   \includegraphics[width=4.2cm]{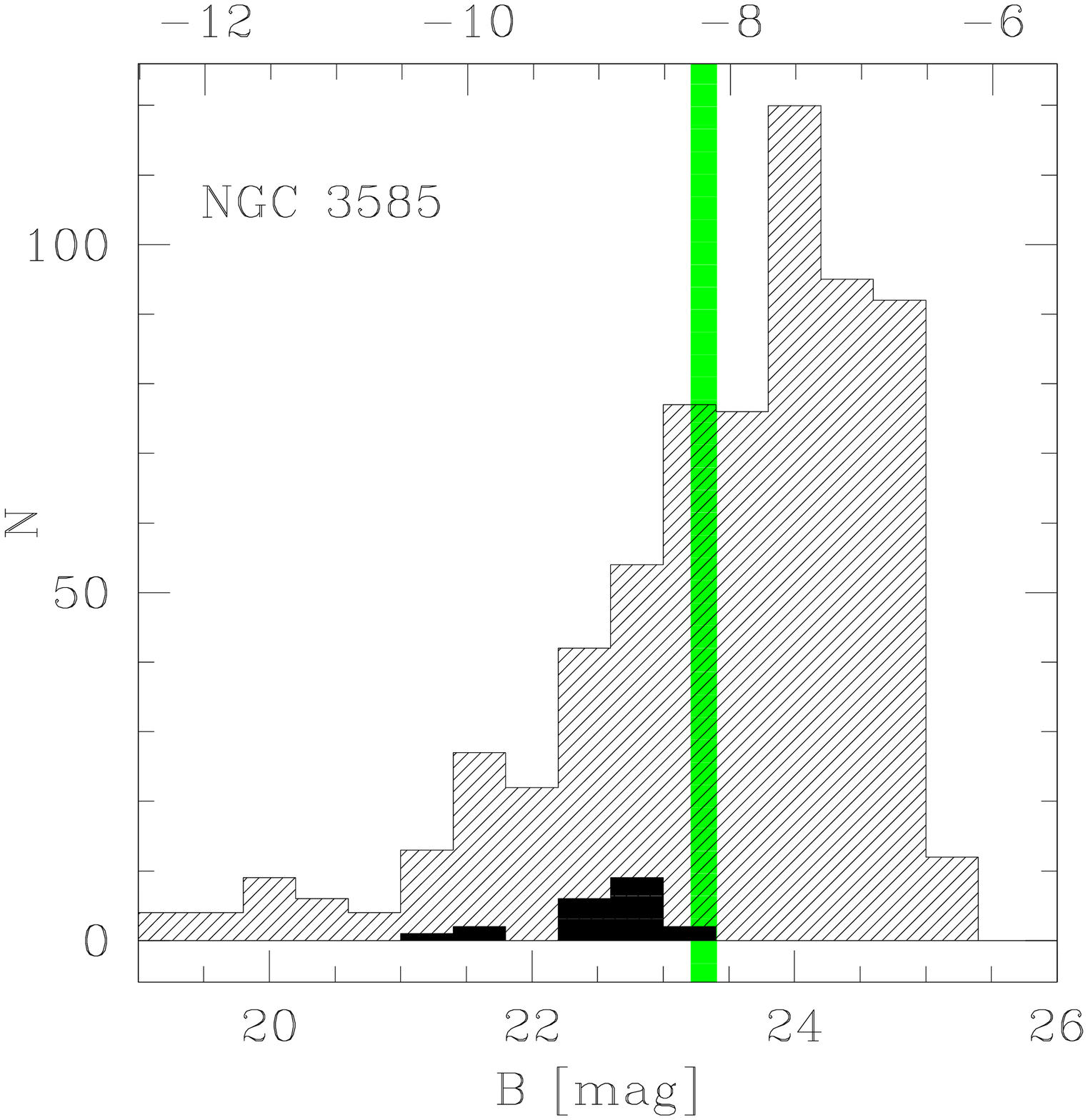}
   \includegraphics[width=4.2cm]{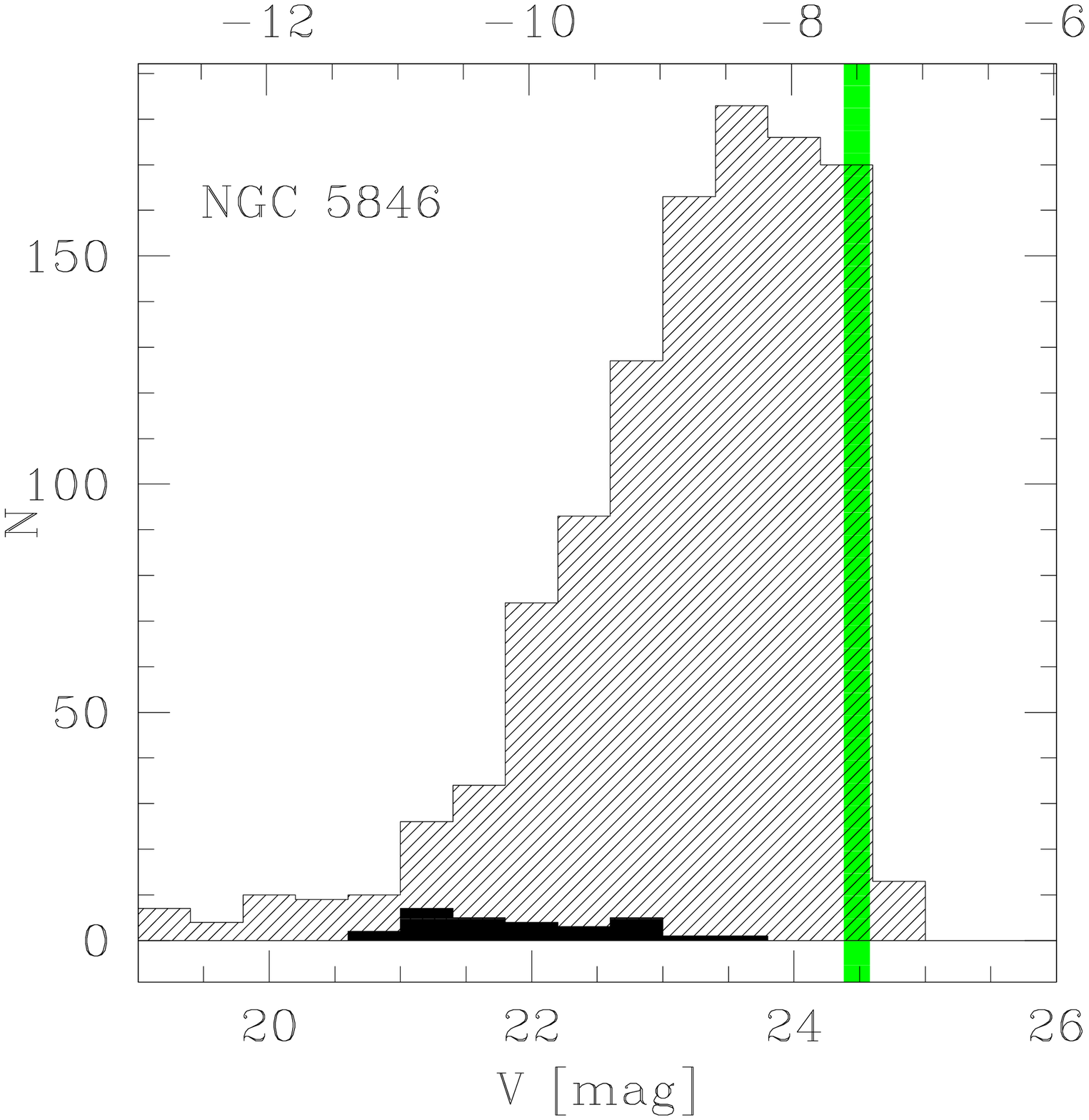}
   \includegraphics[width=4.2cm]{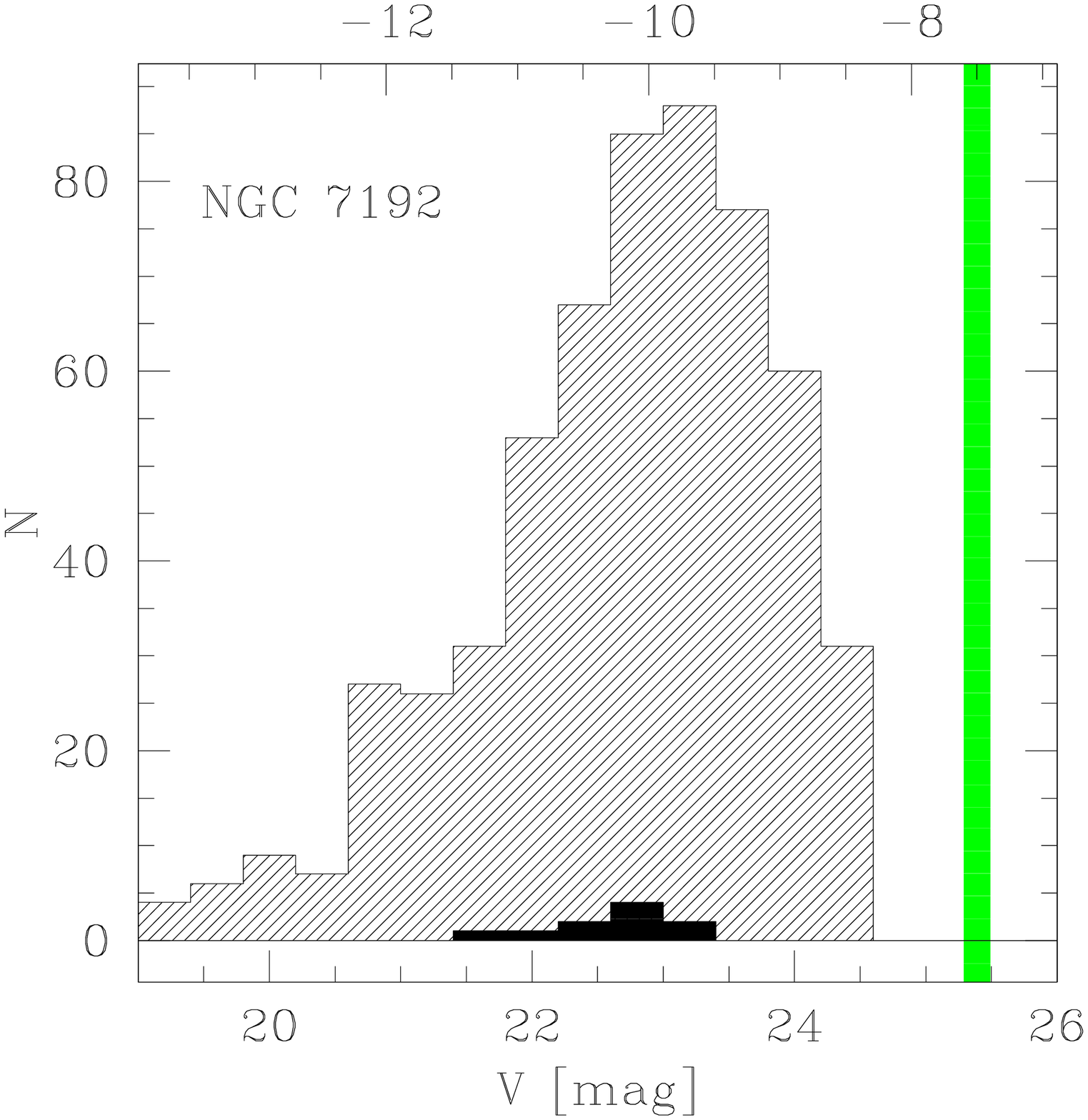} 
   \includegraphics[width=4.2cm]{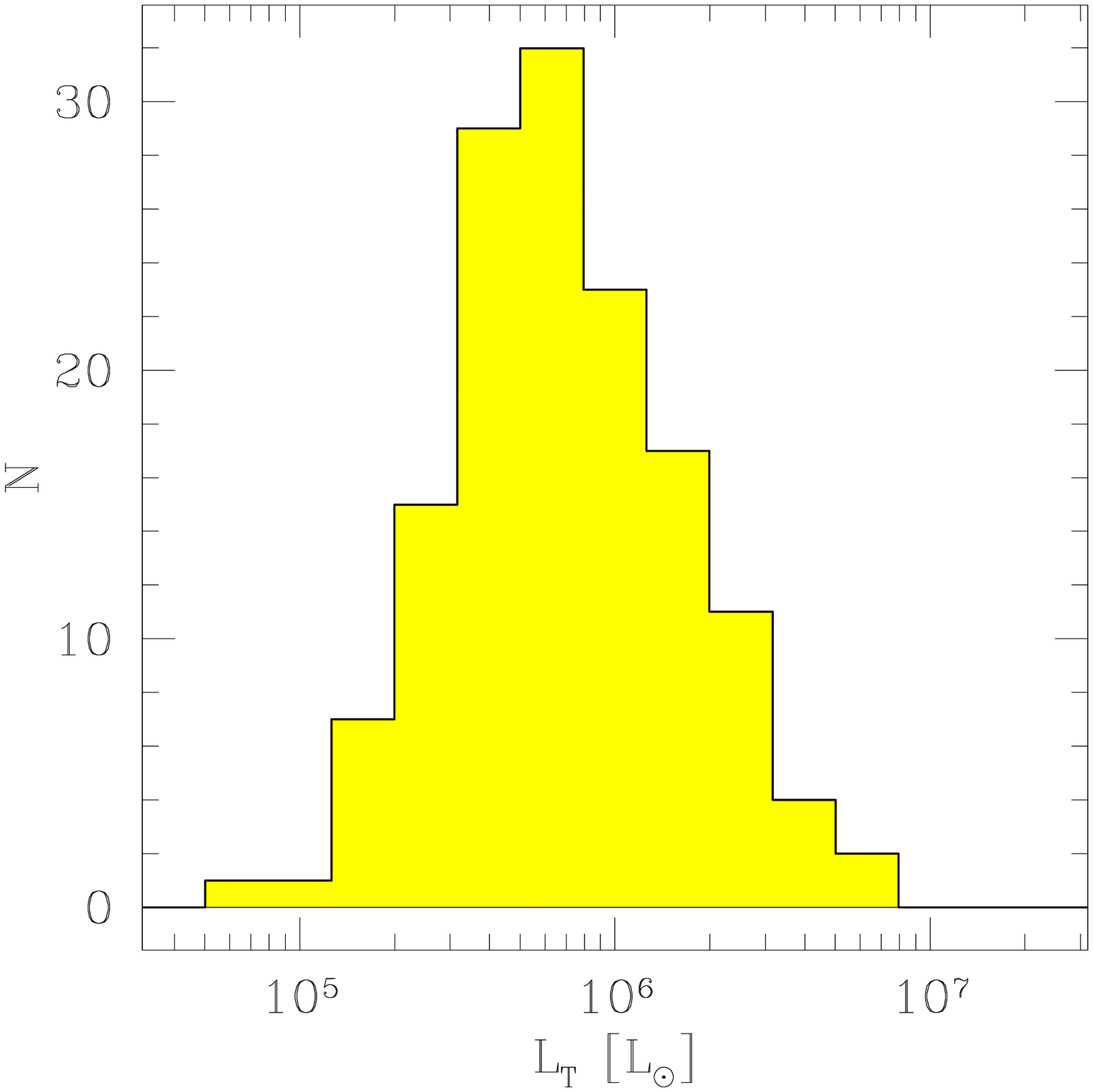}

   \caption{Observed globular cluster luminosity functions
     (GCLFs). Hatched histograms show the magnitude distribution of
     the entire photometric sample. Solid histograms are spectroscopic
     samples. A vertical shaded band indicates the location of the
     GCLF turn over which is found in Local Group globular cluster
     systems and is expected at $M_V\approx-7.4$ to $-7.6$ mag
     \citep{harris01, richtler03}. Note that the panel for NGC~3585
     refers to $B$ band magnitudes. An absolute magnitude scale is
     provided at the upper abscissa of each panel. Note that the GCLFs
     are not corrected for completeness. The lower right panel shows a
     histogram of globular cluster luminosities, $L_{\rm T}$, which
     are sampled by our spectroscopic data.}
\label{ps:vhisto}
\end{figure}

The sampling of the globular cluster luminosity function (GCLF) is
illustrated in Figure~\ref{ps:vhisto}. Our spectroscopic samples
represent, strictly speaking, only small fractions of the entire
globular cluster population and are biased towards high cluster
masses. However, from the Milky Way and photometric studies, no large
variations of globular cluster properties are expected with mass.
Typical values of the sampled fraction of the entire GCLF down to the
faintest cluster in the spectroscopic data set vary between 1.5 and
8.5\%. All spectroscopic sub-samples are biased towards bright
magnitudes and probe the bright end of the GCLF. Taking into account
metallicity and age variations inside a globular cluster system our
data will be biased towards young globular clusters if present.

Spectra of globular clusters with a total sampled luminosity of less
than $\sim10^5 L_\odot$ are likely to be dominated by stochastic
fluctuations of the number of bright stars \citep[e.g.][]{renzini88,
renzini98, puzia02c}. To convince ourselves that enough light is
sampled by the slit we estimate the total luminosity $L_{\rm T}$ of
our sample globular clusters from the photometry. We use the distance
modulus of \cite{tonry01}, foreground reddening maps of
\cite{schlegel98}, and bolometric corrections from \cite{maraston98}
in the equation
\begin{equation}
L_T=BC_I\cdot10^{-0.4\cdot(m_I-(m-M)-M_\odot-A_I)}
\end{equation}
As $I$ band photometry is available for all globular cluster systems
(see Tab.~\ref{tab:jourphot}) we use $I$ magnitudes for our
calculation. The absolute $I$ magnitude of the sun ($M_{I,
\odot}$=3.94) was taken from \cite{cox00}.

The lower right panel in Figure \ref{ps:vhisto} shows the distribution
of luminosities for the entire globular cluster sample. The mean of
the cluster luminosity distribution is $\langle \log L\rangle =
5.85\pm0.03$ with a dispersion of $\sigma=0.37$ dex. Only one globular
cluster has a total luminosity lower than $10^5 L_\odot$ (GC\#10 in
NGC~3379 with $L=7.2\cdot10^4L_\odot$, see also
Tab.~\ref{tab:n3379gcphot}). In other words, all clusters are far from
the low-luminosity regime where the integrated light can be dominated
by a few bright stars. Using the number-counts of \cite{renzini98} we
find that the total luminosity of each globular cluster is the
integral over at least a few thousand stars.

The estimate mentioned above is based on photometry which measures all
the light emitted by the cluster. The slits, however, sample less
light depending on seeing conditions and mask alignment during the
observations. Only four exposures, two NGC~5846 and two NGC~3585
frames, suffered during one night (25./26.5.2001) from bad seeing
($\sim1.5$\arcsec ). These exposures were assigned a low weighting
factor during the combining process and do not affect the final
spectrum.

\subsection{Calibration of Lick Line Indices}
\label{ln:lick}

\begin{figure*}[!ht]
\centering \includegraphics[width=16.5cm]{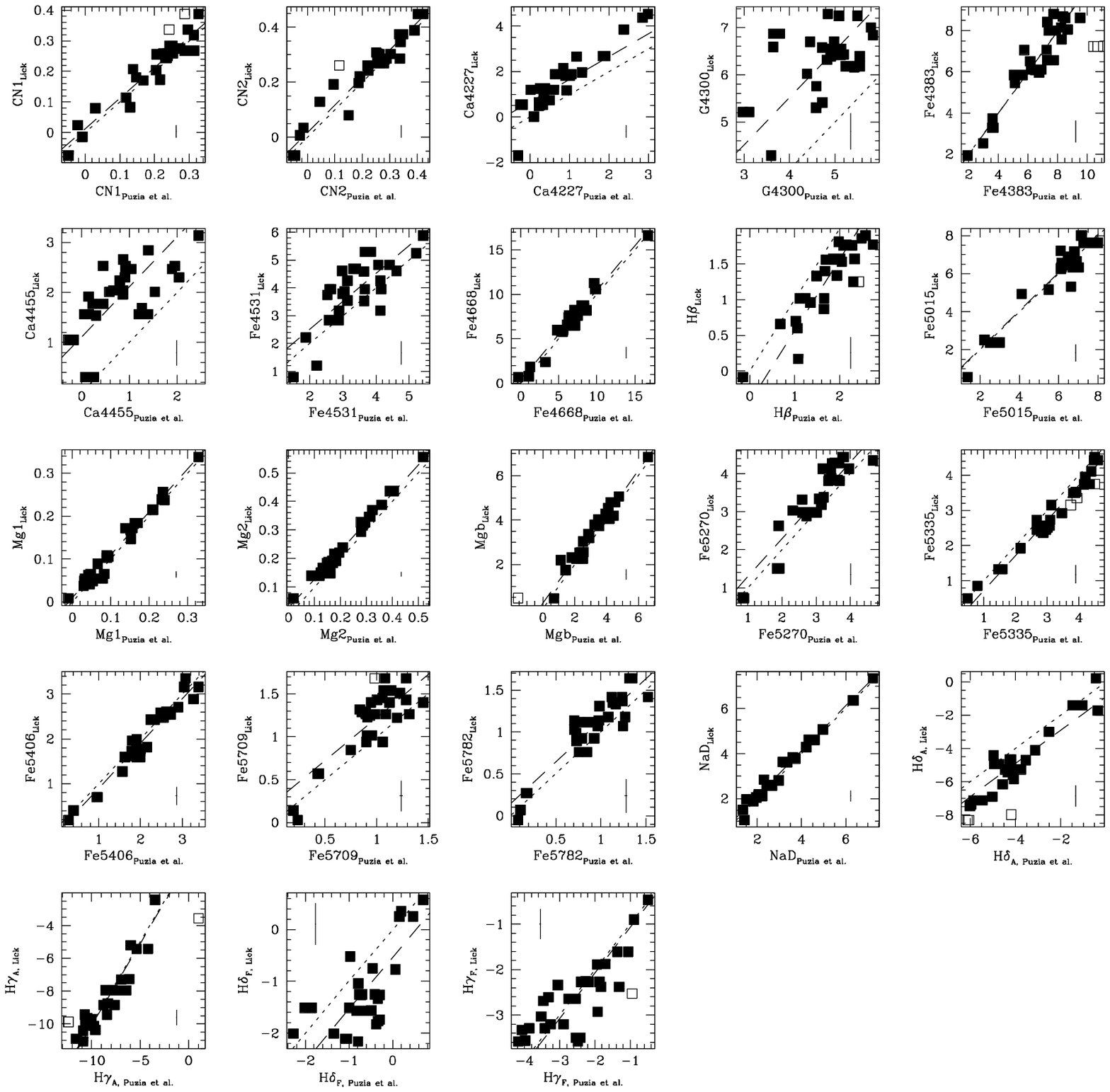}
    \caption{Comparison of Lick index measurements on our smoothed
    standard-star spectra with the published Lick index values for
    corresponding stars taken from \cite{worthey94} and
    \cite{worthey97}. Dashed lines indicate the best least-square fit
    to the data using $\kappa$-$\sigma$ clipping. Dotted lines show
    the one-to-one relation. Data with exceptionally large errors or
    large deviations which were not used in the fitting process are
    shown as open squares. Note that offsets for TiO$_1$ and TiO$_2$
    could not be determined due to the lack of wavelength coverage in
    our standard star spectra. Both these indices remain
    uncorrected. Typical error bars are indicated in a corner of each
    panel.}
\label{ps:idxcomp}
\end{figure*}

Lick indices \citep[for passband definitions see][]{worthey94,
worthey97} are measured on the fluxed and combined globular-cluster
spectra. The wavelength range coverage allows for 17\% of our sample
to have TiO$_1$ index measurements. The redder index TiO$_2$ cannot be
measured for all globular clusters. However, both indices are less
instructive as they are affected by calibration uncertainties and
stochastic fluctuation in the number of mostly contributing cool giant
stars \citep{puzia02c, maraston03, thomas03}. Prior to performing the
line-index measurements, the spectra were smoothed by a
$\lambda$-dependent Gaussian kernel to match the Lick/IDS spectral
resolution \citep{faber85, worthey97, beasley00, puzia02c}. The
transformation to the Lick system, the measurement of Lick indices,
and the error analysis is performed in the same way as described in
\cite{puzia02c}.

In particular, the transformation to the Lick system was
performed in the following way. During our observing runs in period
65, 66, and 67 we observed a total of 31 Lick standard stars which are
used to accurately tune our spectroscopic system to the Lick/IDS
characteristics. All standard-star spectra were observed with the same
slit-size (1.0\arcsec) and were extracted and smoothed in exactly the
same way as the globular cluster spectra. By comparing our
standard-star index measurements with published indices measured on
original Lick/IDS spectra \citep{worthey94, worthey97} we calculate
correction functions of the form
\begin{equation}
\label{eq:idxcorr}
I_{\rm cal} = I_{\rm raw} + \alpha
\end{equation}

where $I_{\rm cal}$ and $I_{\rm raw}$ are the calibrated and the
measured indices, respectively. These functions allow us to reliably
lock each index to the Lick system compensating for minor inaccuracies
during the smoothing process and deviant continuum-slopes, compared
with original Lick spectra, due to our flux-calibration. The
comparison of selected indices between our measurements and the Lick
systems is shown in Figure \ref{ps:idxcomp}. A major fraction of the
scatter is due to the large errors of the Lick/IDS measurements which
are about an order of magnitude larger than our standard-star
values. Most indices require only a small correction while the
calibration of very noisy indices, such as G4300 and Ca4455, remains
uncertain. All globular-cluster indices are corrected with these
zero-point offsets. Table~\ref{tab:indexlicktrafo} summarises the
correction coefficients $\alpha$ used in Equation \ref{eq:idxcorr} and
the r.m.s. of the calibration.

Calibrated indices and their uncertainties for all globular clusters
are presented in the Appendix in Tables~\ref{tab:m1n1380gcindices} to
\ref{tab:m1n7192gcindices}. A few index measurements are influenced by
bad pixels inside the background and/or feature passband due to bad
cosmic-ray interpolation. We discard these index measurements from our
data. Particularly with regard to future age and metallicity
determinations, an age resolution of $\sim1-2$ Gyr requires Balmer
line accuracies $\Delta$H$\beta\la0.05$ \AA\ and $\la0.1$ \AA\ for the
higher-order Balmer line indices, if age is considered as the only
parameter which drives Balmer indices. This is not true in general
because of the metallicity dependence of horizontal branch and turnoff
temperatures \citep{maraston03}. We will take these effects into
account in future analyses (Puzia et al. in preparation). Very few
objects achieve this high index accuracy. A metallicity resolution of
0.1 dex requires $\Delta$[MgFe]\arcmin$\la0.15$ \AA\footnote{A good
metallicity indicator is the composite index [MgFe]\arcmin$=\sqrt{{\rm
Mg}b\cdot (0.72\, {\rm Fe5270} + 0.28\, {\rm Fe5335})}$. It was
designed to be formally independent of [$\alpha$/Fe] abundance ratio
variations \citep{thomas03}.}. About 40\% of our data meets this
criterion.

Clearly, the age resolution needs to be compromised for a large enough
final sample size. Relaxed Balmer-index error cuts at
$\Delta$H$\beta=0.4$ \AA\ and 0.6 \AA\ for the higher-order Balmer
indices guarantee an age resolution of $\Delta t/t\approx0.3$. An
error cut at $\Delta$[MgFe]\arcmin$=0.2$ \AA\ corresponds to a
metallicity resolution $\sim0.25-0.4$ dex, depending on absolute
metallicity. $\sim50$\% of our sample comply with these selection
criteria and allow detailed age/metallicity for individual globular
clusters.

\begin{table}[!ht]
\begin{center}
 \caption{Summary of $\alpha$ coefficients and their rms. The
	corrections for the indices TiO$_1$ and TiO$_2$ could not be
	determined due to the lack of wavelength coverage in our
	standard-star spectra. Hence, both indices remain
	uncorrected.}
\label{tab:indexlicktrafo}
\begin{tabular}[angle=0,width=\textwidth]{lrrc}
\hline\hline
\noalign{\smallskip}
 index & z.p. -- $\alpha$ &  r.m.s. & units\\ 
\noalign{\smallskip}
\hline
\noalign{\smallskip}
CN$_1$     &  0.013 &  0.032 & mag \\
CN$_2$     &  0.017 &  0.035 & mag \\
Ca4227     &  0.664 &  0.587 & \AA \\
G4300      &  1.517 &  0.745 & \AA \\
Fe4384     &$-0.019$&  0.637 & \AA \\
Ca4455     &  1.112 &  0.554 & \AA \\
Fe4531     &  0.521 &  0.716 & \AA \\
Fe4668     &  0.471 &  0.594 & \AA \\
H$\beta$   &$-0.435$&  0.279 & \AA \\
Fe5015     &  0.061 &  0.515 & \AA \\
Mg$_1$     &  0.010 &  0.013 & mag \\
Mg$_2$     &  0.028 &  0.014 & mag \\
Mg$b$      &  0.234 &  0.304 & \AA \\
Fe5270     &  0.295 &  0.370 & \AA \\
Fe5335     &$-0.281$&  0.197 & \AA \\
Fe5406     &$-0.101$&  0.187 & \AA \\
Fe5709     &  0.223 &  0.208 & \AA \\
Fe5782     &  0.140 &  0.157 & \AA \\
NaD        &  0.118 &  0.193 & \AA \\
TiO$_1$    & \dots  & \dots  & mag \\
TiO$_2$    & \dots  & \dots  & mag \\
H$\delta_A$&$-0.899$&  0.699 & \AA \\
H$\gamma_A$&$-0.098$&  0.773 & \AA \\
H$\delta_F$&$-0.539$&  0.609 & \AA \\
H$\gamma_F$&$-0.071$&  0.575 & \AA \\
\noalign{\smallskip}
\hline
\end{tabular}
\end{center}
\end{table}

\subsection{Representative Spectra}
\label{ln:spectra}

\begin{figure*}[!ht]
\centering 
\includegraphics[width=16.0cm]{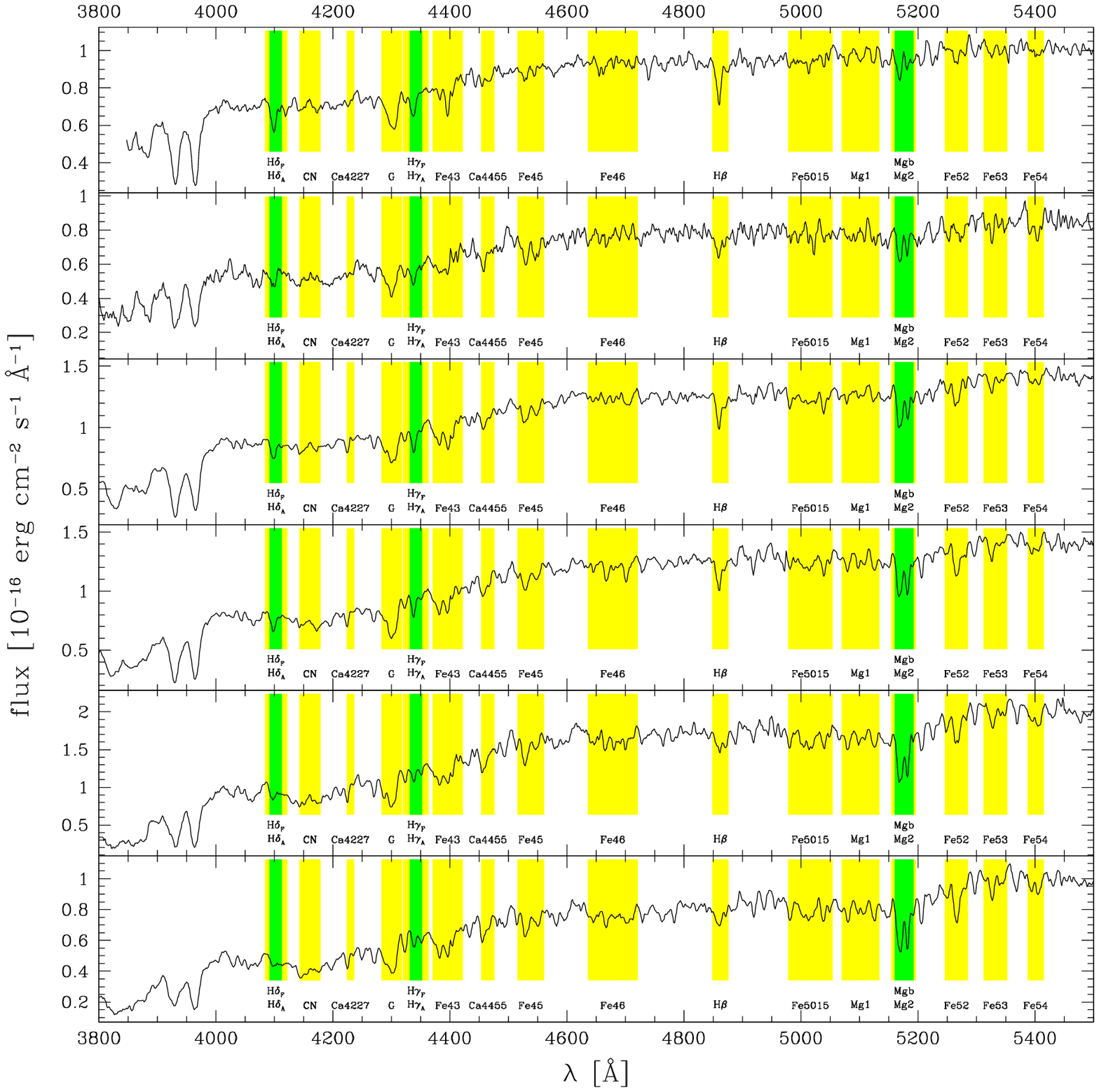}

 \caption{Representative spectra from our final globular cluster
	sample. The relative Mg$b$ index strength increases from the
	upper to the lower panel. Note the anti-correlation in the
	strength of some prominent spectral features such as Balmer
	lines and the Mg$b$ feature at $\sim5180$ \AA. All spectra are
	taken from the sample of globular clusters in NGC~5846 to
	demonstrate the influence of increasing metallicity on the
	continuum flux in the blue (note the changing ordinate scale).
	Feature passbands of measured Lick indices are shaded and
	labeled accordingly. Where two index passbands overlap the
	narrower is shaded darker and the label is elevated.}
\label{ps:spectra}
\end{figure*}

Figure~\ref{ps:spectra} shows representative spectra of NGC~5846
globular clusters of increasing Mg$b$ index strength. The sequence
shows nicely the anti-correlation between the strength of the Balmer
line series and Mg and Fe metal lines from low to high metallicities.
Note the influence of increasing metallicity on the continuum which
lowers the continuum flux in the blue part of the spectrum. In general,
most our spectra are of relatively high quality with an average S/N of
$\ga 20$ per \AA\ between 5000 and 5100 \AA.

\section{Globular Cluster Data from the Literature}
\label{ln:litdata}
In the following, we collect published spectroscopic globular-cluster
data which will be used in future papers of this series. We focus only
on high-quality line indices which were measured with the newer
passband definitions of the Lick group \citep{worthey94, worthey97,
trager98} and exclude index data measured with older passband
definitions \citep{burstein84}.

Some data are measured with the new passband definitions of
\cite{trager98}. However, most Lick-index SSP model predictions are
based the fitting functions of \cite{worthey94} and
\cite{worthey97}. Their passband definitions differ for the indices
CN$_1$, CN$_2$, Ca4227, G4300, Fe4383, Ca4455, Fe4531, C$_2$4668
(former Fe4668), Fe5709, Fe5782, NaD, TiO$_1$, and TiO$_2$ from the
passband definitions of \cite{trager98}. To compare data in a
homogeneous system, we calculate transformations to the Worthey
passband system for indices which were measured with \cite{trager98}
passband definitions. Table~\ref{tab:Tr98Wo94offsets} summarises
offsets for each index which are given in the sense
\begin{equation}
\Delta I = I_{\rm Tr98} - I_{\rm Wo94}.
\end{equation}
Note that passband definitions for all Balmer line indices and the
widely used Mg and Fe indices Mg$_2$, Mg$b$, Fe5270, and Fe5335 do not
change between the two systems.

\begin{table}[!ht]
\begin{center}
 \caption{Summary of index offsets between the Worthey et al. and
	Trager et al. passband system. Most offsets were calculated
	using our standard-star spectra and are given in the sense
	$\Delta I = I_{\rm Tr98} - I_{\rm Wo94}$ with the rms of the
	transformation. Offsets for TiO$_1$ were calculated using our
	globular cluster data since the standard star spectra do not
	cover the full wavelength range. A TiO$_2$ offset could not be
	determined due to the lack of wavelength coverage in both our
	standard-star and globular-cluster spectra.}
\label{tab:Tr98Wo94offsets}
\begin{tabular}[angle=0,width=\textwidth]{lrrc}
\hline\hline
\noalign{\smallskip}
 index & $\Delta I$ &  r.m.s. & units\\ 
\noalign{\smallskip}
\hline
\noalign{\smallskip}
CN$_1$     &  0.010 &  0.025 & mag \\
CN$_2$     &  0.014 &  0.020 & mag \\
Ca4227     &  0.330 &  0.297 & \AA \\
G4300      &  0.936 &  0.692 & \AA \\
Fe4384     &  0.166 &  0.501 & \AA \\
Ca4455     &  0.509 &  0.225 & \AA \\
Fe4531     &  0.334 &  0.127 & \AA \\
Fe4668     &  0.211 &  0.175 & \AA \\
Fe5709     &  0.211 &  0.251 & \AA \\
Fe5782     &  0.166 &  0.177 & \AA \\
NaD        &  0.001 &  0.200 & \AA \\
TiO$_1$    &$-0.003$&  0.017 & mag \\
\noalign{\smallskip}
\hline
\end{tabular}
\end{center}
\end{table}

\begin{figure}[!ht]
\centering 
\includegraphics[width=8.9cm]{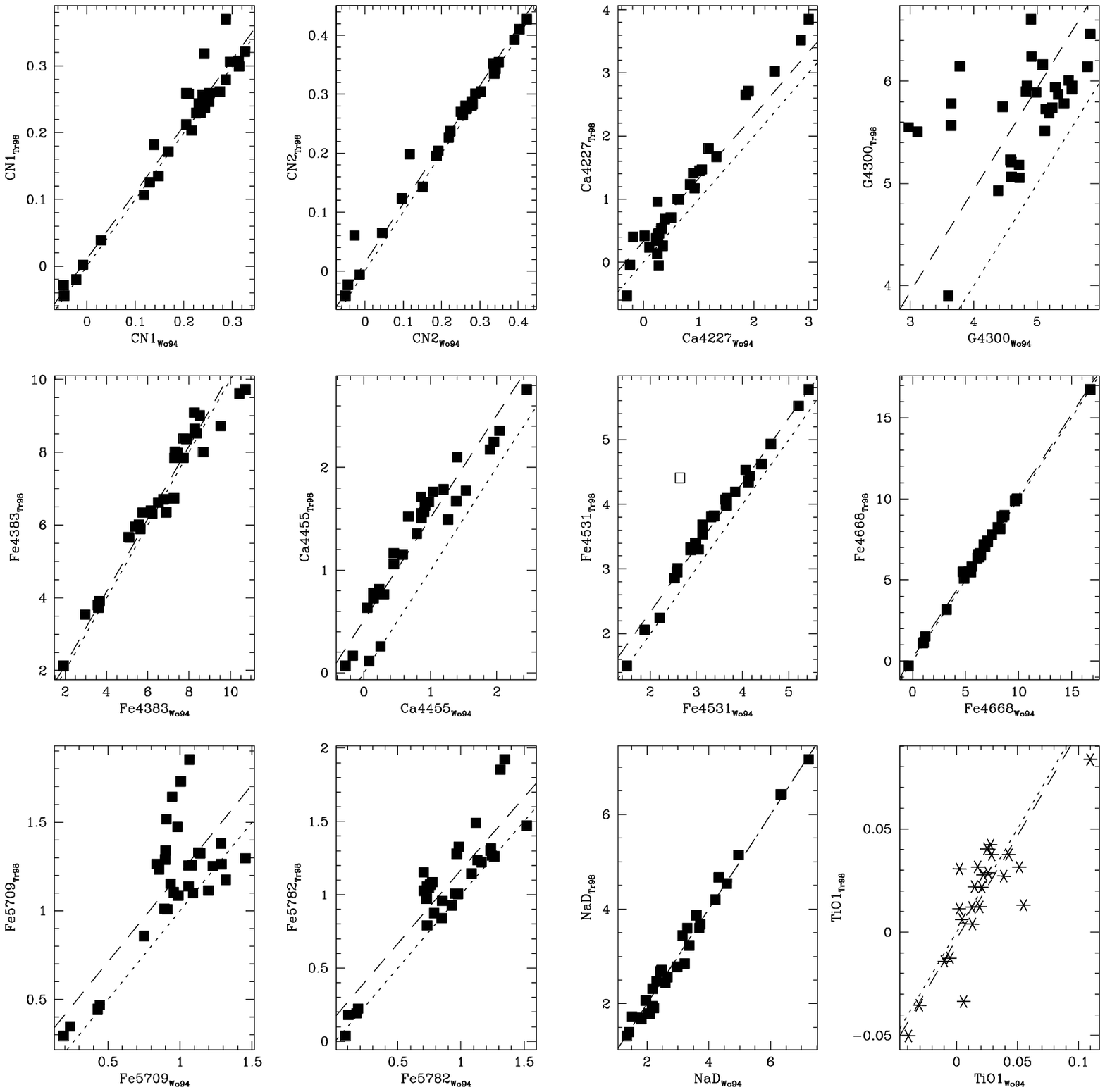}
 \caption{Comparison of Lick index measurements performed with Worthey
 et al. and Trager et al. passband definitions for which passband
 definitions differ between the two systems. Spectra of Lick standard
 stars were used, except for TiO$_1$ where we used globular cluster
 data. Offsets between the two systems are summarised in
 Table~\ref{tab:Tr98Wo94offsets}.}
\label{ps:Tr98Wo94offsets}
\end{figure}

We consider only data which have sufficiently high S/N. A short
description of each data set is given below. Age and metallicity
estimates for each globular cluster sample are taken from the papers
the data were published in.

\subsection{Elliptical Galaxies}
\begin{description}
\item [{\it NGC~1023}] - Lick indices (CN$_2$, H$\beta$, G4300,
Ca4227, Fe5270, Fe5335, and Mg$_2$) for 9 globular clusters were
measured by \cite{larsen02a} with the passband definitions of
\cite{worthey94} and \cite{worthey97} on spectra taken with LRIS
attached to the Keck I telescope. Some spectra have a S/N on the lower
limit to be useful for our future analyses. The sample spans a wide
range of cluster ages with two very young globular clusters at
$\sim500$ Myr to objects at $\sim 15$ Gyr. Metallicities range from
[Fe/H]$\approx-2.0$ to solar values.

\item [{\it NGC~1399}] - \cite{forbes01} measure a sub-set of Lick
line indices (H$\gamma_A$, H$\beta$, Mg$_2$, Mg$b$, and
$\langle$Fe$\rangle$) defined in \cite{trager98} and \cite{worthey97}
for 10 globular clusters on high-S/N spectra taken with LRIS on Keck
I. The majority of the sample are old ($\sim11$ Gyr) globular
clusters. Two clusters are likely to have intermediate ages around 2
Gyr. A broad range in metallicity is covered by the clusters with
[Fe/H] from $\sim-2.3$ to $\sim+0.4$ dex.

\item [{\it NGC 3610}] - Eight globular clusters were observed by
\cite{strader02} using the LRIS instrument on the Keck I
telescope. The relatively faint magnitudes of the clusters and the
short exposure time resulted in medium-S/N spectra on which Lick
indices (H$\delta_A$, H$\gamma_A$, H$\beta$, CN$_2$, Ca4227, G4300,
Fe5270, Fe5335, Mg$_2$, and Mg$b$) were measured using passband
definitions of \cite{trager98} and \cite{worthey97}. Except for one
intermediate-age globular cluster ($\sim3$ Gyr) with a super-solar
metallicity the entire sample appears old covering a wide metallicity
range from [Fe/H]$\approx-2.3$ to 0.0 dex.

\item [{\it NGC 4365}] - \cite{larsen03} measure a subset of Lick
indices (H$\beta$, H$\gamma_A$, H$\delta_A$, Fe5270, Fe5335, Mg$_2$,
and Mg$b$) for 14 globular clusters nine of which are likely to be
intermediate-age metal-rich objects ($-0.4\la$[Z/H]$\la0.0$, $2-5$
Gyr). The remaining clusters are consistent with old ($10-15$ Gyr)
stellar populations covering a wide range in metallicity
($-2.5\la$[Z/H]$\la0.0$). All indices were measured using passband
definitions of \cite{worthey94} and \cite{worthey97}. The data were
taken with LRIS attached to the Keck I telescope.

\end{description}

\subsection{Lenticular/S0 Galaxies}
\begin{description}
\item [{\it NGC~3115}] - High-quality spectra of 17 globular clusters
have been taken by \cite{kuntschner02} with the FORS2 instrument
at the VLT. The full set of 25 Lick line indices was measured with
passband definitions of \cite{worthey97} and \cite{trager98}. The
authors find a coeval old ($\sim12$ Gyr) set of globular clusters
which covers a wide range in metallicity from $-1.5$ dex up to solar
values.

\item [{\it NGC~4594}] - 14 globular cluster spectra of medium S/N
have been obtained by \cite{larsen02b} with the LRIS on the Keck I
telescope. Lick indices (H$\delta_A$, H$\gamma_A$, H$\beta$, Fe5270,
Fe5335, Mg$_2$, and Mg$b$) using \cite{worthey94} and \cite{worthey97}
passband definitions were measured. The sample contains globular
clusters with ages between 10 and 15 Gyr and a large metallicity
spread from very metal-poor to super-solar abundance clusters.
\end{description}

\subsection{Late-Type Galaxies}
\begin{description}

\item [{\it Milky Way}] - The full set of 25 Lick indices was measured
on high-quality spectra for 12 galactic globular clusters by
\cite{puzia02c} using passband definitions of \cite{worthey94},
\cite{worthey97}, and \cite{trager98}. The data were obtained with the
Boller \&\ Chivens Spectrograph of ESO's 1.52~m on La Silla. As the
Milky Way globular clusters consist of old stellar populations all
sample clusters have ages in the range $10-15$ Gyr. Their
metallicities range from [Fe/H]$=-1.48$ to $-0.17$ dex. This sample is
augmented by the data set of \cite{trager98} which adds 12 old
metal-poor globular clusters with metallicities from [Fe/H]$=-2.29$ up
to $-0.73$ dex. The \citeauthor{trager98} data provide line indices
for 21 passbands defined in their own work. Index measurements for
higher-order Balmer line indices which are defined by \cite{worthey97}
are documented in \cite{kuntschner02} and are added to the
\citeauthor{trager98} data set. Where the Puzia et al. and the Trager
et al./Kuntschner et al. data set have objects in common we prefer data of 
\cite{puzia02c} over the other two because of systematically smaller
uncertainties.

\item [{\it M31}] - \cite{trager98} measure Lick line indices defined
in the same work for 18 globular clusters. Index measurements for
higher-order Balmer lines are provided by \cite{kuntschner02} and
added to the former data.

\item [{\it M33}] - The Lick indices G4300, H$\beta$, Mg$_2$, Fe5270,
and Fe5335 were measured by \cite{chandar02} with the passband
definitions of \cite{worthey94} for 21 globular clusters. The sample
clusters have metallicities from [Fe/H]$=-2.0$ to $-0.5$ dex and ages
from a few Gyr to $\sim15$ Gyr. The data were taken with the HYDRA
multifiber spectrograph at WIYN 3.5m telescope (KPNO).

\item [{\it M81}] - \cite{schroder02} measure Lick indices with
passbands defined in \cite{trager98} for 16 globular cluster
candidates. Their data were obtained with the LRIS instrument. Most of
the objects are consistent with stellar populations spanning ages
between 8 and 17 Gyr and metallicities between [Fe/H]$=-2.3$ and solar
values. One globular cluster candidate appears to be of intermediate
age ($\sim3$ Gyr) and intermediate metallicity
([Fe/H]$\approx-1.0$). As M~81 has a negative systemic radial velocity
\citep[$-34\pm4$ km s$^{-1}$][]{RC3} the selection of globular
clusters remains rather uncertain. \cite{schroder02} reduce the
ambiguities by restricting the sample to objects with small projected
radii. Furthermore, they compare the strength of Ca I and H$\delta$
absorption lines with photometric colours and exclude stars using the
technique of \cite{perelmuter95}.

\item [{\it LMC}] - Lick indices of 24 globular clusters have been
measured on high-S/N spectra by \cite{beasley02}. The authors use
passband definitions of \cite{trager98} to measure their 16 bluest
indices (CN$_1$ to Fe5406) and definitions of \cite{worthey97} to
measure 4 higher-order Balmer line indices. Their sample spans
metallicities from [Fe/H]$=-2.1$ up to solar values with globular
cluster ages of a few million years up to old objects of $\sim15$
Gyr. The observations were performed with the FLAIR instrument at the
1.2 m UK Schmidt telescope (AAO).

\item [{\it Fornax}] - \cite{strader03} provide a sub-set of Lick
index measurements (CN$_1$, CN$_2$, Ca4227, G4300, H$\beta$, Mg$_2$,
Mg$b$, Fe5270, Fe5335, H$\gamma_A$, and H$\delta_A$) with passbands
defined by \cite{trager98} and \cite{worthey97} for 4 globular
clusters in the Fornax dwarf galaxy. The clusters appear to be
metal-poor ([Fe/H]$\approx-1.8$) and old ($\sim15$ Gyr) with one
cluster being younger by $\sim2-3$ Gyr. The objects were
observed with the LRIS instrument.
\end{description}

\section{Summary}
\label{ln:summary}
We present a homogeneous set of Lick indices for 143 extragalactic
globular clusters in seven early-type galaxies located in different
environments. The indices were measured on high-quality VLT spectra and
are currently the largest homogeneous spectroscopic data set of
extragalactic globular cluster systems.

The candidate pre-selection for follow-up spectroscopy was confirmed
to work very efficiently. Inside one effective radius the success
rates are between $\sim80-100$\% for galaxies located at high galactic
latitudes ($|b|\ga40^o$).

We provide a method to reduce the number of contaminating fore- and
background objects during the candidate selection. A combination of
near-infrared and optical colours in a $I-K$ vs. $B-K$ colour-colour
diagram allows to disentangle foreground stars and background galaxies
from the globular cluster population very efficiently. Fractional
contamination can be reduced to $\la10$\%.

We fit surface brightness and surface density profiles to the galaxy
light and the globular cluster system and find that globular cluster
systems have in general comparable or more extended profiles than the
galaxy light. By dividing the clearly multi-modal globular cluster
populations in blue and red sub-samples, we find that both have
similar profile slopes. A brute-force division of the remaining
single-peak systems reveals that the red globular cluster
sub-population is more concentrated than its blue counterpart. This
interesting point will be further analysed in a future paper.

Using the radial velocity information of our globular cluster samples
we measure dynamical masses for the seven host galaxies which have
total masses between $\sim8.8\cdot10^{10}M_\odot$ and
$\sim1.2\cdot10^{12}M_\odot$.

The accuracy of index measurements allows an age resolution $\Delta
t/t\approx0.3$ and a metallicity resolution in the range
$\sim0.25-0.4$ dex depending on the absolute metallicity. Hence,
$\sim50$\% of our data allows detailed age/metallicity determinations
for individual globular clusters.
\begin{acknowledgements}
    We thank Rupali Chandar for providing an electronic list of her
    globular cluster line index measurements and Karl Gebhardt for
    helping with his analysis software. THP gratefully acknowledges
    the support by the German \emph{Deut\-sche
    For\-schungs\-ge\-mein\-schaft, DFG\/} project number
    Be~1091/10--2.
\end{acknowledgements}

\bibliographystyle{apj}


\appendix
\section{Spectra of Objects \#10 and \#15 in the NGC~3115 Globular
  Cluster Sample}
\label{ln:app1015}

\begin{figure*}[!ht]
\centering 
   \includegraphics[width=\textwidth]{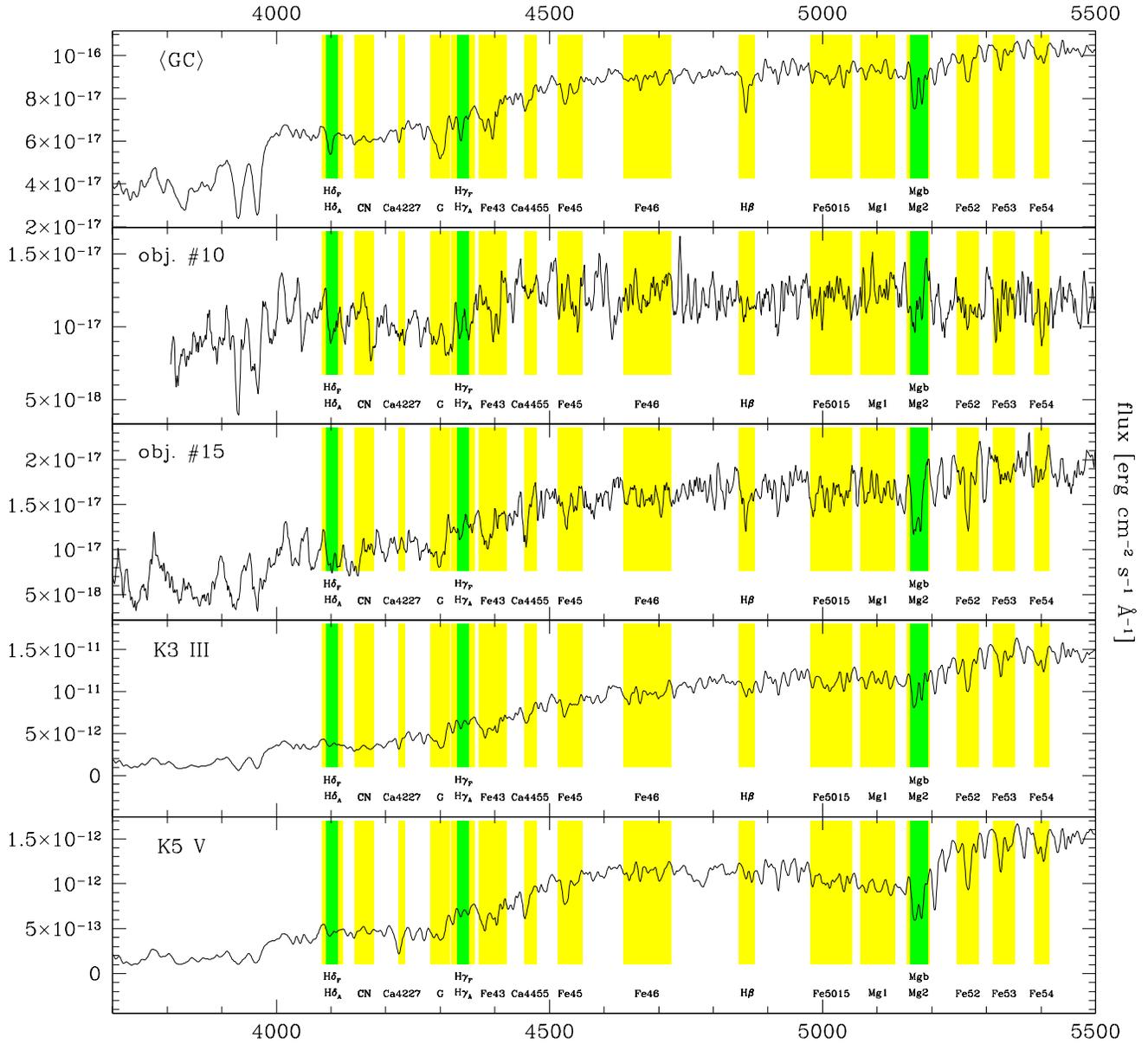}

   \caption{Spectra from top to bottom: mean globular cluster spectrum
	of NGC~5846 globular clusters, object \#10 and \#15 in
	NGC~3115, a K3 giant, and a K5 dwarf spectrum. The latter two
	spectra were taken from our sample of Lick standard stars. All
	spectra are in rest frame and were smoothed to the Lick system
	resolution. Light shaded regions indicate feature passbands of
	Lick indices. Dark shaded regions indicate narrow passbands
	which overlap with broader passbands. The label for the narrow
	index is elevated.}
\label{ps:obj1015}
\end{figure*}

Spectra of stars and globular clusters can partly be disentangled by
means of relative line strengths of spectral features such as Balmer
lines and Ca I (4227 \AA) \citep{perelmuter95}. In globular cluster
spectra the intensity of Balmer lines dominates in general the
strength of the Ca I feature. This ratio approaches unity at high
metallicities. On the other hand, dwarfs later than $\sim$G3 V show a
strong Ca I feature relative to Balmer lines. This is illustrated in
Figure~\ref{ps:obj1015} where the spectra of objects \#10 and \#15 are
compared with an average globular cluster spectrum (using bona-fide
globular clusters in NGC~5846) and two stellar spectra of a late-type
dwarf (K5 V) and a cool giant (K3 III).

The Ca I feature is not detectable in both object spectra (\#10 and
\#15). However, judging by the relative strengths of the Mg$b$ feature
and the Balmer line H$\beta$ both spectra are closer to the mean
globular cluster spectrum than to both stellar spectra. This is less
obvious for object \#15 than for the spectrum of object \#10.

The two objects have rather red colours: $I-K=1.02$ and $B-K=2.62$ for
\#10, and $I-K=2.04$ and $B-K=4.40$ for \#15. The colours of object
\#10 are consistent with late-type G dwarfs. Object \#15, on the other
hand, is too red for its spectrum to be a dwarf. In this case strong
molecular absorption bands would be detectable which is not the
case. Its colours in combination with the type of the spectrum are
rather consistent with a K giant \citep{cox00}.

As an additional test we cross-correlate both object spectra with the
two stellar spectra and the mean globular cluster spectrum and use the
height of the cross-correlation peak (CCP) as a measure of
similarity. It is important to note that the two stellar spectra and
the mean globular cluster spectrum, which are considered as templates,
have similarly high S/N values. For object \#10, the test yields the
highest CCP for the globular cluster spectrum. The spectra of object
\#15 and the K giant are most alike by means of this test. Both
results are, however, not significantly different from the
cross-correlations with the remaining template spectra.

\section{Lick line index measurements}
\label{ln:lickmeasurements}
The following tables contain all globular cluster line index
measurements including statistical and systematic radial velocity
errors. All tables are arranged by slit-mask observation.

\include{n1380m1tab1}
\addtocounter{table}{-1}
\newpage
\include{n1380m1tab2}

\newpage
\include{n1380m2tab1}
\addtocounter{table}{-1}
\newpage
\include{n1380m2tab2}

\newpage
\include{n2434m1tab1}

\newpage
\include{n3115m1tab1}
\addtocounter{table}{-1}
\newpage
\include{n3115m1tab2}

\newpage
\include{n3379m1tab1}
\addtocounter{table}{-1}
\newpage
\include{n3379m1tab2}

\newpage
\include{n3585m1tab1}
\newpage
\include{n3585m1tab2}

\include{n5846m1tab1}
\addtocounter{table}{-1}
\newpage
\include{n5846m1tab2}

\newpage
\include{n7192m1tab1}

\section{Photometric measurements}
\label{ln:photmeasurements}

\include{n1380_n2434.final_BVI}
\include{n3115_n3379_n3585.final_BVRIK}
\include{n5846.final_BVRIK}

\end{document}

%% file: n1380m1tab1.tex
\begin{table*}[h!]
  \caption{Lick indices CN$_1$ -- Mg$_2$ for {\bf mask 1} of {\bf NGC~1380} globular
    cluster observations including statistical and systematic
    errors.The set of indices uses the passband definitions of
    \cite{worthey94} and for the higher-order Balmer lines the
    definitions of \cite{worthey97}.}

 \label{tab:m1n1380gcindices}
        {\tiny
        \begin{center}
\begin{tabular}{lrrrrrrrrrrrrrrrrrrrrrr}
         \hline
         \noalign{\smallskip}

cluster$^{\mathrm{a}}$&CN$_1$ &CN$_2$ & Ca4227& G4300 & Fe4383& Ca4455& Fe4531& Fe4668&H$\beta$&Fe5015&  Mg$_1$ & Mg$_2$ \\
                      &  mag  &  mag  &  \AA  &  \AA  &  \AA  &  \AA  &  \AA  &  \AA  &  \AA  &  \AA  &  mag   &mag\\
\noalign{\smallskip}
\hline
\noalign{\smallskip}
     1380m1GC04& $ 0.061$& $-0.075$& $ 0.44$& $ 4.00$& $-4.50$& $ 1.60$& $ 3.66$& $-0.54$& $-0.92$& $ 1.37$&   \dots &   \dots  \\
$\Delta\cal{B}$& $ 0.004$& $ 0.006$& $ 0.22$& $ 0.24$& $ 0.30$& $ 0.31$& $ 0.34$& $ 0.42$& $ 0.43$& $ 0.46$&   \dots &   \dots  \\
$\Delta v_r$   & $ 0.108$& $ 0.023$& $ 0.15$& $ 0.13$& $ 0.30$& $ 0.16$& $ 0.11$& $ 0.58$& $ 0.08$& $ 0.22$&   \dots &   \dots  \\
     1380m1GC07& $ 0.056$& $ 0.063$& $ 1.43$& $ 6.20$& $-1.09$& $ 2.40$& $ 3.25$& $ 2.92$& $ 1.20$& $ 5.54$& $ 0.088$& $ 0.260$ \\
$\Delta\cal{B}$& $ 0.002$& $ 0.003$& $ 0.09$& $ 0.10$& $ 0.12$& $ 0.12$& $ 0.13$& $ 0.16$& $ 0.16$& $ 0.18$& $ 0.005$& $ 0.005$ \\
$\Delta v_r$   & $ 0.011$& $ 0.027$& $ 0.16$& $ 0.20$& $ 1.41$& $ 0.53$& $ 0.91$& $ 0.85$& $ 0.13$& $ 0.71$& $ 0.012$& $ 0.007$ \\
\hline
\end{tabular}
\end{center}
}
\begin{list}{}{}
\item[$^{\mathrm{a}}$] $\Delta\cal{B}$: Bootstraped 1$\sigma$
  statistical error, $\Delta v_r$: systematic uncertainty due to
  radial velocity errors.
\end{list}
\end{table*}

%% file: n1380m1tab2.tex
\begin{table*}[h!]
      \caption{-- continued. Lick indices Mg$b$ -- H$\gamma_F$ for
        {\bf mask 1} of the {\bf NGC~1380} globular clusters. Note that index
        TiO$_2$ is missing due to the limited wavelength coverage of
        the spectra.}
        {\tiny
        \begin{center}
        \begin{tabular}{lrrrrrrrrrrrrrrrrrrrrrrr}
         \noalign{\smallskip}
         \hline
         \noalign{\smallskip}
cluster$^{\mathrm{a}}$& Mg$b$ & Fe5270& Fe5335& Fe5406& Fe5709& Fe5782&  NaD  &TiO$_1$&H$\delta_A$&H$\gamma_A$&H$\delta_F$&H$\gamma_F$ \\
                      &  \AA  &  \AA  &  \AA  &  \AA  &  \AA  &  \AA  &  \AA  &  mag  & \AA  &  \AA  &  \AA  &  \AA   \\
\noalign{\smallskip}
\hline
\noalign{\smallskip}
     1380m1GC04&   \dots&   \dots&   \dots&   \dots&   \dots&   \dots&   \dots&   \dots & $ 3.22$& $ 2.03$& $ 3.04$& $ 1.54$ \\
$\Delta\cal{B}$&   \dots&   \dots&   \dots&   \dots&   \dots&   \dots&   \dots&   \dots & $ 0.49$& $ 0.50$& $ 0.51$& $ 0.51$ \\
$\Delta v_r$   &   \dots&   \dots&   \dots&   \dots&   \dots&   \dots&   \dots&   \dots & $ 0.05$& $ 0.12$& $ 0.47$& $ 0.21$ \\
     1380m1GC07& $ 4.18$& $ 2.56$& $ 1.68$& $ 1.33$& $ 1.01$& $ 0.90$& $ 2.00$& $ 0.022$& $-1.73$& $-6.98$& $-0.35$& $-2.03$ \\
$\Delta\cal{B}$& $ 0.20$& $ 0.20$& $ 0.21$& $ 0.21$& $ 0.21$& $ 0.21$& $ 0.22$& $ 0.006$& $ 0.23$& $ 0.24$& $ 0.24$& $ 0.24$ \\
$\Delta v_r$   & $ 0.27$& $ 0.16$& $ 0.14$& $ 0.30$& $ 0.12$& $ 0.05$& $ 0.42$& $ 0.012$& $ 0.05$& $ 1.25$& $ 0.51$& $ 0.10$ \\
\hline
\end{tabular}
\end{center}
}
\begin{list}{}{}
\item[$^{\mathrm{a}}$] $\Delta\cal{B}$: Bootstraped 1$\sigma$
  statistical error, $\Delta v_r$: systematic uncertainty due to
  radial velocity errors.
\end{list}
\end{table*}

%% file: n1380m2tab1.tex
\begin{table*}[h!]
  \caption{Lick indices CN$_1$ -- Mg$_2$ for {\bf mask 2} of {\bf NGC~1380} globular
    clusters.}
 \label{tab:m2n1380gcindices}
        {\tiny
        \begin{center}
        \begin{tabular}{lrrrrrrrrrrrrrrrrrrrrrr}
         \hline
         \noalign{\smallskip}

cluster$^{\mathrm{a}}$&CN$_1$ &CN$_2$ & Ca4227& G4300 & Fe4383& Ca4455& Fe4531& Fe4668&H$\beta$&Fe5015&  Mg$_1$ & Mg$_2$ \\
                      &  mag  &  mag  &  \AA  &  \AA  &  \AA  &  \AA  &  \AA  &  \AA  &  \AA  &  \AA  &  mag   &mag\\
\noalign{\smallskip}
\hline
\noalign{\smallskip}
     1380m2GC03& $ 0.227$& $ 0.232$& $ 1.82$& $ 3.04$& $-1.50$& $ 0.75$& $-1.43$& $-3.26$& $ 1.41$& $ 6.10$& $ 0.054$& $ 0.189$ \\
$\Delta\cal{B}$& $ 0.007$& $ 0.011$& $ 0.32$& $ 0.36$& $ 0.45$& $ 0.46$& $ 0.50$& $ 0.59$& $ 0.60$& $ 0.65$& $ 0.018$& $ 0.018$ \\
$\Delta v_r$   & $ 0.027$& $ 0.028$& $ 0.04$& $ 0.28$& $ 0.79$& $ 0.08$& $ 0.11$& $ 0.94$& $ 0.92$& $ 0.46$& $ 0.002$& $ 0.003$ \\
     1380m2GC05&   \dots &   \dots &   \dots&   \dots&   \dots&   \dots& $ 0.52$& $ 3.73$& $-0.86$& $ 9.77$& $ 0.136$& $ 0.308$ \\
$\Delta\cal{B}$&   \dots &   \dots &   \dots&   \dots&   \dots&   \dots& $ 0.96$& $ 1.06$& $ 1.07$& $ 1.13$& $ 0.036$& $ 0.036$ \\
$\Delta v_r$   &   \dots &   \dots &   \dots&   \dots&   \dots&   \dots& $ 5.92$& $ 0.87$& $ 1.35$& $ 0.39$& $ 0.001$& $ 0.006$ \\
\hline
\end{tabular}
\end{center}
}
\begin{list}{}{}
\item[$^{\mathrm{a}}$] $\Delta\cal{B}$: Bootstraped 1$\sigma$
  statistical error, $\Delta v_r$: systematic uncertainty due to
  radial velocity errors.
\end{list}
\end{table*}

%% file: n1380m2tab2.tex
\begin{table*}[h!]
      \caption{-- continued. Lick indices Mg$b$ -- H$\gamma_F$ for
        {\bf mask 1} of the {\bf NGC~1380} globular clusters. Note that index
        TiO$_2$ is missing due to the limited wavelength coverage of
        the spectra.}
        {\tiny
        \begin{center}
        \begin{tabular}{lrrrrrrrrrrrrrrrrrrrrrrr}
         \noalign{\smallskip}
         \hline
         \noalign{\smallskip}
cluster$^{\mathrm{a}}$& Mg$b$ & Fe5270& Fe5335& Fe5406& Fe5709& Fe5782&  NaD  &TiO$_1$&H$\delta_A$&H$\gamma_A$&H$\delta_F$&H$\gamma_F$ \\
                      &  \AA  &  \AA  &  \AA  &  \AA  &  \AA  &  \AA  &  \AA  &  mag  & \AA  &  \AA  &  \AA  &  \AA   \\
\noalign{\smallskip}
\hline
\noalign{\smallskip}
     1380m2GC03& $ 4.13$& $ 3.98$& $-0.23$& $ 2.81$& $ 0.83$& $ 0.67$& $ 0.47$& $ 0.017$& $-2.76$& $ 1.94$& $-1.61$& $ 0.37$ \\
$\Delta\cal{B}$& $ 0.68$& $ 0.69$& $ 0.70$& $ 0.71$& $ 0.72$& $ 0.72$& $ 0.73$& $ 0.021$& $ 0.78$& $ 0.81$& $ 0.83$& $ 0.84$ \\
$\Delta v_r$   & $ 0.23$& $ 0.76$& $ 0.59$& $ 1.10$& $ 0.25$& $ 0.26$& $ 2.79$& $ 0.018$& $ 0.24$& $ 0.78$& $ 0.52$& $ 0.18$ \\
     1380m2GC05& $ 5.08$& $ 0.42$& $ 0.96$& $-3.45$& $ 1.26$& $-0.51$&   \dots&   \dots &   \dots&   \dots&   \dots&   \dots \\
$\Delta\cal{B}$& $ 1.19$& $ 1.21$& $ 1.22$& $ 1.24$& $ 1.25$& $ 1.26$&   \dots&   \dots &   \dots&   \dots&   \dots&   \dots \\
$\Delta v_r$   & $ 1.29$& $ 0.33$& $ 0.65$& $ 1.44$& $ 0.45$& $ 1.30$&   \dots&   \dots &   \dots&   \dots&   \dots&   \dots \\
\hline
\end{tabular}
\end{center}
}
\begin{list}{}{}
\item[$^{\mathrm{a}}$] $\Delta\cal{B}$: Bootstraped 1$\sigma$
  statistical error, $\Delta v_r$: systematic uncertainty due to
  radial velocity errors.
\end{list}
\end{table*}

%% file: n2434m1tab1.tex
\begin{table*}[h!]
  \caption{Lick indices CN$_1$ -- Mg$_2$ for {\bf mask 1} {\it and} {\bf mask 2} of
    {\bf NGC~2434} globular clusters.}
 \label{tab:m1n2434gcindices}
        {\tiny
        \begin{center}
        \begin{tabular}{lrrrrrrrrrrrrrrrrrrrrrr}
         \hline
         \noalign{\smallskip}
cluster$^{\mathrm{a}}$&CN$_1$ &CN$_2$ & Ca4227& G4300 & Fe4383& Ca4455& Fe4531& Fe4668&H$\beta$&Fe5015&  Mg$_1$ & Mg$_2$ \\
                      &  mag  &  mag  &  \AA  &  \AA  &  \AA  &  \AA  &  \AA  &  \AA  &  \AA  &  \AA  &  mag   &mag\\
\noalign{\smallskip}
\hline
\noalign{\smallskip}
     2434m1GC11& $ 0.094$& $ 0.106$& $ 0.15$& $ 5.54$& $ 1.08$& $ 0.60$& $ 4.12$& $ 0.12$& $ 2.23$& $ 2.94$& $ 0.024$& $ 0.083$ \\
$\Delta\cal{B}$& $ 0.002$& $ 0.003$& $ 0.11$& $ 0.12$& $ 0.15$& $ 0.15$& $ 0.17$& $ 0.20$& $ 0.21$& $ 0.22$& $ 0.006$& $ 0.006$ \\
$\Delta v_r$   & $ 0.043$& $ 0.058$& $ 0.15$& $ 0.83$& $ 0.48$& $ 0.15$& $ 0.54$& $ 0.23$& $ 0.10$& $ 0.41$& $ 0.022$& $ 0.028$ \\
     2434m1GC19& $ 0.026$& $ 0.013$& $ 1.01$& $ 5.25$& $ 4.48$& $ 2.43$& $ 3.10$& $ 1.87$& $ 1.55$& $ 7.75$& $ 0.087$& $ 0.219$ \\
$\Delta\cal{B}$& $ 0.002$& $ 0.004$& $ 0.12$& $ 0.14$& $ 0.15$& $ 0.16$& $ 0.17$& $ 0.20$& $ 0.20$& $ 0.22$& $ 0.006$& $ 0.006$ \\
$\Delta v_r$   & $ 0.009$& $ 0.023$& $ 0.10$& $ 0.15$& $ 0.39$& $ 0.40$& $ 1.00$& $ 0.18$& $ 0.25$& $ 3.36$& $ 0.015$& $ 0.008$ \\
\hline
\end{tabular}
\end{center}
}
\begin{list}{}{}
\item[$^{\mathrm{a}}$] $\Delta\cal{B}$: Bootstraped 1$\sigma$
  statistical error, $\Delta v_r$: systematic uncertainty due to
  radial velocity errors.
\end{list}
\end{table*}

\addtocounter{table}{-1}
\begin{table*}
      \caption{-- continued. Lick indices Mg$b$ -- H$\gamma_F$ for
        {\bf mask 1} {\it and} {\bf mask 2} of {\bf NGC~2434} globular clusters.}
        {\tiny
        \begin{center}
        \begin{tabular}{lrrrrrrrrrrrrrrrrrrrrrrr}
         \noalign{\smallskip}
         \hline
         \noalign{\smallskip}
cluster$^{\mathrm{a}}$& Mg$b$ & Fe5270& Fe5335& Fe5406& Fe5709& Fe5782&  NaD  &TiO$_1$&H$\delta_A$&H$\gamma_A$&H$\delta_F$&H$\gamma_F$ \\
                      &  \AA  &  \AA  &  \AA  &  \AA  &  \AA  &  \AA  &  \AA  &  mag  & \AA  &  \AA  &  \AA  &  \AA   \\
\noalign{\smallskip}
\hline
\noalign{\smallskip}
     2434m1GC11& $ 1.61$& $ 0.72$& $ 0.85$& $-0.29$& $ 0.67$&   \dots&   \dots&   \dots & $ 2.14$& $ 0.03$& $ 1.56$& $ 1.07$ \\
$\Delta\cal{B}$& $ 0.25$& $ 0.25$& $ 0.26$& $ 0.26$& $ 0.26$&   \dots&   \dots&   \dots & $ 0.27$& $ 0.28$& $ 0.29$& $ 0.29$ \\
$\Delta v_r$   & $ 0.26$& $ 0.24$& $ 0.75$& $ 0.18$& $ 0.04$&   \dots&   \dots&   \dots & $ 2.02$& $ 0.85$& $ 0.35$& $ 0.22$ \\
     2434m1GC19& $ 2.94$& $ 2.56$& $ 1.27$& $ 1.14$& $ 0.96$& $ 0.08$& $ 2.09$& $ 0.026$& $ 2.05$& $-3.62$& $ 1.62$& $-2.78$ \\
$\Delta\cal{B}$& $ 0.24$& $ 0.24$& $ 0.24$& $ 0.25$& $ 0.25$& $ 0.25$& $ 0.25$& $ 0.007$& $ 0.26$& $ 0.27$& $ 0.28$& $ 0.28$ \\
$\Delta v_r$   & $ 0.45$& $ 0.10$& $ 0.09$& $ 0.10$& $ 0.12$& $ 0.04$& $ 0.42$& $ 0.003$& $ 0.14$& $ 1.12$& $ 0.08$& $ 1.69$ \\
\hline
\end{tabular}
\end{center}
}
\begin{list}{}{}
\item[$^{\mathrm{a}}$] $\Delta\cal{B}$: Bootstraped 1$\sigma$
  statistical error, $\Delta v_r$: systematic uncertainty due to
  radial velocity errors.
\end{list}
\end{table*}

%% file: n3115m1tab1.tex
\begin{table*}[h!]
  \caption{Lick indices CN$_1$ -- Mg$_2$ for {\bf mask 1} of {\bf NGC~3115} globular
    clusters.}

 \label{tab:m1n3115gcindices}
        {\tiny
        \begin{center}
        \begin{tabular}{lrrrrrrrrrrrrrrrrrrrrrr}
         \hline
         \noalign{\smallskip}

cluster$^{\mathrm{a}}$&CN$_1$ &CN$_2$ & Ca4227& G4300 & Fe4383& Ca4455& Fe4531& Fe4668&H$\beta$&Fe5015&  Mg$_1$ & Mg$_2$ \\
                      &  mag  &  mag  &  \AA  &  \AA  &  \AA  &  \AA  &  \AA  &  \AA  &  \AA  &  \AA  &  mag   &mag\\
\noalign{\smallskip}
\hline
\noalign{\smallskip}
     3115m1GC03& $ 0.153$& $ 0.218$& $-1.50$& $ 4.32$& $ 6.17$& $ 2.87$& $ 2.85$& $ 8.25$& $ 2.43$& $ 2.65$& $ 0.104$& $ 0.201$ \\
$\Delta\cal{B}$& $ 0.017$& $ 0.025$& $ 0.90$& $ 0.97$& $ 1.05$& $ 1.07$& $ 1.13$& $ 1.29$& $ 1.30$& $ 1.37$& $ 0.038$& $ 0.039$ \\
$\Delta v_r$   & $ 0.012$& $ 0.060$& $ 0.20$& $ 1.27$& $ 2.15$& $ 1.25$& $ 2.57$& $ 0.86$& $ 1.11$& $ 3.32$& $ 0.046$& $ 0.026$ \\
     3115m1GC04& $-0.071$& $-0.073$& $ 0.62$& $ 4.20$& $ 2.26$& $ 1.84$& $ 2.69$& $ 1.98$& $ 1.80$& $ 2.40$& $ 0.018$& $ 0.111$ \\
$\Delta\cal{B}$& $ 0.006$& $ 0.008$& $ 0.29$& $ 0.33$& $ 0.42$& $ 0.43$& $ 0.45$& $ 0.53$& $ 0.54$& $ 0.58$& $ 0.015$& $ 0.016$ \\
$\Delta v_r$   & $ 0.023$& $ 0.030$& $ 0.16$& $ 0.09$& $ 2.39$& $ 0.28$& $ 0.07$& $ 0.13$& $ 0.52$& $ 0.16$& $ 0.009$& $ 0.007$ \\
\hline
\end{tabular}
\end{center}
}
\begin{list}{}{}
\item[$^{\mathrm{a}}$] $\Delta\cal{B}$: Bootstraped 1$\sigma$
  statistical error, $\Delta v_r$: systematic uncertainty due to
  radial velocity errors.
\end{list}
\end{table*}

%% file: n3115m1tab2.tex
\begin{table*}[h!]
      \caption{-- continued. Lick indices Mg$b$ -- H$\gamma_F$ for
        {\bf mask 1} of {\bf NGC~3115} globular clusters.}
        {\tiny
        \begin{center}
        \begin{tabular}{lrrrrrrrrrrrrrrrrrrrrrrr}
         \noalign{\smallskip}
         \hline
         \noalign{\smallskip}
cluster$^{\mathrm{a}}$& Mg$b$ & Fe5270& Fe5335& Fe5406& Fe5709& Fe5782&  NaD  &TiO$_1$&H$\delta_A$&H$\gamma_A$&H$\delta_F$&H$\gamma_F$ \\
                      &  \AA  &  \AA  &  \AA  &  \AA  &  \AA  &  \AA  &  \AA  &  mag  & \AA  &  \AA  &  \AA  &  \AA   \\
\noalign{\smallskip}
\hline
\noalign{\smallskip}
     3115m1GC03& $ 2.85$& $ 4.76$& $ 1.87$& $ 0.54$& $ 1.65$& $-1.77$&   \dots&   \dots & $-1.13$& $-1.52$& $-2.99$& $ 0.10$ \\
$\Delta\cal{B}$& $ 1.46$& $ 1.48$& $ 1.50$& $ 1.51$& $ 1.53$& $ 1.54$&   \dots&   \dots & $ 1.60$& $ 1.67$& $ 1.72$& $ 1.73$ \\
$\Delta v_r$   & $ 0.58$& $ 0.59$& $ 0.91$& $ 1.15$& $ 0.70$& $ 0.68$&   \dots&   \dots & $ 1.36$& $ 1.02$& $ 1.88$& $ 0.50$ \\
     3115m1GC04& $ 0.97$& $ 2.29$& $-1.33$& $ 2.03$& $ 0.77$& $ 0.28$& $-0.45$& $ 0.029$& $ 2.50$& $ 0.79$& $ 2.45$& $ 2.66$ \\
$\Delta\cal{B}$& $ 0.62$& $ 0.63$& $ 0.65$& $ 0.65$& $ 0.66$& $ 0.66$& $ 0.68$& $ 0.019$& $ 0.72$& $ 0.75$& $ 0.75$& $ 0.76$ \\
$\Delta v_r$   & $ 0.40$& $ 0.34$& $ 1.17$& $ 0.52$& $ 0.23$& $ 0.17$& $ 0.80$& $ 0.007$& $ 1.49$& $ 0.30$& $ 0.16$& $ 0.42$ \\
\hline
\end{tabular}
\end{center}
}
\begin{list}{}{}
\item[$^{\mathrm{a}}$] $\Delta\cal{B}$: Bootstraped 1$\sigma$
  statistical error, $\Delta v_r$: systematic uncertainty due to
  radial velocity errors.
\end{list}
\end{table*}

%% file: n3379m1tab1.tex
\begin{table*}[h!]
  \caption{Lick indices CN$_1$ -- Mg$_2$ for {\bf mask 1} of {\bf NGC~3379} globular
    clusters.}

 \label{tab:m1n3379gcindices}
        {\tiny
        \begin{center}
        \begin{tabular}{lrrrrrrrrrrrrrrrrrrrrrr}
         \hline
         \noalign{\smallskip}

cluster$^{\mathrm{a}}$&CN$_1$ &CN$_2$ & Ca4227& G4300 & Fe4383& Ca4455& Fe4531& Fe4668&H$\beta$&Fe5015&  Mg$_1$ & Mg$_2$ \\
                      &  mag  &  mag  &  \AA  &  \AA  &  \AA  &  \AA  &  \AA  &  \AA  &  \AA  &  \AA  &  mag   &mag\\
\noalign{\smallskip}
\hline
\noalign{\smallskip}
     3379m1GC05& $ 0.022$& $ 0.046$& $ 0.84$& $ 4.26$& $ 1.01$& $ 1.42$& $ 3.15$& $ 0.94$& $ 2.50$& $ 1.65$& $ 0.013$& $ 0.092$ \\
$\Delta\cal{B}$& $ 0.001$& $ 0.002$& $ 0.05$& $ 0.06$& $ 0.07$& $ 0.08$& $ 0.09$& $ 0.11$& $ 0.11$& $ 0.12$& $ 0.003$& $ 0.003$ \\
$\Delta v_r$   & $ 0.026$& $ 0.046$& $ 0.03$& $ 0.22$& $ 0.17$& $ 0.24$& $ 0.57$& $ 0.25$& $ 0.19$& $ 0.49$& $ 0.005$& $ 0.002$ \\
     3379m1GC09& $-0.023$& $-0.018$& $ 0.97$& $ 2.84$& $ 0.86$& $ 1.24$& $ 0.87$& $ 1.43$& $ 1.92$& $ 1.81$& $ 0.004$& $ 0.068$ \\
$\Delta\cal{B}$& $ 0.001$& $ 0.001$& $ 0.03$& $ 0.04$& $ 0.05$& $ 0.05$& $ 0.06$& $ 0.08$& $ 0.08$& $ 0.09$& $ 0.002$& $ 0.002$ \\
$\Delta v_r$   & $ 0.014$& $ 0.021$& $ 0.05$& $ 0.90$& $ 0.31$& $ 0.13$& $ 0.40$& $ 0.15$& $ 0.17$& $ 0.26$& $ 0.000$& $ 0.000$ \\
\hline
\end{tabular}
\end{center}
}
\begin{list}{}{}
\item[$^{\mathrm{a}}$] $\Delta\cal{B}$: Bootstraped 1$\sigma$
  statistical error, $\Delta v_r$: systematic uncertainty due to
  radial velocity errors.
\end{list}
\end{table*}

%% file: n3379m1tab2.tex
\begin{table*}[h!]
      \caption{-- continued. Lick indices Mg$b$ -- H$\gamma_F$ for
        {\bf mask 1} of {\bf NGC~3379} globular clusters.}
        {\tiny
        \begin{center}
        \begin{tabular}{lrrrrrrrrrrrrrrrrrrrrrrr}
         \noalign{\smallskip}
         \hline
         \noalign{\smallskip}
cluster$^{\mathrm{a}}$& Mg$b$ & Fe5270& Fe5335& Fe5406& Fe5709& Fe5782&  NaD  &TiO$_1$&H$\delta_A$&H$\gamma_A$&H$\delta_F$&H$\gamma_F$ \\
                      &  \AA  &  \AA  &  \AA  &  \AA  &  \AA  &  \AA  &  \AA  &  mag  & \AA  &  \AA  &  \AA  &  \AA   \\
\noalign{\smallskip}
\hline
\noalign{\smallskip}
     3379m1GC05& $ 1.31$& $ 1.00$& $ 0.56$& $ 0.69$& $ 1.20$& $ 0.31$& $ 0.80$& $ 0.004$& $ 0.73$& $-0.60$& $ 1.75$& $ 1.01$ \\
$\Delta\cal{B}$& $ 0.13$& $ 0.13$& $ 0.13$& $ 0.13$& $ 0.14$& $ 0.14$& $ 0.14$& $ 0.004$& $ 0.15$& $ 0.15$& $ 0.15$& $ 0.15$ \\
$\Delta v_r$   & $ 0.08$& $ 0.07$& $ 0.08$& $ 0.15$& $ 0.10$& $ 0.03$& $ 0.06$& $ 0.004$& $ 0.21$& $ 0.16$& $ 0.08$& $ 0.17$ \\
     3379m1GC09& $ 0.77$& $ 1.55$& $ 0.47$& $ 0.45$& $ 0.28$& $ 0.08$& $ 0.87$&   \dots & $ 2.09$& $ 1.20$& $ 2.07$& $ 1.31$ \\
$\Delta\cal{B}$& $ 0.09$& $ 0.10$& $ 0.10$& $ 0.10$& $ 0.10$& $ 0.10$& $ 0.10$&   \dots & $ 0.11$& $ 0.11$& $ 0.11$& $ 0.11$ \\
$\Delta v_r$   & $ 0.15$& $ 0.05$& $ 0.06$& $ 0.09$& $ 0.08$& $ 0.05$& $ 0.14$&   \dots & $ 0.04$& $ 0.33$& $ 0.15$& $ 0.19$ \\
\hline
\end{tabular}
\end{center}
}
\begin{list}{}{}
\item[$^{\mathrm{a}}$] $\Delta\cal{B}$: Bootstraped 1$\sigma$
  statistical error, $\Delta v_r$: systematic uncertainty due to
  radial velocity errors.
\end{list}
\end{table*}

%% file: n3585m1tab1.tex
\begin{table*}[h!]
  \caption{Lick indices CN$_1$ -- Mg$_2$ for {\bf mask 1} of {\bf NGC~3585} globular
    clusters.}

 \label{tab:m1n3585gcindices}
        {\tiny
        \begin{center}
        \begin{tabular}{lrrrrrrrrrrrrrrrrrrrrrr}
         \hline
         \noalign{\smallskip}

cluster$^{\mathrm{a}}$&CN$_1$ &CN$_2$ & Ca4227& G4300 & Fe4383& Ca4455& Fe4531& Fe4668&H$\beta$&Fe5015&  Mg$_1$ & Mg$_2$ \\
                      &  mag  &  mag  &  \AA  &  \AA  &  \AA  &  \AA  &  \AA  &  \AA  &  \AA  &  \AA  &  mag   &mag\\
\noalign{\smallskip}
\hline
\noalign{\smallskip}
     3585m1GC01& $ 0.110$& $ 0.133$& $ 1.40$& $ 7.37$& $ 4.43$& $ 0.27$& $ 3.53$& $ 6.07$& $ 0.86$& $ 5.05$& $ 0.132$& $ 0.294$ \\
$\Delta\cal{B}$& $ 0.002$& $ 0.003$& $ 0.11$& $ 0.12$& $ 0.15$& $ 0.15$& $ 0.16$& $ 0.18$& $ 0.18$& $ 0.19$& $ 0.006$& $ 0.006$ \\
$\Delta v_r$   & $ 0.004$& $ 0.015$& $ 0.17$& $ 0.88$& $ 1.47$& $ 0.41$& $ 0.03$& $ 1.57$& $ 0.25$& $ 0.52$& $ 0.013$& $ 0.008$ \\
     3585m1GC03& $-0.015$& $ 0.015$& $ 1.12$& $ 3.18$& $ 0.24$& $ 1.36$& $ 2.06$& $ 1.45$& $ 1.77$&   \dots&   \dots &   \dots \\
$\Delta\cal{B}$& $ 0.001$& $ 0.001$& $ 0.03$& $ 0.04$& $ 0.04$& $ 0.05$& $ 0.05$& $ 0.06$& $ 0.06$&   \dots&   \dots &   \dots \\
$\Delta v_r$   & $ 0.012$& $ 0.030$& $ 0.09$& $ 0.92$& $ 0.36$& $ 0.34$& $ 0.21$& $ 0.32$& $ 0.07$&   \dots&   \dots &   \dots \\
\hline
\end{tabular}
\end{center}
}
\begin{list}{}{}
\item[$^{\mathrm{a}}$] $\Delta\cal{B}$: Bootstraped 1$\sigma$
  statistical error, $\Delta v_r$: systematic uncertainty due to
  radial velocity errors.
\end{list}
\end{table*}

%% file: n3585m1tab2.tex
\begin{table*}[h!]
      \caption{-- continued. Lick indices Mg$b$ -- H$\gamma_F$ for
        {\bf mask 1} of {\bf NGC~3585} globular clusters.}
        {\tiny
        \begin{center}
        \begin{tabular}{lrrrrrrrrrrrrrrrrrrrrrrr}
         \noalign{\smallskip}
         \hline
         \noalign{\smallskip}
cluster$^{\mathrm{a}}$& Mg$b$ & Fe5270& Fe5335& Fe5406& Fe5709& Fe5782&  NaD  &TiO$_1$&H$\delta_A$&H$\gamma_A$&H$\delta_F$&H$\gamma_F$ \\
                      &  \AA  &  \AA  &  \AA  &  \AA  &  \AA  &  \AA  &  \AA  &  mag  & \AA  &  \AA  &  \AA  &  \AA   \\
\noalign{\smallskip}
\hline
\noalign{\smallskip}
     3585m1GC01& $ 4.05$& $ 2.32$& $ 2.64$& $ 2.23$& $ 0.19$& $ 1.66$& $ 5.81$&   \dots & $-2.94$& $-7.52$& $-0.41$& $-2.51$ \\
$\Delta\cal{B}$& $ 0.20$& $ 0.20$& $ 0.20$& $ 0.21$& $ 0.21$& $ 0.21$& $ 0.21$&   \dots & $ 0.23$& $ 0.24$& $ 0.24$& $ 0.25$ \\
$\Delta v_r$   & $ 0.71$& $ 0.97$& $ 0.10$& $ 0.12$& $ 0.13$& $ 0.13$& $ 0.38$&   \dots & $ 0.36$& $ 0.25$& $ 0.59$& $ 0.53$ \\
     3585m1GC03&   \dots&   \dots&   \dots&   \dots&   \dots&   \dots&   \dots&   \dots & $ 1.39$& $ 0.12$& $ 1.80$& $ 0.93$ \\
$\Delta\cal{B}$&   \dots&   \dots&   \dots&   \dots&   \dots&   \dots&   \dots&   \dots & $ 0.24$& $ 0.24$& $ 0.25$& $ 0.25$ \\
$\Delta v_r$   &   \dots&   \dots&   \dots&   \dots&   \dots&   \dots&   \dots&   \dots & $ 0.23$& $ 0.26$& $ 0.13$& $ 0.07$ \\
\hline
\end{tabular}
\end{center}
}
\begin{list}{}{}
\item[$^{\mathrm{a}}$] $\Delta\cal{B}$: Bootstraped 1$\sigma$
  statistical error, $\Delta v_r$: systematic uncertainty due to
  radial velocity errors.
\end{list}
\end{table*}

%% file: n5846m1tab1.tex
\begin{table*}[h!]
  \caption{Lick indices CN$_1$ -- Mg$_2$ for {\bf mask 1} of {\bf NGC~5846} globular
    clusters.}
 \label{tab:m1n5846gcindices}
        {\tiny
        \begin{center}
        \begin{tabular}{lrrrrrrrrrrrrrrrrrrrrrr}
         \hline
         \noalign{\smallskip}

cluster$^{\mathrm{a}}$&CN$_1$ &CN$_2$ & Ca4227& G4300 & Fe4383& Ca4455& Fe4531& Fe4668&H$\beta$&Fe5015&  Mg$_1$ & Mg$_2$ \\
                      &  mag  &  mag  &  \AA  &  \AA  &  \AA  &  \AA  &  \AA  &  \AA  &  \AA  &  \AA  &  mag   &mag\\
\noalign{\smallskip}
\hline
\noalign{\smallskip}
     5846m1GC03& $ 0.067$& $ 0.056$& $ 1.52$& $ 5.03$& $-1.19$& $ 1.14$& $ 2.16$& $-0.86$& $ 2.84$& $ 2.53$& $ 0.020$& $ 0.112$ \\
$\Delta\cal{B}$& $ 0.002$& $ 0.003$& $ 0.09$& $ 0.11$& $ 0.14$& $ 0.14$& $ 0.16$& $ 0.19$& $ 0.20$& $ 0.21$& $ 0.006$& $ 0.006$ \\
$\Delta v_r$   & $ 0.003$& $ 0.032$& $ 0.18$& $ 0.73$& $ 1.51$& $ 0.08$& $ 0.38$& $ 0.39$& $ 0.35$& $ 0.34$& $ 0.005$& $ 0.006$ \\
     5846m1GC04& $ 0.230$& $ 0.269$& $ 1.28$& $ 5.98$& $ 4.91$& $ 1.55$& $ 4.06$& $ 7.49$& $ 1.71$& $ 5.52$& $ 0.159$& $ 0.353$ \\
$\Delta\cal{B}$& $ 0.001$& $ 0.002$& $ 0.05$& $ 0.06$& $ 0.08$& $ 0.08$& $ 0.09$& $ 0.10$& $ 0.10$& $ 0.11$& $ 0.003$& $ 0.003$ \\
$\Delta v_r$   & $ 0.017$& $ 0.024$& $ 0.17$& $ 0.31$& $ 2.93$& $ 0.58$& $ 0.45$& $ 0.13$& $ 0.12$& $ 0.80$& $ 0.019$& $ 0.012$ \\
\hline
\end{tabular}
\end{center}
}
\begin{list}{}{}
\item[$^{\mathrm{a}}$] $\Delta\cal{B}$: Bootstraped 1$\sigma$
  statistical error, $\Delta v_r$: systematic uncertainty due to
  radial velocity errors.
\end{list}
\end{table*}

%% file: n5846m1tab2.tex
\begin{table*}[h!]
      \caption{-- continued. Lick indices Mg$b$ -- H$\gamma_F$ for
        {\bf mask 1} of {\bf NGC~5846} globular clusters.}
        {\tiny
        \begin{center}
        \begin{tabular}{lrrrrrrrrrrrrrrrrrrrrrrr}
         \noalign{\smallskip}
         \hline
         \noalign{\smallskip}
cluster$^{\mathrm{a}}$& Mg$b$ & Fe5270& Fe5335& Fe5406& Fe5709& Fe5782&  NaD  &TiO$_1$&H$\delta_A$&H$\gamma_A$&H$\delta_F$&H$\gamma_F$ \\
                      &  \AA  &  \AA  &  \AA  &  \AA  &  \AA  &  \AA  &  \AA  &  mag  & \AA  &  \AA  &  \AA  &  \AA   \\
\noalign{\smallskip}
\hline
\noalign{\smallskip}
     5846m1GC03& $ 1.81$& $ 1.03$& $-0.01$& $-0.28$& $ 1.30$& $ 0.85$& $ 0.54$& $-0.005$& $-0.17$& $-0.87$& $ 1.57$& $ 1.49$ \\
$\Delta\cal{B}$& $ 0.23$& $ 0.23$& $ 0.24$& $ 0.24$& $ 0.25$& $ 0.25$& $ 0.25$& $ 0.007$& $ 0.27$& $ 0.28$& $ 0.28$& $ 0.28$ \\
$\Delta v_r$   & $ 0.61$& $ 0.21$& $ 0.58$& $ 0.72$& $ 0.15$& $ 0.59$& $ 0.72$& $ 0.002$& $ 0.18$& $ 0.27$& $ 0.21$& $ 0.63$ \\
     5846m1GC04& $ 5.66$& $ 3.06$& $ 2.21$& $ 1.92$&   \dots&   \dots&   \dots&    \dots& $-3.17$& $-5.52$& $-0.40$& $-2.16$ \\
$\Delta\cal{B}$& $ 0.12$& $ 0.12$& $ 0.12$& $ 0.12$&   \dots&   \dots&   \dots&    \dots& $ 0.13$& $ 0.14$& $ 0.14$& $ 0.14$ \\
$\Delta v_r$   & $ 0.27$& $ 0.46$& $ 0.21$& $ 0.18$&   \dots&   \dots&   \dots&    \dots& $ 0.65$& $ 0.91$& $ 0.24$& $ 0.32$ \\
\hline
\end{tabular}
\end{center}
}
\begin{list}{}{}
\item[$^{\mathrm{a}}$] $\Delta\cal{B}$: Bootstraped 1$\sigma$
  statistical error, $\Delta v_r$: systematic uncertainty due to
  radial velocity errors.
\end{list}
\end{table*}

%% file: n7192m1tab1.tex
\begin{table*}[h!]
  \caption{Lick indices CN$_1$ -- Mg$_2$ for {\bf mask 1} of {\bf NGC~7192} globular
    clusters.}

 \label{tab:m1n7192gcindices}
        {\tiny
        \begin{center}
        \begin{tabular}{lrrrrrrrrrrrrrrrrrrrrrr}
         \hline
         \noalign{\smallskip}
cluster$^{\mathrm{a}}$&CN$_1$ &CN$_2$ & Ca4227& G4300 & Fe4383& Ca4455& Fe4531& Fe4668&H$\beta$&Fe5015&  Mg$_1$ & Mg$_2$ \\
                      &  mag  &  mag  &  \AA  &  \AA  &  \AA  &  \AA  &  \AA  &  \AA  &  \AA  &  \AA  &  mag   &mag\\
\noalign{\smallskip}
\hline
\noalign{\smallskip}
     7192m1GC02& $ 0.004$& $-0.050$& $-0.10$& $ 0.93$& $ 4.13$&   \dots&   \dots&   \dots&   \dots&   \dots& $ 0.058$& $-0.048$ \\
$\Delta\cal{B}$& $ 0.006$& $ 0.009$& $ 0.33$& $ 0.36$& $ 0.42$&   \dots&   \dots&   \dots&   \dots&   \dots& $ 0.017$& $ 0.017$ \\
$\Delta v_r$   & $ 0.004$& $ 0.012$& $ 0.98$& $ 1.85$& $ 0.75$&   \dots&   \dots&   \dots&   \dots&   \dots& $ 0.043$& $ 0.038$ \\
     7192m1GC15& $-0.012$& $ 0.020$& $ 1.03$& $ 4.83$& $-0.58$& $ 1.32$& $ 0.70$& $ 1.62$& $ 1.85$& $ 2.84$& $ 0.033$& $ 0.139$ \\
$\Delta\cal{B}$& $ 0.002$& $ 0.003$& $ 0.10$& $ 0.11$& $ 0.14$& $ 0.14$& $ 0.15$& $ 0.17$& $ 0.17$& $ 0.18$& $ 0.005$& $ 0.005$ \\
$\Delta v_r$   & $ 0.024$& $ 0.061$& $ 0.13$& $ 0.71$& $ 1.48$& $ 0.33$& $ 0.33$& $ 0.62$& $ 0.33$& $ 0.71$& $ 0.000$& $ 0.002$ \\
\hline
\end{tabular}
\end{center}
}
\begin{list}{}{}
\item[$^{\mathrm{a}}$] $\Delta\cal{B}$: Bootstraped 1$\sigma$
  statistical error, $\Delta v_r$: systematic uncertainty due to
  radial velocity errors.
\end{list}
\end{table*}

\addtocounter{table}{-1}
\begin{table*}
      \caption{-- continued. Lick indices Mg$b$ -- H$\gamma_F$ for
        {\bf mask 1} of {\bf NGC~7192} globular clusters.}
        {\tiny
        \begin{center}
        \begin{tabular}{lrrrrrrrrrrrrrrrrrrrrrrr}
         \noalign{\smallskip}
         \hline
         \noalign{\smallskip}
cluster$^{\mathrm{a}}$& Mg$b$ & Fe5270& Fe5335& Fe5406& Fe5709& Fe5782&  NaD  &TiO$_1$&H$\delta_A$&H$\gamma_A$&H$\delta_F$&H$\gamma_F$ \\
                      &  \AA  &  \AA  &  \AA  &  \AA  &  \AA  &  \AA  &  \AA  &  mag  & \AA  &  \AA  &  \AA  &  \AA   \\
\noalign{\smallskip}
\hline
\noalign{\smallskip}
     7192m1GC02& $ 1.43$& $ 7.00$& $ 1.18$& $-0.50$&   \dots&   \dots&   \dots&   \dots & $ 3.35$& $-4.52$& $ 1.87$& $-4.43$ \\
$\Delta\cal{B}$& $ 0.84$& $ 0.85$& $ 0.86$& $ 0.86$&   \dots&   \dots&   \dots&   \dots & $ 0.88$& $ 0.91$& $ 0.92$& $ 0.93$ \\
$\Delta v_r$   & $ 0.10$& $ 0.96$& $ 0.34$& $ 0.37$&   \dots&   \dots&   \dots&   \dots & $ 2.05$& $ 1.14$& $ 0.47$& $ 0.92$ \\
     7192m1GC15& $ 3.23$& $ 0.78$& $ 1.71$& $ 0.63$& $ 0.65$& $ 1.06$& $ 1.09$&   \dots & $ 3.28$& $ 2.48$& $ 3.80$& $-0.59$ \\
$\Delta\cal{B}$& $ 0.20$& $ 0.20$& $ 0.20$& $ 0.21$& $ 0.21$& $ 0.21$& $ 0.21$&   \dots & $ 0.22$& $ 0.23$& $ 0.23$& $ 0.24$ \\
$\Delta v_r$   & $ 0.31$& $ 0.23$& $ 0.09$& $ 0.54$& $ 0.49$& $ 0.25$& $ 0.44$&   \dots & $ 0.07$& $ 0.46$& $ 0.36$& $ 2.12$ \\
\hline
\end{tabular}
\end{center}
}
\begin{list}{}{}
\item[$^{\mathrm{a}}$] $\Delta\cal{B}$: Bootstraped 1$\sigma$
  statistical error, $\Delta v_r$: systematic uncertainty due to
  radial velocity errors.
\end{list}
\end{table*}

%% file: n1380_n2434.final_BVI.tex
\begin{table*}[h!]
  \caption{Photometry for {\bf NGC~1380} globular clusters. $B$ band photometry was performed on HST/WFPC2 data.
           $V$ and $I$ magnitudes were extracted from VLT/FORS2 data.}
 \label{tab:n1380gcphot}
        {\tiny
        \begin{center}
        \begin{tabular}{lcccccr}
         \hline
         \noalign{\smallskip}
cluster&  RA (J2000)  &  DEC (J2000)  & $B$ & $V$ & $I$ & $v_r$ [km/s] \\
\noalign{\smallskip}
\hline
\noalign{\smallskip}
      1380m1GC04&  54.10717& $-$35.03080&        \dots &$22.84\pm0.01$&$21.82\pm0.01$&$2021\pm76$ \\
      1380m1GC07&  54.14715& $-$35.00487&        \dots &$21.48\pm0.01$&$20.26\pm0.01$&$1989\pm15$ \\
      1380m1GC08&  54.11343& $-$35.01822&        \dots &$22.05\pm0.01$&$21.06\pm0.01$&$2069\pm54$ \\
      1380m1GC09&  54.12835& $-$35.00936&        \dots &$22.33\pm0.01$&$21.39\pm0.01$&$1992\pm53$ \\
      1380m1GC11&  54.12429& $-$35.00638&        \dots &$22.17\pm0.01$&$21.17\pm0.01$&$2177\pm42$ \\
      1380m1GC12&  54.12302& $-$35.00412&        \dots &$21.64\pm0.01$&$20.37\pm0.01$&$1759\pm19$ \\
      1380m1GC15&  54.11587& $-$35.00054&        \dots &$21.54\pm0.01$&$20.56\pm0.01$&$2011\pm42$ \\
      1380m1GC16&  54.11230& $-$34.99823&        \dots &$23.27\pm0.02$&$20.73\pm0.01$&$2037\pm95$ \\
      1380m1GC19&  54.11019& $-$34.99161&        \dots &$21.53\pm0.01$&$20.33\pm0.01$&$1891\pm27$ \\
      1380m1GC23&  54.13079& $-$34.97301&        \dots &$22.32\pm0.01$&$21.03\pm0.01$&$1911\pm27$ \\
      1380m1GC24&  54.12244& $-$34.97604&        \dots &$22.55\pm0.01$&$21.39\pm0.01$&$1753\pm79$ \\
      1380m1GC25&  54.11998& $-$34.97541&        \dots &$23.56\pm0.02$&$22.21\pm0.02$&$1872\pm54$ \\
      1380m1GC27&  54.11623& $-$34.97274&        \dots &$23.13\pm0.02$&$22.13\pm0.02$&$1845\pm119$ \\
      1380m1GC28&  54.11594& $-$34.97202&        \dots &$21.94\pm0.01$&$20.06\pm0.01$&$1555\pm41$ \\
      1380m1GC30&  54.11164& $-$34.96936&$22.60\pm0.03$&$21.32\pm0.01$&$19.65\pm0.01$&$1840\pm72$ \\
      1380m1GC33&  54.10018& $-$34.96597&        \dots &$21.97\pm0.01$&$20.77\pm0.01$&$2010\pm24$ \\
      1380m1GC36&  54.11802& $-$34.94870&$22.50\pm0.02$&$21.89\pm0.01$&$20.82\pm0.01$&$1969\pm28$ \\
      1380m1GC37&  54.10672& $-$34.94987&$23.31\pm0.08$&$22.36\pm0.01$&$21.15\pm0.01$&$1813\pm44$ \\
      1380m1GC39&  54.10054& $-$34.94803&$23.88\pm0.14$&$22.83\pm0.01$&$21.85\pm0.01$&$1936\pm57$ \\
      1380m1GC45&  54.10862& $-$34.93067&$23.59\pm0.05$&$22.67\pm0.01$&$21.62\pm0.01$&$1647\pm67$ \\
      1380m1GC46&  54.10822& $-$34.92786&$23.28\pm0.04$&$22.30\pm0.01$&$21.05\pm0.01$&$1686\pm31$ \\
\noalign{\smallskip}
      1380m2GC03&  54.11109& $-$35.01416&        \dots &$22.32\pm0.01$&$21.24\pm0.01$&$1798\pm93$ \\
      1380m2GC05&  54.09937& $-$35.00162&        \dots &$23.00\pm0.01$&$21.91\pm0.02$&$1953\pm66$ \\
      1380m2GC11&  54.11286& $-$35.00132&        \dots &$22.31\pm0.01$&$21.08\pm0.01$&$1784\pm46$ \\
      1380m2GC13&  54.10945& $-$34.99274&        \dots &$23.38\pm0.02$&$22.10\pm0.02$&$2031\pm53$ \\
      1380m2GC14&  54.11019& $-$34.99161&        \dots &$21.53\pm0.01$&$20.33\pm0.01$&$1785\pm31$ \\
      1380m2GC15&  54.11664& $-$34.99397&        \dots &$22.42\pm0.01$&$21.26\pm0.01$&$1885\pm46$ \\
      1380m2GC19&  54.11410& $-$34.98408&        \dots &$20.19\pm0.01$&$19.05\pm0.01$&$2140\pm19$ \\
      1380m2GC21&  54.11310& $-$34.98104&        \dots &$21.46\pm0.01$&$20.64\pm0.01$&$1733\pm50$ \\
      1380m2GC22&  54.11684& $-$34.98021&        \dots &$22.35\pm0.01$&$21.20\pm0.01$&$1503\pm87$ \\
      1380m2GC23&  54.11171& $-$34.97392&        \dots &$22.12\pm0.01$&$21.38\pm0.01$&$2194\pm48$ \\
      1380m2GC26&  54.10950& $-$34.96517&$23.66\pm0.07$&$22.98\pm0.01$&$21.73\pm0.01$&$1663\pm71$ \\
      1380m2GC27&  54.12712& $-$34.97306&        \dots &$21.01\pm0.01$&$19.93\pm0.01$&$1919\pm21$ \\
      1380m2GC30&  54.11450& $-$34.95576&$21.55\pm0.01$&$20.87\pm0.01$&$19.88\pm0.01$&$1907\pm35$ \\
      1380m2GC33&  54.11982& $-$34.95058&        \dots &$22.01\pm0.01$&$20.77\pm0.01$&$1771\pm48$ \\
      1380m2GC34&  54.11879& $-$34.94580&$23.63\pm0.05$&$22.74\pm0.01$&$21.58\pm0.01$&$1503\pm68$ \\
      1380m2GC35&  54.12341& $-$34.94678&$22.92\pm0.03$&$21.98\pm0.01$&$20.80\pm0.01$&$1799\pm35$ \\
      1380m2GC36&  54.12584& $-$34.94553&$23.22\pm0.04$&$22.26\pm0.01$&$20.99\pm0.01$&$1709\pm35$ \\
      1380m2GC37&  54.12193& $-$34.94023&$23.53\pm0.05$&$22.70\pm0.01$&$21.66\pm0.01$&$1855\pm57$ \\
      1380m2GC38&  54.13341& $-$34.94582&$22.53\pm0.02$&$21.62\pm0.01$&$20.51\pm0.01$&$2075\pm22$ \\
      1380m2GC42&  54.12449& $-$34.92257&        \dots &$21.96\pm0.01$&$20.75\pm0.01$&$2163\pm59$ \\
      1380m2GC43&  54.13475& $-$34.92701&        \dots &$22.32\pm0.01$&$21.18\pm0.01$&$1891\pm47$ \\
      1380m2GC45&  54.14384& $-$34.92875&        \dots &$21.65\pm0.01$&$20.67\pm0.01$&$1799\pm40$ \\
\noalign{\smallskip}
\hline
\end{tabular}
\end{center}
}
\end{table*}

\begin{table*}[h!]
  \caption{Photometry for {\bf NGC~2434} globular clusters. $B$ band photometry was performed on HST/WFPC2 data. 
           $V$ and $I$ magnitudes were extracted from VLT/FORS2 data.}
 \label{tab:n2434gcphot}
        {\tiny
        \begin{center}
        \begin{tabular}{lcccccc}
         \hline
         \noalign{\smallskip}
cluster&  RA (J2000)  &  DEC (J2000)  & $B$ & $V$ & $I$ & $v_r$ [km/s] \\
\noalign{\smallskip}
\hline
\noalign{\smallskip}
      2434m1GC11& 113.72887& $-$69.31087&        \dots &$21.18\pm0.01$&$19.89\pm0.01$&$1431\pm45$ \\
      2434m1GC19& 113.73373& $-$69.28346&$23.16\pm0.03$&$22.03\pm0.01$&$20.66\pm0.01$&$1425\pm30$ \\
      2434m1GC20& 113.69241& $-$69.31384&        \dots &$20.39\pm0.01$&$19.14\pm0.01$&$1288\pm58$ \\
\noalign{\smallskip}
      2434m2GC29$^{\mathrm{a}}$& 113.71639& $-$69.28687&$23.56\pm0.07$&$22.50\pm0.05$&$21.07\pm0.04$&$1507\pm56$ \\
      2434m2GC37$^{\mathrm{b}}$& 113.70513& $-$69.25395&        \dots &$22.01\pm0.01$&        \dots &$1462\pm54$ \\
      2434m2GC40               & 113.73322& $-$69.26787&$23.95\pm0.06$&$22.96\pm0.02$&$21.68\pm0.02$&$1508\pm71$ \\
\noalign{\smallskip}
\hline
\end{tabular}
\end{center}
}
\begin{list}{}{}
\item[$^{\mathrm{a}}$] all passbands from HST/WFPC2 photometry
\item[$^{\mathrm{b}}$] $I$ band data uncertain due to nearby blooming spike. $V$ band data was calibrated assuming $V-I=1.0$.
\end{list}
\end{table*}

%% file: n3115_n3379_n3585.final_BVRIK.tex
\begin{table*}[h!]
  \caption{Photometry for {\bf NGC~3115} globular clusters. Optical photometry was performed on VLT/FORS2 data.
           Near-infrared $K$ band magnitudes were measured on VLT/ISAAC data and taken from \cite{puzia02a}.}
 \label{tab:n3115gcphot}
        {\tiny
        \begin{center}
        \begin{tabular}{lcccccccc}
         \hline
         \noalign{\smallskip}
cluster&  RA (J2000)  &  DEC (J2000)  & $B$ & $V$ & $R$ & $I$ & $K$ & $v_r$ [km/s] \\
\noalign{\smallskip}
\hline
\noalign{\smallskip}
      3115m1GC03& 151.28017&  $-$7.75347&$23.02\pm0.04$&$22.16\pm0.01$&$21.51\pm0.01$&$20.86\pm0.01$&        \dots &$782\pm46$ \\
      3115m1GC04& 151.29340&  $-$7.75755&$22.20\pm0.02$&$21.43\pm0.01$&$20.90\pm0.01$&$20.36\pm0.01$&        \dots &$476\pm25$ \\
      3115m1GC05& 151.29004&  $-$7.74616&$22.91\pm0.03$&$22.27\pm0.01$&$21.78\pm0.02$&$21.21\pm0.02$&        \dots &$537\pm38$ \\
      3115m1GC06& 151.29037&  $-$7.74095&$22.09\pm0.02$&$21.15\pm0.01$&$20.50\pm0.01$&$19.78\pm0.01$&        \dots &$532\pm19$ \\
      3115m1GC08& 151.29779&  $-$7.73471&$22.39\pm0.02$&$21.47\pm0.01$&$20.72\pm0.01$&$20.12\pm0.01$&        \dots &$652\pm22$ \\
      3115m1GC09& 151.29184&  $-$7.72455&$21.86\pm0.01$&$21.07\pm0.01$&$20.52\pm0.01$&$20.03\pm0.01$&        \dots &$544\pm36$ \\
      3115m1GC10& 151.31731&  $-$7.73558&$21.24\pm0.01$&$20.53\pm0.01$&$20.09\pm0.01$&$19.65\pm0.01$&$18.63\pm0.05$&$344\pm48$ \\
      3115m1GC11& 151.31862&  $-$7.73416&$22.03\pm0.02$&$21.28\pm0.01$&$20.76\pm0.01$&$20.26\pm0.01$&$18.72\pm0.05$&$805\pm30$ \\
      3115m1GC12& 151.31346&  $-$7.72581&$21.93\pm0.01$&$21.19\pm0.01$&$20.62\pm0.01$&$20.07\pm0.01$&$18.58\pm0.04$&$688\pm22$ \\
      3115m1GC13& 151.32008&  $-$7.71888&$21.69\pm0.01$&$20.94\pm0.01$&$20.38\pm0.01$&$19.83\pm0.01$&$18.23\pm0.04$&$803\pm25$ \\
      3115m1GC14& 151.31091&  $-$7.70486&$22.21\pm0.02$&$21.26\pm0.01$&$20.62\pm0.01$&$19.94\pm0.01$&$17.90\pm0.03$&$654\pm29$ \\
      3115m1GC15& 151.32661&  $-$7.71223&$22.53\pm0.02$&$21.52\pm0.01$&$20.87\pm0.01$&$20.18\pm0.01$&$18.14\pm0.03$&$285\pm19$ \\
      3115m1GC16& 151.33287&  $-$7.70611&$23.21\pm0.04$&$22.03\pm0.01$&$21.38\pm0.01$&$20.64\pm0.01$&        \dots &$806\pm24$ \\
      3115m1GC17& 151.32942&  $-$7.69605&$22.55\pm0.02$&$21.82\pm0.01$&$21.28\pm0.01$&$20.70\pm0.01$&        \dots &$675\pm34$ \\
      3115m1GC18& 151.33321&  $-$7.69482&$22.75\pm0.03$&$21.99\pm0.01$&$21.45\pm0.01$&$20.90\pm0.01$&        \dots &$764\pm37$ \\
      3115m1GC19& 151.33606&  $-$7.69091&$22.40\pm0.02$&$21.43\pm0.01$&$20.80\pm0.01$&$20.13\pm0.01$&        \dots &$957\pm24$ \\
      3115m1GC21& 151.33311&  $-$7.68339&$23.51\pm0.05$&$22.60\pm0.02$&$21.99\pm0.02$&$21.31\pm0.02$&        \dots &$569\pm50$ \\
      3115m1GC22& 151.34456&  $-$7.68690&$22.43\pm0.02$&$21.51\pm0.01$&$20.89\pm0.01$&$20.30\pm0.01$&        \dots &$826\pm24$ \\
\noalign{\smallskip}
\hline
\end{tabular}
\end{center}
}
\end{table*}

\begin{table*}[h!]
  \caption{Photometry for {\bf NGC~3379} globular clusters. All photometric data were obtained with VLT/FORS2 data.}
 \label{tab:n3379gcphot}
        {\tiny
        \begin{center}
        \begin{tabular}{lccccr}
         \hline
         \noalign{\smallskip}
cluster&  RA (J2000)  &  DEC (J2000)  & $V$ & $I$ & $v_r$ [km/s] \\
\noalign{\smallskip}
\hline
\noalign{\smallskip}
      3379m1GC05& 162.00133&  12.54659&$21.15\pm0.01$&$20.20\pm0.01$&$ 632\pm26$ \\
      3379m1GC09& 161.98579&  12.55754&$20.59\pm0.01$&$19.70\pm0.01$&$1130\pm34$ \\
      3379m1GC10& 161.97459&  12.55823&$22.83\pm0.02$&$22.14\pm0.07$&$1037\pm102$ \\
      3379m1GC11& 161.97401&  12.56302&$21.53\pm0.01$&$20.72\pm0.02$&$ 705\pm35$ \\
      3379m1GC12& 161.96921&  12.56070&$21.87\pm0.01$&$20.86\pm0.02$&$ 867\pm35$ \\
      3379m1GC13& 161.97229&  12.57023&$21.88\pm0.01$&$21.08\pm0.03$&$ 941\pm58$ \\
      3379m1GC14& 161.96004&  12.56337&$21.40\pm0.01$&$20.47\pm0.02$&$ 645\pm29$ \\
      3379m1GC15& 161.93739&  12.55101&$21.75\pm0.01$&$20.58\pm0.02$&$ 976\pm58$ \\
      3379m1GC18& 161.95697&  12.57579&$19.46\pm0.01$&$18.43\pm0.01$&$1099\pm21$ \\
      3379m1GC19& 161.96552&  12.58456&$21.65\pm0.01$&$20.56\pm0.02$&$1080\pm32$ \\
      3379m1GC20$^{\mathrm{a}}$& 161.95975&  12.58526&$21.68\pm0.06$&$20.52\pm0.06$&$1258\pm38$ \\
      3379m1GC22& 161.96140&  12.59146&$21.43\pm0.01$&$20.19\pm0.01$&$ 586\pm27$ \\
      3379m1GC23& 161.96083&  12.59229&$21.11\pm0.01$&$20.09\pm0.01$&$ 791\pm57$ \\
      3379m1GC24& 161.96268&  12.59707&$20.02\pm0.01$&$18.97\pm0.01$&$ 801\pm14$ \\
      3379m1GC26& 161.94557&  12.58963&$21.78\pm0.01$&$20.86\pm0.02$&$ 747\pm48$ \\
      3379m1GC27& 161.95096&  12.59586&$20.90\pm0.01$&$19.91\pm0.01$&$ 911\pm32$ \\
      3379m1GC30& 161.94980&  12.60642&$22.26\pm0.01$&$21.63\pm0.05$&$1119\pm77$ \\
      3379m1GC33& 161.91904&  12.59856&$20.33\pm0.01$&$19.43\pm0.01$&$1310\pm51$ \\
\noalign{\smallskip}
\hline
\end{tabular}
\end{center}
}
\begin{list}{}{}
\item[$^{\mathrm{a}}$] all passbands from HST/WFPC2 photometry
\end{list}
\end{table*}

\begin{table*}[h!]
  \caption{Photometry for {\bf NGC~3585} globular clusters. $B$ and $I$ band photometry was obtained with VLT/FORS2 data.
           The $V$ band measurements were performed on HST/WFPC2 data.}
 \label{tab:n3585gcphot}
        {\tiny
        \begin{center}
        \begin{tabular}{lcccccr}
         \hline
         \noalign{\smallskip}
cluster&  RA (J2000)  &  DEC (J2000)  & $B$ & $V$ & $I$ & $v_r$ [km/s] \\
\noalign{\smallskip}
\hline
\noalign{\smallskip}
      3585m1GC01& 168.37447& $-$26.76739&$23.12\pm0.02$&       \dots  &$20.89\pm0.01$&$1329\pm36$ \\
      3585m1GC03& 168.34747& $-$26.80268&$21.67\pm0.01$&       \dots  &$20.00\pm0.01$&$1355\pm52$ \\
      3585m1GC04& 168.37122& $-$26.75289&$22.73\pm0.01$&       \dots  &$20.51\pm0.01$&$1416\pm27$ \\
      3585m1GC05& 168.35960& $-$26.76735&$23.05\pm0.02$&       \dots  &$20.93\pm0.01$&$1569\pm40$ \\
      3585m1GC07& 168.35529& $-$26.76541&$21.78\pm0.01$&       \dots  &$19.57\pm0.01$&$1338\pm13$ \\
      3585m1GC11& 168.33916& $-$26.76138&$22.74\pm0.01$&       \dots  &$20.92\pm0.01$&$1126\pm43$ \\
      3585m1GC12& 168.33437& $-$26.76063&$21.97\pm0.01$&       \dots  &$19.84\pm0.01$&$1540\pm23$ \\
      3585m1GC13& 168.32205& $-$26.76943&$22.99\pm0.02$&       \dots  &$20.81\pm0.01$&$1358\pm35$ \\
      3585m1GC14& 168.32941& $-$26.75135&$23.21\pm0.02$&       \dots  &$21.12\pm0.01$&$1107\pm36$ \\
      3585m1GC15& 168.32906& $-$26.74574&$22.61\pm0.01$&$22.38\pm0.02$&$20.94\pm0.01$&$1411\pm43$ \\
      3585m1GC16& 168.31792& $-$26.76046&$22.53\pm0.01$&       \dots  &$20.37\pm0.01$&$1103\pm33$ \\
      3585m1GC18& 168.31964& $-$26.74491&$23.20\pm0.02$&$22.94\pm0.04$&$21.28\pm0.01$&$1477\pm41$ \\
      3585m1GC19& 168.30736& $-$26.76420&$22.70\pm0.01$&$22.15\pm0.02$&$20.55\pm0.01$&$1401\pm38$ \\
      3585m1GC20& 168.32455& $-$26.72389&$22.81\pm0.02$&       \dots  &$20.57\pm0.01$&$1697\pm21$ \\
      3585m1GC21& 168.30592& $-$26.74966&$23.02\pm0.02$&$22.68\pm0.03$&$21.26\pm0.01$&$1513\pm51$ \\
      3585m1GC23& 168.30353& $-$26.74171&$23.37\pm0.02$&$23.13\pm0.04$&$21.64\pm0.02$&$1412\pm61$ \\
      3585m1GC24& 168.31082& $-$26.72189&$23.18\pm0.02$&       \dots  &$21.48\pm0.02$&$1419\pm59$ \\
      3585m1GC26& 168.29779& $-$26.73590&$23.20\pm0.02$&$22.52\pm0.03$&$21.03\pm0.01$&$1137\pm30$ \\
      3585m1GC30& 168.29041& $-$26.72560&$23.21\pm0.02$&       \dots  &$21.57\pm0.02$&$1113\pm69$ \\
      3585m1GC33& 168.28152& $-$26.71842&$23.40\pm0.03$&       \dots  &$21.20\pm0.01$&$1326\pm54$ \\
\noalign{\smallskip}
\hline
\end{tabular}
\end{center}
}
\end{table*}

%% file: n5846.final_BVRIK.tex
\begin{table*}[h!]
  \caption{Photometry for {\bf NGC~5846} globular clusters. Optical photometry was performed on VLT/FORS2 data.
           Near-infrared $K$ band magnitudes were measured on VLT/ISAAC data and taken from \cite{hempel02}.}
 \label{tab:n5846gcphot}
        {\tiny
        \begin{center}
        \begin{tabular}{lcccccccr}
         \hline
         \noalign{\smallskip}
cluster&  RA (J2000)  &  DEC (J2000)  & $B$ & $V$ & $R$ & $I$ & $K$ & $v_r$ [km/s] \\
\noalign{\smallskip}
\hline
\noalign{\smallskip}
      5846m1GC03& 226.60867&   1.55027&$23.50\pm0.02$&$22.56\pm0.02$&$22.11\pm0.02$&$21.54\pm0.02$&       \dots  &$1264\pm49$ \\
      5846m1GC04& 226.58614&   1.57500&$22.66\pm0.01$&$21.51\pm0.01$&$20.90\pm0.01$&$20.13\pm0.01$&       \dots  &$1848\pm22$ \\
      5846m1GC05& 226.58652&   1.57532&$23.97\pm0.02$&$23.04\pm0.03$&$22.43\pm0.03$&$21.81\pm0.02$&       \dots  &$1691\pm49$ \\
      5846m1GC06& 226.59790&   1.56952&$24.07\pm0.03$&$23.03\pm0.03$&$22.38\pm0.02$&$21.63\pm0.02$&       \dots  &$1923\pm53$ \\
      5846m1GC08& 226.60419&   1.57618&$22.17\pm0.01$&$21.15\pm0.01$&$20.62\pm0.01$&$19.92\pm0.01$&       \dots  &$1602\pm22$ \\
      5846m1GC10& 226.60513&   1.58633&$24.61\pm0.04$&$23.82\pm0.05$&$23.32\pm0.06$&$22.64\pm0.05$&       \dots  &$1880\pm75$ \\
      5846m1GC11& 226.60857&   1.58853&$23.23\pm0.01$&$22.22\pm0.01$&$21.71\pm0.01$&$21.07\pm0.01$&       \dots  &$2166\pm33$ \\
      5846m1GC16& 226.61493&   1.60071&$22.71\pm0.01$&$21.65\pm0.01$&$21.11\pm0.01$&$20.35\pm0.01$&       \dots  &$1921\pm30$ \\
      5846m1GC17& 226.62686&   1.59675&$22.79\pm0.01$&$21.80\pm0.01$&$21.29\pm0.01$&$20.61\pm0.01$&$18.94\pm0.02$&$1918\pm45$ \\
      5846m1GC18& 226.63960&   1.58716&$22.50\pm0.01$&$21.58\pm0.01$&$21.11\pm0.01$&$20.49\pm0.01$&$19.06\pm0.02$&$2090\pm29$ \\
      5846m1GC20& 226.62782&   1.60650&$23.43\pm0.01$&$22.80\pm0.02$&$22.29\pm0.02$&$21.52\pm0.02$&$19.86\pm0.04$&$1688\pm38$ \\
      5846m1GC21& 226.63219&   1.60597&$22.46\pm0.01$&$21.41\pm0.01$&$20.98\pm0.01$&$20.34\pm0.01$&$18.68\pm0.01$&$1391\pm28$ \\
      5846m1GC22& 226.64700&   1.59792&$23.15\pm0.01$&$22.12\pm0.01$&$21.62\pm0.01$&$20.92\pm0.01$&$19.31\pm0.02$&$1849\pm36$ \\
      5846m1GC23& 226.62381&   1.62037&$23.26\pm0.01$&$22.35\pm0.01$&$21.88\pm0.02$&$21.30\pm0.01$&$20.01\pm0.05$&$1576\pm50$ \\
      5846m1GC24& 226.63728&   1.61347&$22.26\pm0.01$&$21.18\pm0.01$&$20.64\pm0.01$&$19.91\pm0.01$&$18.10\pm0.01$&$1459\pm18$ \\
      5846m1GC25& 226.63792&   1.61555&$24.16\pm0.03$&$23.10\pm0.03$&$22.53\pm0.03$&$21.83\pm0.02$&$20.18\pm0.06$&$1401\pm62$ \\
      5846m1GC26& 226.63879&   1.61653&$23.44\pm0.02$&$22.58\pm0.02$&$22.11\pm0.02$&$21.56\pm0.02$&$20.20\pm0.06$&$1858\pm61$ \\
      5846m1GC27& 226.62816&   1.62800&$22.74\pm0.01$&$21.60\pm0.01$&$21.00\pm0.01$&$20.24\pm0.01$&$18.18\pm0.01$&$1932\pm22$ \\
      5846m1GC28& 226.63023&   1.62971&$22.56\pm0.01$&$21.49\pm0.01$&$20.93\pm0.01$&$20.21\pm0.01$&$18.39\pm0.01$&$2018\pm31$ \\
      5846m1GC29& 226.64372&   1.62213&$23.67\pm0.02$&$22.72\pm0.02$&$22.27\pm0.02$&$21.60\pm0.02$&$20.17\pm0.05$&$1600\pm46$ \\
      5846m1GC30& 226.64409&   1.62263&$22.44\pm0.01$&$21.43\pm0.01$&$20.91\pm0.01$&$20.22\pm0.01$&$18.53\pm0.01$&$1732\pm25$ \\
      5846m1GC31& 226.65656&   1.61735&$23.44\pm0.02$&$22.37\pm0.01$&$21.79\pm0.01$&$21.04\pm0.01$&       \dots  &$2009\pm26$ \\
      5846m1GC32& 226.64847&   1.62785&$22.52\pm0.01$&$21.64\pm0.01$&$21.15\pm0.01$&$20.53\pm0.01$&       \dots  &$2073\pm34$ \\
      5846m1GC33& 226.65421&   1.62574&$24.35\pm0.03$&$23.57\pm0.04$&$23.06\pm0.04$&$22.42\pm0.04$&       \dots  &$1543\pm100$ \\
      5846m1GC36& 226.65306&   1.63870&$23.06\pm0.01$&$21.97\pm0.01$&$21.38\pm0.01$&$20.63\pm0.01$&       \dots  &$1838\pm26$ \\
      5846m1GC37& 226.65140&   1.64578&$23.82\pm0.02$&$22.81\pm0.02$&$22.21\pm0.02$&$21.50\pm0.02$&       \dots  &$1644\pm49$ \\
      5846m1GC38& 226.65915&   1.64533&$22.47\pm0.01$&$21.55\pm0.01$&$21.03\pm0.01$&$20.41\pm0.01$&       \dots  &$1537\pm41$ \\
      5846m1GC39& 226.66132&   1.64462&$21.99\pm0.01$&$20.82\pm0.01$&$20.23\pm0.01$&$19.42\pm0.01$&       \dots  &$1426\pm23$ \\
\noalign{\smallskip}
\hline
\end{tabular}
\end{center}
}
\end{table*}

\begin{table*}[h!]
  \caption{Photometry for {\bf NGC~7192} globular clusters. Optical photometry was performed on VLT/FORS2 data.
           Near-infrared $K$ band magnitudes were measured on VLT/ISAAC data and taken from \cite{hempel02}.}
 \label{tab:n7192gcphot}
        {\tiny
        \begin{center}
        \begin{tabular}{lcccccccr}
         \hline
         \noalign{\smallskip}
cluster&  RA (J2000)  &  DEC (J2000)  & $B$ & $V$ & $R$ & $I$ & $K$ & $v_r$ [km/s] \\
\noalign{\smallskip}
\hline
\noalign{\smallskip}
      7192m1GC02& 331.66461& $-$64.35384&$24.20\pm0.05$&$23.17\pm0.03$&$22.73\pm0.02$&$22.34\pm0.04$&       \dots  &$3108\pm96$ \\
      7192m1GC15& 331.70432& $-$64.33788&$23.16\pm0.02$&$22.42\pm0.02$&$21.85\pm0.01$&$21.41\pm0.02$&$19.91\pm0.04$&$2785\pm43$ \\
      7192m1GC16& 331.70166& $-$64.33168&$22.79\pm0.01$&$21.97\pm0.01$&$21.39\pm0.01$&$20.86\pm0.01$&$19.30\pm0.02$&$2956\pm28$ \\
      7192m1GC18& 331.70337& $-$64.32433&$23.65\pm0.03$&$22.97\pm0.03$&$22.36\pm0.01$&$21.93\pm0.03$&$20.46\pm0.07$&$2917\pm60$ \\
      7192m1GC19& 331.70413& $-$64.32026&$23.65\pm0.03$&$23.41\pm0.04$&$22.96\pm0.02$&$22.19\pm0.03$&$20.46\pm0.07$&$2895\pm62$ \\
      7192m1GC20& 331.71234& $-$64.32143&$22.69\pm0.01$&$21.89\pm0.01$&$21.25\pm0.01$&$20.69\pm0.01$&       \dots  &$2713\pm26$ \\
      7192m1GC21& 331.72278& $-$64.32507&$23.77\pm0.04$&$22.84\pm0.02$&$22.15\pm0.01$&$21.55\pm0.02$&       \dots  &$2915\pm31$ \\
      7192m1GC22& 331.70593& $-$64.30412&$23.86\pm0.04$&$23.09\pm0.03$&$22.44\pm0.01$&$21.97\pm0.03$&$20.58\pm0.07$&$2975\pm56$ \\
      7192m1GC23& 331.72015& $-$64.31181&$23.15\pm0.02$&$22.42\pm0.02$&$21.85\pm0.01$&$21.43\pm0.02$&       \dots  &$2599\pm55$ \\
      7192m1GC31& 331.75983& $-$64.30913&$23.44\pm0.03$&$22.76\pm0.02$&$22.16\pm0.01$&$21.76\pm0.02$&       \dots  &$3040\pm55$ \\
\noalign{\smallskip}
\hline
\end{tabular}
\end{center}
}
\end{table*}